\documentclass[a4paper,11pt]{article}
\pdfoutput=1 
\usepackage{jheppub}
\usepackage{amssymb,amsmath,verbatim,mathtools,needspace,enumitem,graphicx,physics,mathrsfs}
\usepackage{gensymb}
\usepackage{appendix}
\usepackage{tensor}
\usepackage{array}
\usepackage[normalem]{ulem}
\usepackage[dvipsnames, usenames]{xcolor}
\usepackage{xr-hyper}
\definecolor{linkcolor}{rgb}{0.0,0.3,0.5}
\definecolor{myred}{RGB}{179, 27, 27}
\usepackage[utf8]{inputenc}
\usepackage{hyperref}
\hypersetup{colorlinks=true,citecolor=myred,linkcolor=myred,urlcolor=myred,breaklinks}
\usepackage[all]{hypcap}
\usepackage[T1]{fontenc}
\usepackage{orcidlink}

\newcommand{\tho}{\text{\th}}
\newcommand{\edt}{\text{\dh}}
\newcommand{\thop}{\text{\th}^\prime}
\newcommand{\edtp}{{\text{\dh}^\prime}}
\newcommand{\cA}{\mathcal{A}}

\newcommand{\cK}{\mathcal{K}}

\newcommand{\GHPweight}{\overset{\circ}{=}}
\newcommand{\Rpp}{\overset{\scriptscriptstyle ++}{\mathcal{R}}}
\newcommand{\Rpm}{\overset{\scriptscriptstyle +-}{\mathcal{R}}}

\title{Perturbations of Plane Waves and \\ Quadratic Quasinormal Modes on the Lightring}

\author[a]{Kwinten Fransen\orcidlink{0000-0002-0919-9971}}
\author[b,c]{David Pereñiguez\orcidlink{0000-0002-1007-4551}}
\author[b]{Jaime Redondo-Yuste\orcidlink{0000-0003-3697-0319}}
\affiliation[a]{Walker Burke Institute for Theoretical Physics, California Institute of Technology, Pasadena, CA 91125, USA}
\affiliation[b]{Center of Gravity, Niels Bohr Institute, Blegdamsvej 17, 2100, Copenhagen, Denmark}
\affiliation[c]{William H. Miller III Department of Physics and Astronomy, Johns Hopkins
University, 3400 North Charles Street, Baltimore, Maryland, 21218, USA }

\emailAdd{kfransen@caltech.edu}
\emailAdd{david.pereniguez@nbi.ku.dk}
\emailAdd{jaime.redondo.yuste@nbi.ku.dk}

\abstract{We study second order gravitational perturbations on plane wave spacetimes from both the metric and curvature perturbation points of view. For the former, we explicitly use the isometries of the background to introduce tensor oscillator harmonics, which render Einstein's equations algebraic around symmetric plane waves. For the latter, we formulate the first and second order Teukolsky equations in a Geroch-Held-Penrose covariant way. Both approaches are useful in their own right, and together with our discussion on gauge freedom, they provide a foundation for the study of higher-order gravitational dynamics around plane wave spacetimes. Taking the perspective that these plane wave spacetimes arise from Penrose limits, we subsequently use these results to explore the nonlinear gravitational dynamics close to black hole lightrings. Specifically, we define and discuss quadratic quasinormal mode ratios, observe that they satisfy emergent selection rules, and make publicly available a code to compute them.
}

\begin{document} 
\maketitle
\flushbottom

\section{Introduction}

Gravity at its most interesting is highly non-linear. One effective strategy to explore these non-linearities is to consider small deformations of simple, highly symmetric solutions and proceed perturbatively. 
This method is especially effective when a problem features several distinct perturbative regimes. In such cases, the fully non-linear analysis, though indispensable for quantitative results, often reduces to interpolating between these regimes where the underlying physics is well understood.

The binary black hole problem is a prime example of observational interest~\cite{LIGOScientific:2016aoc,TianQin:2015yph,ET:2019dnz,Badurina:2019hst,LISA:2022yao}. Our understanding of this problem largely comes from combining different approaches during the inspiral, including perturbation theory around flat space~\cite{Goldberger:2004jt,Bern:2019nnu,Porto:2016pyg} and a post-Newtonian expansion~\cite{Blanchet:2013haa}, with perturbation theory around isolated black hole backgrounds to describe the post-merger emission~\cite{Buonanno:2006ui,Cardoso:2008bp,Barack:2018yvs,Pound:2021qin, Berti:2025hly}. Numerical simulations then provide a way to interpolate between these two regimes and build complete waveform models~\cite{Baumgarte:2002jm,Husa:2015iqa,Blackman:2017pcm,Rashti:2024yoc}.

Capturing non-linear effects perturbatively is most straightforward when the linearized problem is itself simple or under excellent control.
On this account, the perturbation theories on curved space that are best understood
are those on Lorentzian symmetric spaces: Minkowski~\cite{Elvang:2015rqa, Travaglini:2022uwo}, Anti-de Sitter (AdS)~\cite{Liu:1998th,Aharony:2016dwx,Rastelli:2017udc,Giombi:2017hpr,Carmi:2018qzm,Jepsen:2019svc,Eberhardt:2020ewh,Rostworowski:2017ruj, Rostworowski:2017tcx, Dias:2016ewl, Dias:2017tjg, Choptuik:2017cyd}, and de Sitter (dS) spacetimes~\cite{Arkani-Hamed:2015bza,Benincasa:2022gtd}. Cahen-Wallach plane waves form another class of Lorentzian symmetric spaces~\cite{cahen1970lorentzian}. More general plane wave spacetimes, including the Cahen-Wallach family, also 
%These as well as more general plane wave spacetimes, 
allow for a particularly tractable linear perturbation theory~\cite{Gibbons:1975jb,mason1989ward,Horowitz:1990sr}.

By contrast, the linearized analysis on black hole backgrounds is already technically challenging and rich in structure. As a result, progress on non-linear corrections has been on-going for more than two decades~\cite{Gleiser:1995gx, Gleiser:1998rw, Nicasio:1998aj,Campanelli:1998jv, Abramo:1999gh,Ioka:2007ak,Nakano:2007cj,Okuzumi:2008ej,Brizuela:2009qd, Pazos:2010xf,Loutrel:2020wbw, Ripley:2020xby}. The observational prospects for gravitational waves drive most current advances including: the self-force program for extreme mass-ratio inspirals~\cite{Barack:2018yvs,Pound:2021qin,Wardell:2021fyy,Albertini:2022rfe,Spiers:2023cip}, improved ringdown modeling close to the peak gravitational wave strain~\cite{Sberna:2021eui,Redondo-Yuste:2023ipg, May:2024rrg, Zhu:2024dyl} (see also~\cite{Berti:2025hly} and references therein), and the excitation of quadratic quasinormal modes (QQNMs)~\cite{Cheung:2022rbm, Mitman:2022qdl, Redondo-Yuste:2023seq, Perrone:2023jzq, Bucciotti:2024jrv, Bucciotti:2024zyp, Bourg:2024jme, Bourg:2025lpd, Zhu:2024rej, Bucciotti:2025rxa, Ma:2024qcv, Khera:2024bjs}.

QQNMs represent the leading non-linear corrections to the quasinormal modes -- the characteristic resonances governing black hole relaxation to equilibrium. It has long been known that, for asymptotically flat black holes, quasinormal modes are in close relation with the bound null geodesics (the ``lightring''), 
%have a close relation to the bound null geodesics of the ``lightring'', 
a correspondence  which can be made precise in the eikonal or high (real) frequency limit~\cite{Press:1971wr,Goebel:1972,Mashhoon:1985cya,Schutz:1985km, Nollert:1999ji, Cardoso:2008bp, Berti:2025hly}. It is then a natural question how this correspondence extends to QQNMs.

From the perspective of observations, the high-frequency limit provides a link between dynamical gravitational wave emission during the ringdown and lightring measurements through electromagnetic observations with the Event Horizon Telescope~\cite{EventHorizonTelescope:2022xqj,Chael:2021rjo} or the proposed Black Hole Explorer~\cite{Johnson:2024ttr, Lupsasca:2024xhq}. While typical binary mergers lack strong high-frequency gravitational wave content~\cite{Davis:1971gg}, special configurations -- such as hierarchical triples -- may enhance these features~\cite{Cardoso:2021vjq}. The remarkable effectiveness of the eikonal approximation even for low-lying quasinormal modes motivates that certain qualitative features may persist also for QQNMs in the sector relevant for gravitational wave observations.  
%Regardless, the eikonal approximation works well even for relatively low-lying quasinormal modes and one may hope that certain qualitative features persist also for quadratic quasinormal modes in the sector relevant for gravitational wave observations.  

More recently, it has been appreciated that a version of the eikonal limit and the corresponding quasinormal modes can be captured by the physics of plane waves~\cite{Fransen:2023eqj,Kapec:2024lnr}. The reason is that these plane waves appear in the limit of a geometry near null geodesics; a construction formalized by Penrose~\cite{penrose1976any}. In particular, the Penrose limit near the lightring of a black hole maps the local curvature dynamics -- directly tied to high-frequency quasinormal modes -- to the dynamics of perturbations around a plane wave.

Moreover, for equatorial lightrings, the plane waves capturing the eikonal quasinormal modes are precisely in the class of Lorentzian symmetric spaces. It is thus reasonable to expect that a perturbation theory can be developed comparable in detail to those on AdS and dS backgrounds which, at least to some approximation, can be connected to an interesting aspect of black hole mergers. It is this that we set out to do here more systematically.

Beyond QQNMs, plane wave spacetimes have proven useful in diverse contexts, suggesting a broader value for a systematic perturbative framework. Such applications include vacuum polarization in curved spacetime~\cite{Hollowood:2008kq,Hollowood:2009qz,Stanley:2011kz}, the physics of tidal excitations of relativistic particles or strings~\cite{Giddings:2007bw,Dodelson:2020lal,Martinec:2020cml,Heidmann:2025yzd} as well as caustics and lensing phenomena~\cite{Harte:2012jg,Harte:2012uw,Harte:2015ila}.

Various applications in high energy physics, notably string theory on non-trivial backgrounds~\cite{Lee:2002rm,Spradlin:2002rv,Lee:2002vz,Pankiewicz:2002gs,Pankiewicz:2003pg,Son:2003zv,Eberhardt:2018exh} and specific limits of the AdS/CFT correspondence~\cite{Berenstein:2002jq,Plefka:2003nb,Mann:2003qp,Russo:2004kr} already implicitly contain a wealth of information on higher order gravitational interactions in plane wave spacetimes, albeit in more specific contexts. Similarly, perturbative non-linearities have been explored using scattering amplitude techniques in, for instance,~\cite{Adamo:2017nia,Adamo:2017sze,Adamo:2018mpq,Adamo:2022rmp}. However, at least from the perspective of QQNMs these studies make some sub-optimal choices, such as restricting to so-called ``sandwich'' plane waves and using a ``coherent state'' basis (instead of a ``number'' basis), which prevent a direct application to the quasinormal mode problem, although it would be interesting to revisit them with an eye on bringing them closer to observational black hole physics. 

Naturally, other simplified approaches to the problem of higher-order black hole perturbation theory have also been taken in the literature. For instance, re-expanding around flat space after setting up an EMRI effective field theory framework~\cite{Cheung:2023lnj,Wilson-Gerow:2023syq,Cheung:2024byb,Wilson-Gerow:2025xhr}, making a flat space approximation to a near-horizon region~\cite{Gaddam:2020rxb,Gaddam:2020mwe,Gaddam:2022pnb}, or restricting to the near-horizon region of near-extremal Kerr~\cite{Yang:2014tla}. The latter approaches in particular have much in common with our Penrose limit perspective as they rely on the simplifications due to an emergent symmetry while still capturing an asymptotic branch of quasinormal modes. However, contrary to the near-horizon, near-extremal limit, the lightring Penrose limit can be taken for arbitrary black holes.

In this work, we develop the framework to study second-order perturbations around plane waves. In particular we propose a partial gauge and frame fixing based on the geometry of algebraically type N spacetimes which we dub geodesic, parallel, and transverse (GPT) gauge. This gauge and frame fixing guarantees that the amplitudes of QQNMs on the Weyl scalar are physically meaningful, and invariant under the residual gauge freedom. We also present a complementary approach for second-order perturbations restricted to homogeneous plane waves, based on the construction of metric harmonics. Both frameworks are used to study the excitation of QQNMs in homogeneous plane waves arising as Penrose limits of the equatorial lightring of Kerr. We compute the ratio describing the excitability of QQNMs for arbitrary mode-coupling combinations, including all overtone indices, and uncover a set of selection rules, inherited from the algebraic properties of the background. Special cases of these ratios have also been discussed recently in Refs.~\cite{Kehagias:2025ntm,Perrone:2025zhy}. 

The remainder of the paper is structured as follows. First, in section~\ref{sec:background}, we review plane wave spacetimes, and discuss how certain classes of plane waves emerge as approximate descriptions of lightrings for black hole spacetimes. We emphasize that the approximation in particular captures an eikonal branch of quasinormal modes. Next, in section~\ref{sec:curvature} we discuss perturbation theory up to second order, including a novel geometric gauge and frame-fixing approach on plane waves. Then we further restrict to homogeneous plane waves, specifically those describing equatorial lightring physics. We discuss mode solutions for this case, establishing their relationship with the simple and inverted harmonic oscillator in section~\ref{sec:homogeneous}, as well as discussing an alternative approach based on metric perturbations. Finally, we discuss the excitation of QQNMs, first for a nonlinear scalar toy model, and then for the full case of Einstein gravity in section~\ref{sec:QQNMs}.

\textbf{Conventions.} We use geometric units, $G=c=1$, and mostly follow the conventions in the monograph by Penrose and Rindler~\cite{Penrose:1985bww}. In particular we follow a mostly minus metric signature $(+,-,-,-)$ but, contrary to~\cite{Penrose:1985bww}, we take the Riemann tensor's sign given by $[\nabla_{a},\nabla_{b}]V_{c}=R_{abcd}V^{d}$. The latter choice ensures this sign convention matches other well-established references (such as the original papers on the Newman-Penrose and Geroch-Held-Penrose formalisms~\cite{Newman:1961qr,Geroch:1973am}) and monographs (such as Wald's~\cite{Wald:1984rg} and Chandrasekar's books~\cite{Chandrasekhar:1985kt}).

%%%%%%%%%%%%%%%%%%%%%%%%%%%%%%%%%%%%%%%%%%%%%%%%%%%%%%%%%%%%%%
\section{Plane Waves}\label{sec:background}
%%%%%%%%%%%%%%%%%%%%%%%%%%%%%%%%%%%%%%%%%%%%%%%%%%%%%%%%%%%%%%

In this work we consider nonlinear fluctuations about a family of background spacetimes that belong to the class of \textit{plane waves}. In Brinkmann coordinates, the line element reads 
\begin{equation}\label{ppWave}
    ds^2 = -A_{ij}(u)x^i x^j du^2 + 2dudv - dx^2-dy^2 \, .
\end{equation}
where $x^i, x^j \in \left\lbrace x,y \right\rbrace$ and $A_{ij}$ is an arbitrary symmetric matrix depending on the coordinate $u$. A more general discussion of parallel, plane-fronted waves can be found in~\cite{Stephani:2003tm}. That the spacetime metric \eqref{ppWave} indeed describes a type of parallel, plane-fronted wave can be seen as follows. First, the metric takes the form of flat space (in double null coordinates) plus a gravitational deformation $\sim du^{2}$. In addition, the null vector field $\ell^{a}=\left(\partial_{v}\right)^{a}$ is covariantly-constant (or \textit{parallel}),
\begin{equation}
\nabla_{a}\ell_{b}=0\, ,
\end{equation}
so it is also a Killing vector field ($\nabla_{(a}\ell_{b)}=0$). This shows that the gravitational deformation $\sim du^{2}$ propagates at the speed of light along $\ell^{a}$. Finally, eq.~\eqref{ppWave} is plane-fronted in the sense that transverse surfaces of constant $u,v$ are flat. It should be stressed, though, that the spacetimes~\eqref{ppWave} are not asymptotically flat, since the gravitational deformation extends infinitely along $\ell^{a}$ (a discussion on the causal structure of $pp$-waves can be found in~\cite{Hubeny:2002zr}). The quadratic dependence in $x$ and $y$ of the deformation $\sim du^{2}$ is sometimes called the \textit{plane-wave} condition. It allows us to interpret the spacetime \eqref{ppWave} quite literally as a two-dimensional (potentially time-dependent) harmonic oscillator. This is seen easily in the case that $A_{ij}(u)=A_{ij}$ is a constant matrix. In that case, as it is also symmetric, without loss of generality it can also be taken to be diagonal. Denoting the eigenvalues to be $-\Omega^2$ and $\Lambda^2$, the metric has the following (complex-valued) Killing vectors,  
\begin{align}\label{eqn:genisometries}
a_{\pm}=\frac{e^{\pm i \Omega u}}{\sqrt{2 \Omega}}\left(\pm i \Omega x \partial_{v}+\partial_{x}\right)\,,\quad b_{\pm}=\frac{e^{\mp \Lambda u}}{\sqrt{2 \Lambda}}\left(\mp \Lambda y \partial_{v}+\partial_{y}\right)\, , \quad \bold{1}=i\partial_{v}\, .
\end{align}
These Killing vectors span the Heisenberg algebras of the simple and inverted harmonic oscillators (reviewed in appendix~\ref{app:SHOIHO}), 
\begin{equation}
[a_{-},a_{+}]=\bold{1}\, , \quad [b_{-},b_{+}]=i \bold{1}\, ,
\end{equation}
where all other commutators vanish. In the case of arbitrary $A_{ij}(u)$ one loses the isometry generated by $\partial_{u}$, but it can be shown that \eqref{ppWave} still enjoys a Heisenberg Killing algebra (see e.g.~\cite{Stephani:2003tm}).
We will distinguish three subcases within the geometry \eqref{ppWave},
\begin{enumerate}[label=(\roman*)]
    \item \textit{General}: $\delta^{ij} A_{ij}(u) \neq 0$ and $A_{ij}(u)\ne\text{constant}$.
    \item \textit{Vacuum}: $\delta^{ij}A_{ij}(u) = 0$,
    \item \textit{Homogeneous}: $A_{ij}(u)=\text{constant}$.\footnote{Note that this class should more properly be referred to as symmetric (or Cahen-Wallach~\cite{cahen1970lorentzian}), homogeneous plane waves being slightly more general~\cite{Blau:2002js}. However, we believe that the analysis we do for this case would generalize to the broader class of homogeneous plane waves, although at an algebraic cost which we avoid in this work.}
\end{enumerate}
Whenever the vacuum condition holds, then eq.~\eqref{ppWave} is a Ricci-flat geometry for any $u$-dependence of $A_{ij}(u)$. The homogeneous case will still be of interest even when it is not a solution of vacuum Einstein's equations, as it involves an additional Killing vector containing the $u$-direction. However, it is the vacuum and homogeneous case with $\partial_u$ a Killing vector that will be of most relevance in this work.

In general, plane-wave spacetimes emerge naturally in several physical scenarios. Most importantly for our purposes, the geometries \eqref{ppWave} appear as \textit{Penrose limits}~\cite{Blau:2006ar} as reviewed next.

%%%%%%%%%%%%%%%%%%%%%%%%%%%%%%%%%%%%%%%%%%%%%%%%%%%%%%%%%%%%%%
\subsection{Penrose Limits}\label{sec:PenroseLimits}
%%%%%%%%%%%%%%%%%%%%%%%%%%%%%%%%%%%%%%%%%%%%%%%%%%%%%%%%%%%%%%

In a precise sense, the spacetime in the neighborhood of a null geodesic $\gamma$ resembles a gravitational plane wave. This observation follows after performing a specific ``zoom-in'' of the spacetime close to $\gamma$, known as a \textit{Penrose limit}~\cite{penrose1976any}. This is introduced by first writing the spacetime metric in Penrose coordinates $(u,V, Y^i)$ adapted to the geodesic,\footnote{Here we follow~\cite{Blau:2006ar}, where one can also find a description of the Penrose limit based on null Fermi normal coordinates.}
\begin{equation}
    d\text{s}^2 = 2du dV + a(u,V,Y^{i}) dV^2 + 2b_i(u,V,Y^{i})  dV dY^i + g_{ij}(u,V,Y^{i}) dY^idY^j \, , \qquad (i,j=1,2)\, ,
\end{equation}
where $\partial_{u}=\nabla V$ is tangent to a twist-free null geodesic congruence. The latter can always be taken such that $\gamma$ corresponds to the curve $(u,V=0,Y^i=0)$. Next, the idea consists in performing a ``zoom-in'' towards $\gamma$ while ``blowing-up'' the spacetime metric in such a way that one recovers a finite, well-defined geometry. Precisely, this is realised by performing an inhomogeneous, constant coordinate re-scaling
\begin{equation}
    (u,V,Y^i) = (u, \lambda^2 \tilde{V}, \lambda \tilde{Y}^i) \, , \quad \lambda\in\mathbb{R} \, ,
\end{equation}
and then taking the limit
\begin{equation}\label{eqn:Rosen}
    d s^2_{\gamma} \coloneqq \lim_{\lambda\to 0}\lambda^{-2}d\text{s}^2 = 2dud\tilde{V} + g_{ij}(u,0,0)d\tilde{Y}^i d\tilde{Y}^j \, .
\end{equation}
We say that $d s^2_{\gamma}$, obtained this way, is the Penrose limit of the original metric $d\text{s}^{2}$ along $\gamma$. Given that the Ricci tensor is invariant under constant conformal transformations, $R_{ab}(g)=R_{ab}(\lambda^{2}g)$, it is guaranteed that if $d \text{s}^2$ is Ricci-flat, then so are its Penrose limits $d s^2_{\gamma}$. The coordinates $(u,\tilde{V},\tilde{Y}^{i})$ are known as \textit{Einstein--Rosen} coordinates for the plane wave. These have the advantage of exhibiting manifestly many of the symmetries of spacetime but, in general, they are not globally defined. Instead, we choose to work in \textit{Brinkmann coordinates} $(u,v,x^{i})$, which preserve $\gamma$ as the curve $(u,v=0,x^{i}=0)$, are global coordinates, and cast the metric in the form of eq.~\eqref{ppWave} (see e.g.~\cite{Blau:2002js} for an explicit construction)
\begin{equation}
    ds^2_{\gamma} = -A_{ij}(u)x^i x^j du^2 + 2dudv - dx^2-dy^2 \, .
\end{equation}
Furthermore, it can be shown that in these coordinates the symmetric matrix $A_{ij}(u)$ is given by~\cite{Blau:2003dz} 
\begin{equation}\label{eq:intrinsicPL}
    A_{ij}(u)=\left. \text{R}_{abcd}E^{a}_{i}E^{b}_{+}E^{c}_{j}E^{d}_{+}\right|_{\gamma(u)}\, ,
\end{equation}
where $\text{R}_{abcd}$ is the Riemann tensor of the \textit{original} spacetime, and $(E_{+}^{a},E^{a}_{-},E^{a}_{i})$ is a null frame where $E^{a}_{+}$ is tangent to a null congruence containing $\gamma$ and $E_{-}^{a},E^{a}_{i}$ are parallelly-propagated along $E^{a}_{+}$. Equation~\eqref{eq:intrinsicPL} shows explicitly and covariantly the amount of information retained about the original spacetime, which is the same tidal information as encoded in the null geodesic deviation equation. In addition, it allows one to think of the Penrose limit formally as a map $(d\text{s}^{2},\gamma)\mapsto A_{ij}(u)$ and yields a powerful way of computing it (especially using the 2-spinor formulation of \eqref{eq:intrinsicPL}, see~\cite{Tod:2019urw,Chawla:2024mse}). For example, the Penrose limit along an equatorial null geodesic in a Kerr black hole reads~\cite{Chawla:2024mse} (see also~\cite{Fransen:2023eqj,Hollowood:2009qz,Papadopoulos:2020qik,Kubiznak:2008zs}),
\begin{equation}\label{eq:PLKerrEquator}
    ds^{2}=-\frac{3 M b^{2}}{r(u)^{5}}\left(y^{2}-x^{2}\right)du^{2}+2dudv-dx^{2}-dy^{2}\,,
\end{equation}
where $M$ is the black hole mass, $r(u)$ is a solution to the Kerr geodesic equation for the radial coordinate, and $b=L/E$ is the geodesic's ``impact parameter'', with $E,L$ its conserved energy and azimuthal angular momentum. In general, this is an inhomogeneous plane wave, but it becomes homogeneous if one chooses the equatorial geodesic to be one of the lightrings, where $r(u)=constant$. In that situation, the coordinates $x$ and $y$ can be thought of as measuring the distance away from the original geodesic along the polar and radial directions, respectively. This case is discussed in detail in Section~\ref{sec:homogeneous}, and is of special importance due to the correspondence between these lightrings and the eikonal regime of QNMs.

%%%%%%%%%%%%%%%%%%%%%%%%%%%%%%%%%%%%%%%%%%%%%%%%%%%%%%%%%%%%%%
\subsection{Geroch-Held-Penrose (GHP) description}\label{sec:GHPppWave}
%%%%%%%%%%%%%%%%%%%%%%%%%%%%%%%%%%%%%%%%%%%%%%%%%%%%%%%%%%%%%%

According to Petrov's classification, the plane wave spacetime~\eqref{ppWave} is of type $N$ where the unique principal null direction (PND) is defined by the parallel null vector $\partial_{v}$. Consequently, the geometry~\eqref{ppWave} has a very simple description in terms of the Geroch-Held-Penrose (GHP) formalism~\cite{Geroch:1973am}, which we review in Appendix~\ref{app:GHPsum}. Indeed, in terms of the null frame 
\begin{equation}\label{NPframe}
    \ell = \partial_v \, , \quad n = \partial_u +\frac{A_{ij}(u)}{2}x^i x^j\partial_v \, , \quad m = \left(\partial_x + i\partial_y\right)/\sqrt{2} \, ,
\end{equation}
all spin coefficients and Weyl scalars vanish except for
\begin{equation}\label{NP_plane_wave}
    \kappa' = \frac{1}{2\sqrt{2}}(\partial_x - i \partial_y)\left(A_{ij}(u)x^i x^j\right) \, ,\qquad \Psi_4 = -\frac{1}{4}(\partial_x - i \partial_y)^2\left(A_{ij}(u)x^i x^j\right) \, ,
\end{equation}
and their complex conjugates. In particular, this means that the directional GHP derivatives are just the ordinary directional derivatives along the null frame \eqref{NPframe}, 
\begin{equation}
    \tho=\ell^{a}\nabla_{a}\,, \quad \tho'=n^{a}\nabla_{a}\,,\quad \eth=m^{a}\nabla_{a}\, ,\quad \eth'=\bar{m}^{a}\nabla_{a}\, .
\end{equation}
Using them, one can easily verify the GHP-covariant expressions,
\begin{equation}\label{eq:GHPCoefs1}
\begin{aligned}
    \tho \kappa' = 0\, , \qquad  \tho \Psi_{4} = 0\, , \qquad \eth' \kappa' = -\Psi_4\, ,
\end{aligned}
\end{equation}
which will be useful in the next sections. Other useful expressions that only hold in the particular frame~\eqref{NPframe} are
\begin{equation}\label{eq:GHPCoefs2}
\begin{aligned}
    \kappa' &= \frac{1}{2}\eth'\left(A_{ij}(u)x^i x^j\right) \, ,\qquad \Psi_4 = -\frac{1}{2}\eth'^{2}\left(A_{ij}(u)x^i x^j\right) \, , \qquad \eth  \kappa' = \frac{1}{2}\delta^{ij} A_{ij}(u)\, ,
\end{aligned}
\end{equation}
and the wave operator is 
\begin{equation}
    \square \equiv \nabla^a \nabla_a = 2(\tho'\tho-\eth'\eth) \, .
\end{equation}
As is often the case in algebraically special spacetimes, the fact that most GHP quantities of the background vanish will drastically simplify the analysis of gravitational fluctuations. This is therefore the approach to perturbation theory that we take next. 

%%%%%%%%%%%%%%%%%%%%%%%%%%%%%%%%%%%%%%%%%%%%%%%%%%
\section{Perturbation theory}\label{sec:curvature}
%%%%%%%%%%%%%%%%%%%%%%%%%%%%%%%%%%%%%%%%%%%%%%%%%%
  
Throughout this section we will consider vacuum plane waves, thus setting $\delta^{ij} A_{ij}(u) = 0$ in eq.~\eqref{ppWave}. Our goal is to describe the propagation of linear and quadratic gravitational fluctuations on such backgrounds. The approach presented here focuses on solving for the fluctuations of the spacetime curvature, rather than the fluctuations of the metric itself, given that these capture the physical degrees of freedom of the gravitational field. In section~\ref{sec:metric} we will instead show how to address the problem from a pure metric perspective in the special case of homogeneous background plane waves. 

Morally, our formalism here can be seen as an adaptation of Teukolsky's original derivation~\cite{Teukolsky:1973ha} in Kerr's space, as well as the work on quadratic fluctuations by Campanelli and Lousto~\cite{Campanelli:1998jv}. However, the derivation proceeds in a different fashion, starting from non-perturbative curvature wave equations and retaining GHP-covariance at each step in the perturbative expansion. In addition, the gauge choices are adapted to the case of plane waves, whose geometric properties often have no analogue in Kerr's space. 

General relativity is the unique theory of gravity where the spacetime curvature (in $4$ dimensions) is described (in vacuum) by only one fully-symmetric 2-spinor $\Psi_{ABCD}$. This is the spinor analogue of Weyl's tensor, and in vacuum it satisfies the non-linear wave equation~\cite{Penrose:1960eq}\footnote{We take a spinorial starting point as it simplifies the structure of the equations but no familiarity with this formulation is needed in the remainder.}
\begin{equation}\label{Penrose_Wave_Equation}
    \Box \Psi_{ABCD} - 6\tensor{\Psi}{^E^F_{(A}_B}\Psi_{CD)EF} = 0 \, .
\end{equation}
In the case of type N spacetimes such as parallel plane waves, and specifically eq.~\eqref{ppWave}, the second term vanishes $\tensor{\Psi}{^E^F_{(A}_B}\Psi_{CD)EF}=0$, so the Weyl spinor satisfies the genuine linear wave equation $ \Box \Psi_{ABCD}=0$. In a general vacuum spacetime, the various projections of eq.~\eqref{Penrose_Wave_Equation} into a
spin dyad $(o^{A},\iota^{A})$ yield wave equations for the usual Weyl scalars $\{\Psi_{i}\}_{i=0}^{4}$, (see Appendix~\ref{app:GHPsum} for further details). In particular, we find
\begin{enumerate}
\item \textit{Equation for $\Psi_{0}$}:
\begin{equation}\label{Nonlinear_Teukolsky_Equation}
    \mathcal{O}_0\Psi_0 + \mathcal{O}_1\Psi_1 + \mathcal{O}_2\Psi_2= 0 \, , 
\end{equation}
with 
\begin{equation}
    \begin{aligned}
            \mathcal{O}_0 =&\tho' \tho - \eth' \eth - \bar{\rho}' \tho - 5 \rho \tho' + \bar{\tau} \eth + 5 \tau \eth' + 4 \sigma \sigma' - 4 \kappa \kappa' - 10 \Psi_2  ,\\
            \mathcal{O}_1=&4 \tho' \kappa - 4 \eth' \sigma - 4(\bar{\rho}' - 2\rho')\kappa + 4(\bar{\tau} - 2\tau')\sigma + 10 \Psi_1,\\
            \mathcal{O}_2=&4 \kappa \eth-4 \sigma \tho  - 12 \kappa \tau + 12 \rho \sigma.
    \end{aligned}
\end{equation}
In obtaining eq.~\eqref{Nonlinear_Teukolsky_Equation}, we first project eq.~\eqref{Penrose_Wave_Equation} on $o^{A}o^{B}o^{C}o^{D}$, then commute GHP operators so that non-primed ones act first, eliminate first derivatives of $\Psi_{1}$ by using Bianchi identities~\eqref{eq:GHPBianchi}, and finally eliminate first derivatives of $\rho$ and $\tau$ by using the Ricci-flatness equations~\eqref{eq:GHPEinstein}.

\item \textit{Equation for $\Psi_{4}$}:
\begin{equation}\label{Nonlinear_Teukolsky_Equation4}
    \mathcal{O}'_0\Psi_4 + \mathcal{O}'_1\Psi_3 + \mathcal{O}'_2\Psi_2= 0 \, , 
\end{equation}
with 
\begin{equation}
    \begin{aligned}
            \mathcal{O}'_0 =&\tho'\tho - \eth'\eth - (4\rho' + \bar{\rho}')\tho - \rho\tho' + (4\tau' + \bar{\tau})\eth + \tau\eth' + 4\rho\rho' - 4\tau\tau' - 2\Psi_2 ,\\
            \mathcal{O}'_1=&4\tho\kappa' - 4\eth\sigma' - 4(\bar{\rho} - 2\rho)\kappa' + 4(\bar{\tau}' - 2\tau)\sigma' + 10\Psi_3,\\
            \mathcal{O}'_2=&4\kappa'\eth'-4\sigma'\tho'  - 12\kappa'\tau' + 12\rho'\sigma'.
    \end{aligned}
\end{equation}
Equation~\eqref{Nonlinear_Teukolsky_Equation4} can be obtained via a similar procedure, now projecting eq.~\eqref{Penrose_Wave_Equation} on $\iota^{A}\iota^{B}\iota^{C}\iota^{D}$, or just by priming eq.~\eqref{Nonlinear_Teukolsky_Equation}. That $\mathcal{O}_{0}'$ is not manifestly the prime of $\mathcal{O}_{0}$ is merely due to our convention of writing the non-primed GHP derivatives acting first.\footnote{In equation \eqref{Nonlinear_Teukolsky_Equation4} we are correcting a typo in the operator $\mathcal{O}_{1}'$ that appears in Refs.~\cite{Stewart:1974uz} and~\cite{Bini:2002jx}.}
\end{enumerate}
Equations~\eqref{Nonlinear_Teukolsky_Equation} and~\eqref{Nonlinear_Teukolsky_Equation4} were first obtained from a different approach by Stewart and Walker~\cite{Stewart:1974uz} and subsequently in Ref.~\cite{Bini:2002jx}, where equations for the remaining Weyl scalars were provided, too (these can also be obtained directly from a suitable projection of \eqref{Penrose_Wave_Equation}). These equations hold exactly on any Ricci-flat spacetime, and rely on neither symmetry assumptions nor perturbative expansions.

Our goal is to solve these equations perturbatively, assuming that the spacetime consists of a generic gravitational deformation of the vacuum plane wave \eqref{ppWave}. To that end, we assume that any quantity $\mathcal{Q}$ admits a formal perturbative expansion 
\begin{equation}\label{eq:pertexp}
\mathcal{Q}=\mathcal{Q}^\circ+\epsilon\dot{\mathcal{Q}}+\frac{1}{2!}\epsilon^{2}\ddot{\mathcal{Q}}+...
\end{equation}
Here $\mathcal{Q}^\circ$ denotes the value of $\mathcal{Q}$ in the plane wave background, $\epsilon \ll 1$ is a perturbative book-keeping parameter (that we omit in general) and the over-dot notation stands for the operators $\dot{\mathcal{Q}}=d\mathcal{Q}/d\epsilon\rvert_{\epsilon=0},\ddot{\mathcal{Q}}=d^2\mathcal{Q}/d\epsilon^2\rvert_{\epsilon=0}$, etc, and should not be confused with a total space or time derivative of $\mathcal{Q}$. We will sometimes abuse the notation and write $\mathcal{Q}^\circ=\mathcal{Q}$ if there is no risk of confusion. 

For concreteness we will focus on eq.~\eqref{Nonlinear_Teukolsky_Equation}, and solve for $\Psi_{0}$ to first and second order in the perturbative expansion, $\dot{\Psi}_{0}$ and $\ddot{\Psi}_{0}$ respectively. Taking into account that the only non-vanishing GHP quantities of our background are $\kappa'$ and $\Psi_4$ given by~\eqref{NP_plane_wave}, eq.~\eqref{Nonlinear_Teukolsky_Equation} expanded at first- and second-order readily gives 
\begin{align}\label{eq:formalfirst}
    \text{First order:}&\quad \left(\tho'\tho-\eth'\eth\right)\dot{\Psi}
_{0}=0\, ,\\\label{eq:formalsecond}
\text{Second order:}&\quad \left(\tho'\tho-\eth'\eth\right)\ddot{\Psi}_{0}=-2\left(\dot{\mathcal{O}}_{0}\dot{\Psi}_{0}+\dot{\mathcal{O}}_{1}\dot{\Psi}_{1}+\dot{\mathcal{O}}_{2}\dot{\Psi}_{2}\right)\, .
\end{align}
At first order, one encounters a decoupled equation for $\dot{\Psi}_{0}$, which is analogous to Teukolsky's equation for $\Psi_{0}$~\cite{Teukolsky:1973ha}. In the second-order equation, the only second-order quantity that appears is $\ddot{\Psi}_{0}$, while the remaining terms are quadratic combinations of first-order quantities that act as source terms for $\ddot{\Psi}_{0}$. Such source terms contain the first variations of the spin coefficients ($\dot{\kappa},\dot{\rho}$, etc), curvature scalars ($\dot{\Psi}_{0},\dot{\Psi}_{1}$, etc), as well as first variations of the directional GHP derivatives ($\dot{\tho},\dot{\eth}$, etc) that we will introduce below as well-defined GHP derivative operators. Hence, solving the second order problem for $\ddot{\Psi}_{0}$ requires not only the knowledge of $\dot{\Psi}_{0}$ at first order, but of several other GHP quantities at first order.

The structure of the remainder of this section is as follows. We start with a discussion of gauge freedom at first and second order in perturbation theory in~\ref{sec:gauge}. In particular, we will introduce a gauge that we refer to as the \textit{geodesic, parallel, and transverse gauge}. Next, in section~\ref{sec:first} we show that the first order problem can be solved entirely only from the knowledge of a solution to the scalar wave equation. This relies on metric reconstruction techniques in algebraically special spaces~\cite{Chrzanowski:1976jy, Wald:1978vm,Campanelli:1998jv, Green:2019nam, Toomani:2021jlo,Loutrel:2020wbw, Hollands:2024iqp}. In section~\ref{sec:second} we discuss the second-order problem, providing an explicit GHP-covariant expression for the source term in eq.~\eqref{eq:formalsecond}. The resulting equation at second order, valid for quadratic fluctuations about the vacuum plane wave eq.~\eqref{ppWave}, is one of our main results and is the basis of our analysis of quadratic quasinormal modes in section~\ref{sec:QQNMs}. 

%%%%%%%%%%%%%%%%%%%%%%%%%%%%%%%%%%%%%%%%%%%%%%%%%%
\subsection{Gauge considerations}\label{sec:gauge}
%%%%%%%%%%%%%%%%%%%%%%%%%%%%%%%%%%%%%%%%%%%%%%%%%%

In the GHP formulation of GR, the usual gauge symmetry due to spacetime diffeomorphisms is enhanced to also account for Lorentz rotations of the null-tetrad. These factorize in three distinct classes,
\begin{equation}\label{eq:Lorentzframerots}
    \begin{aligned}
       \text{Class I:}&\quad \ell \to \ell \, ,  \quad n \to n + a \bar{m} + \bar{a} m +a \bar{a} \ell \ , \quad m \to m+ a \ell \ , \\
       \text{Class II:}&  \quad \ell \to \ell + b \bar{m} + \bar{b} m +b\bar{b} n \ , \quad n \to n \, , \quad m \to m+ b n \ , \\
        \text{Class III:}&\quad \ell \to \Lambda \ell \ , \quad n \to \Lambda^{-1} n \ , \quad m \to e^{i \theta} m \ , \quad \left(\text{with} \quad \Lambda \in \mathbb{R}_{\times}\, , \ \ \theta \in \mathbb{R}\right)\, , 
    \end{aligned}
\end{equation}
where class III corresponds to GHP transformations. Thus, the gauge parameters are a vector field $\xi^{a}$ for diffeomorphisms, two complex functions $a$ and $b$ parametrising rotations of class I and II, and two real functions $\Lambda$ and $\theta$ (with $\Lambda\ne0$) parametrising class III frame transformations. In general, all GHP quantities transform non-trivially under these gauge symmetries (see e.g.~\cite{Chandrasekhar:1985kt}). 

The gauge parameters that respect our choice of background variables and preserve the perturbative scheme \eqref{eq:pertexp} are those admitting an expansion of the form 
\begin{equation}
    \begin{aligned}
        \xi^{a}=&\epsilon \dot{\xi}^{a}+\frac{1}{2!}\epsilon^{2}\ddot{\xi}^{a}+...\, ,\quad a=\epsilon \dot{a}+\frac{1}{2!}\epsilon^{2}\ddot{a}+...\,, \quad b=\epsilon \dot{b}+\frac{1}{2!}\epsilon^{2}\ddot{b}+...\,, \\ \Lambda=&1+\epsilon \dot{\Lambda}+\frac{1}{2!}\epsilon^{2}\ddot{\Lambda}+...\,, \quad \theta=\epsilon \dot{\theta}+\frac{1}{2!}\epsilon^{2}\ddot{\theta}+...
    \end{aligned}
\end{equation}
At linear order, the gauge transformations of the fluctuations of the metric, GHP frame and Weyl scalars read
\begin{align}\label{eq:gaugemetric}
    h_{ab}&\mapsto h_{ab} + 2\nabla_{(a}\dot{\xi}_{b)}\, ,\\
    \dot{\ell}_{a}&\mapsto \dot{\ell}_{a}+\pounds_{\dot{\xi}}\ell_{a}+\dot{\bar{b}}m_{a}+\dot{b}\bar{m}_{a}+\dot{\Lambda}\ell_a\, ,\\
    \dot{n}_{a}&\mapsto \dot{n}_{a}+\pounds_{\dot{\xi}}n_{a}+\dot{\bar{a}}m_{a}+\dot{a}\bar{m}_{a}-\dot{\Lambda}n_a\, ,\\
    \dot{m}_{a}&\mapsto \dot{m}_{a}+\pounds_{\dot{\xi}}m_{a}+\dot{a}\ell_{a}+\dot{b}n_{a}+i\dot{\theta}m_a\, ,\\ \label{eq:psi3gauge}
    \dot{\Psi}_{3}&\mapsto \dot{\Psi}_{3}+\dot{b} \Psi_{4}\, , \\   \label{eq:psi4gauge}\dot{\Psi}_{4}&\mapsto\dot{\Psi}_{4}+\pounds_{\dot{\xi}}\Psi_{4}-2\left(\dot{\Lambda}+i\dot{\theta}\right)\Psi_{4}\, ,
\end{align}
where we have used that all Weyl scalars save for $\Psi_4$ vanish for the type N background, and, in the case of symmetric plane waves, one additionally has $\pounds_{\dot{\xi}}\Psi_{4}=0$. 

The Weyl scalar $\Psi_{0}$ that we focus on in this work is fully gauge-invariant at first order (so are $\Psi_{1}$ and $\Psi_{2}$). However, at second and higher orders it transform non-trivially under the gauge symmetries (see e.g.~\cite{Spiers:2023mor} for a discussion of gauge-symmetry beyond linear order in perturbation theory). At second order, we find 
\begin{equation}\label{eq:ddPsi0gauge}
    \ddot{\Psi}_{0}\mapsto \ddot{\Psi}_{0}+2\pounds_{\dot{\xi}}\dot{\Psi}_{0}+8 \dot{b}\dot{\Psi}_{1}+4(\dot{\Lambda}+i\dot{\theta})\dot{\Psi}_{0}\, .
\end{equation}
In particular, only first-order gauge parameters contribute to the transformation, but not second-order ones $(\ddot{\xi}^{a},\ddot{a},...)$. This is due to the fact that $\dot{\Psi}_{0}$ is itself gauge-invariant. Our approach to deal with gauge symmetry in this work consists in fixing partially the \textit{linear} gauge freedom, by imposing some conditions that are both physically and mathematically well-defined. This ensures that several quantities we compute related to $\ddot{\Psi}_{0}$ are physically meaningful, and invariant under the residual gauge freedom. This applies specially to the QQNM excitation ratios computed in section~\ref{sec:QQNMs}. The aim of this section is to introduce our choice of linear gauge. 

Consider a spacetime that is a vacuum deformation of our plane wave. We chose our frame so that 
$\ell^{a}$ is tangent to a twist free geodesic null congruence, and $n^{a}$ and $m^{a}$ are parallelly-propagated along $\ell^{a}$, that is,
\begin{equation}\label{eq:nonpertframe}
    \ell_{a}=\nabla_{a}u\, , \quad \ell^{a}\nabla_{a}\ell^{b}=\ell^{a}\nabla_{a}n^{b}=\ell^{a}\nabla_{a}m^{b}=0\, ,
\end{equation}
where $u$ is a function satisfying $\nabla^{a}u\nabla_{a}u=0$ (which implies $\ell^{a}\nabla_{a}\ell_{b}=0$ automatically). Notice this frame can always be constructed, at least locally. Thus, eqs.~\eqref{eq:nonpertframe} are exact statements that are valid to all orders, in particular they hold for our background frame~\eqref{NPframe}, and imply the vanishing of several spin coefficients and their fluctuations. With this choice, we will be able to impose three geometric conditions that define our linear gauge, up to certain residual freedom. We describe them next, showing that each condition can always be met for any solution to the linear equations, and identifying the remaining gauge transformations that preserve them.
\begin{enumerate}
    \item \textit{Geodesic condition}: Our first gauge choice is to set
\begin{equation}\label{eq:geodcond}
    \dot{\ell}_{a}=0\, .
\end{equation}
That this is always possible can be seen as follows. Linearising the first equation in~\eqref{eq:nonpertframe}, we find $\dot{\ell}_{a}=\nabla_{a}\dot{u}$, and this can be set to zero through a diffeomorphism with $\dot{\xi}_{v}=-\dot{u}$ (where $\dot{\xi}_{v}=\ell^{a}\dot{\xi}_{a}$). Notice that the choice~\eqref{eq:geodcond} is equivalent to taking the function $u$, which satisfies $\nabla^{a}u\nabla_{a}u=0$ nonperturbatively, as a coordinate. This fact motivates calling eq.~\eqref{eq:geodcond} the geodesic condition.

The gauge parameters that leave~\eqref{eq:geodcond} invariant satisfy
\begin{equation}
    0=\pounds_{\dot{\xi}}\ell_{a}+\dot{\bar{b}}m_{a}+\dot{b}\bar{m}_{a}+\dot{\Lambda}\ell_a\, ,
\end{equation}
and projecting this equation along the null frame, we find 
\begin{equation}\label{eq:parametersgeodesicconditions}
    \partial_{v}\dot{\xi}_{v}=0\, ,\quad \dot{b}=m^{a}\nabla_{a}\left(\dot{\xi}_{v}\right) \, ,\quad \dot{\Lambda}=-n^{a}\nabla_{a}\left(\dot{\xi}_{v}\right)\,.
\end{equation}
That is, the generators of class II and class III rotations $\dot{b}$ and $\dot{\Lambda}$ cannot act independently, but only together with a diffeomorphism whose component $\dot{\xi}_{v}$ is $v$-independent. Finally, we remark that in this gauge $h_{vv}=-(g^{ab}\ell_{a}\ell_{b})\dot{}=0$.

\item \textit{Transverse condition}: The next gauge choice is
\begin{equation}\label{eq:transcond}
    \ell^{a}h_{ab}=0\, .
\end{equation}
To show that this gauge can always be achieved, consider acting on any solution $h_{ab}$ with a diffeomorphism whose component $\dot{\xi}_{v}$ is constant, while the remaining components are generic and denoted $\dot{\xi}_{I}\equiv\left\{\dot{\xi}_{u},\dot{\xi}_{x},\dot{\xi}_{y}\right\}$. This satisfies \eqref{eq:parametersgeodesicconditions} so it preserves \eqref{eq:geodcond}. Then, imposing that the transformed metric satisfies \eqref{eq:transcond} one immediately finds 
\begin{equation}
    0=\ell^{a}(h_{ab}+2\nabla_{(a}\dot{\xi}_{b)})\quad \Longrightarrow \quad \dot{\xi}_{I}=-\int h_{vI}dv\,, \quad \text{with}\quad I=\{u,x,y\}\, .
\end{equation}
In obtaining this, it is useful to notice that in our coordinates $\Gamma^{a}_{bv}=0$, and that since \eqref{eq:geodcond} holds then $h_{vv}=0$. Now, imposing that the gauge parameters leave invariant \eqref{eq:geodcond} (so $\partial_{v}\dot{\xi}_{v}=0$) and \eqref{eq:transcond}, it readily follows that  
\begin{equation}\label{eq:parameterstransverscond}
   \dot{\xi}_{I}=-v\partial_{I}\dot{\xi}_{v}+\dot{C}_{I}\, , \quad \partial_{v}\dot{C}_{I}=0 \,, \quad \text{with}\quad I=\{u,x,y\}\, ,
\end{equation}
where $\dot{C}_{I}$ are $v$-independent functions that arise as integration constants. Thus, the diffeomorphisms allowed by \eqref{eq:geodcond} and \eqref{eq:transcond} are entirely fixed by the $v$-independent functions $\dot{\xi}_{v}$ and $\dot{C}_{I}$ through \eqref{eq:parameterstransverscond}, although we notice that in general $\dot{\xi}_{I}$ can exhibit a linear dependence on $v$. We also remark that in this gauge $\left(\ell^{a}\right)\dot{}=\left(g^{ab}\ell_{b}\right)\dot{}=-h^{ab}\ell_{b}=0$. For a discussion on the transverse gauge condition in more general (type $II$) spaces we refer the reader to~\cite{Price:2006ke}. 

\item \textit{Parallel condition}: The final requirement follows by consistency with our non-perturbative choice eq.~\eqref{eq:nonpertframe}, that the frame is parallelly propagated along $\ell_{a}$. At linear order, that statement reads 
\begin{equation}\label{eq:parallelcond}
    \left(\ell^{b}\nabla_{b}X_{a}\right)\dot{}=\ell^{b}\nabla_{b}\dot{X}_{a}-\dot{\Gamma}^{c}_{\ ab}X_{c}\ell^{b}=0\, ,
\end{equation}
where we used that $\left(\ell^{a}\right)\dot{}=0$ and defined
\begin{equation}
    \dot{\Gamma}^{a}_{\ bc}\equiv\frac{1}{2}g^{ad}\left(\nabla_{b}h_{cd}+\nabla_{c}h_{bd}-\nabla_{d}h_{bc}\right)\, , \quad X_{a}\equiv\left\{\ell_{a},n_{a},m_{a}\right\}\, .
\end{equation}
Assuming that the geodesic and transverse conditions (eqs.~\eqref{eq:geodcond} and~\eqref{eq:transcond}) hold, it can be verified by direct computation that
\begin{equation}\label{frame2}
\begin{aligned}
    \dot{n}^{a}&=\frac{1}{2}\left(-h_{nn}\ell^{a}+h_{n\bar{m}}m^{a}+h_{nm}\bar{m}^{a}\right)\, ,\\
    \dot{m}^{a}&=\frac{1}{2}\left(-h_{nm}\ell^{a}+h_{m\bar{m}}m^{a}+h_{m m}\bar{m}^{a}\right)\, ,
\end{aligned}    
\end{equation}
is parallely propagated along $\ell^a$, i.e., it satisifies eq.~\eqref{eq:parallelcond}, and it is also a well-defined perturbed GHP frame since it satisfies 
\begin{equation}
\begin{aligned}
 h_{ab}=2\left(\ell_{(a}\dot{n}_{b)}-\dot{m}_{(a}\bar{m}_{b)}-m_{(a}\dot{\bar{m}}_{b)}\right)\,.
\end{aligned}
\end{equation}
We notice that this frame can also be obtained by applying to the frame in Ref.~\cite{Campanelli:1998jv} a linear class II rotation with parameter $\dot{a}=(1/2)h_{nm}$. Finally, requiring that the gauge parameters leave invariant the parallel condition eq.~\eqref{eq:parallelcond} only imposes two new constraints,
\begin{equation}\label{eq:parametersparallelcond}
    \partial_{v}\dot{a}=\partial_{v}\dot{\theta}=0\, ,
\end{equation}
as can be readily checked. 
\end{enumerate}
In sum, we have found that at linear order it is always possible to impose the geodesic, parallel and transverse conditions (equations~\eqref{eq:geodcond}, \eqref{eq:transcond} and~\eqref{eq:parallelcond} respectively),
and we should call this the geodesic, parallel and transverse (GPT) gauge. We have also shown that the gauge parameters generating the residual gauge freedom are fixed by the $v$-independent (but otherwise arbitrary) functions $\dot{\xi}_{v},\dot{C}_{I},\dot{a},\dot{\theta}$, through equations~\eqref{eq:parametersgeodesicconditions} and~\eqref{eq:parameterstransverscond}.

The fact that the GPT gauge is based on a frame parallelly-propagated along a null geodesic ensures that the associated curvature scalars are physically meaningful (the choice of a non-inertial null frame could induce unphysical features on the corresponding Weyl scalars). In addition, although the GPT gauge still allows a significant amount of gauge freedom, we will see that the corresponding transformation of $\ddot{\Psi}_{0}$ (given in~\eqref{eq:ddPsi0gauge}) leaves invariant the observables we will focus on in section~\ref{sec:QQNMs}.

%%%%%%%%%%%%%%%%%%%%%%%%%%%%%%%%%
\subsection{First-order perturbations}\label{sec:first}
%%%%%%%%%%%%%%%%%%%%%%%%%%%%%%%%%

In this section we construct a complete linear solution in the GPT gauge. The metric perturbation is obtained by adapting to type $N$ spaces the reconstruction procedure derived in ref.~\cite{Wald:1978vm} for the type $D$ case (given that the techniques are very similar to the original reference, we give the details in Appendix~\ref{app:recmetN} and here we simply report the result). However, our approach to handle variations of GHP quantities differs from previous approaches (e.g.~\cite{Campanelli:1998jv}), so we elaborate more on them here. In addition, we provide a simple new expression relating the fluctuations of the Weyl scalars, which can be seen as a generalisation to our type-$N$ spaces of the Teukolsky--Starobinsky identities.

As explained in the Appendix~\ref{app:recmetN}, given a GHP scalar $\Psi_{H}\GHPweight(-4,0)$ that satisfies the massless wave equation 
\begin{equation}\label{eq:HertzEq}
    \Bigl(\thop\tho-\edtp\edt\Bigr)\Psi_H = 0\, ,
\end{equation}
one can generate a solution for the metric perturbation in the transverse gauge \eqref{eq:transcond} by simply acting on $\Psi_{H}$ with differential operators. We refer to $\Psi_{H}$ as the \textit{Hertz potential}. In turn, from the reconstructed metric perturbation and benefiting from the work done in Section~\ref{sec:gauge}, the complete solution for the GHP quantities in the GPT gauge can be obtained in terms of $\Psi_{H}$. For general plane waves, eq.~\eqref{eq:HertzEq} does not generically admit analytic solutions in Brinkmann coordinates\footnote{Specifically, they would rely on the solutions of an arbitrary second order linear ordinary differential equation.}, but it does in the homogeneous waves as discussed in Section~\ref{sec:homogeneous}. However, the reconstruction procedure works regardless of the nature of the solution to eq.~\eqref{eq:HertzEq}. We list the solution next, where the quantities that vanish are omitted:
\begin{itemize}
    \item \textit{Metric perturbation}: 
\begin{equation}\label{eq:recmet2}
    \begin{aligned}
        h_{ab}=&-\ell_{a}\ell_{b}\left(\eth^{2}\Psi_{H}+\eth'^{2}\bar{\Psi}_{H}\right)+2\ell_{(a}m_{b)}\tho\eth\Psi_{H}+2\ell_{(a}\bar{m}_{b)}\tho\eth'\bar{\Psi}_{H}\\
        &- m_{a}m_{b} \tho^{2}\Psi_{H}-\bar{m}_{a}\bar{m}_{b}\tho^{2}\bar{\Psi}_{H}\,,
    \end{aligned}
\end{equation}
which satisfies the transverse condition $\ell^ah_{ab}=0$.

\item \textit{Frame perturbations}:  
\begin{equation}\label{eq:frame}
  \dot{n}_a = \frac{1}{2}\Bigl(\ell_a\edt^2\Psi_H-m_a\tho\edt\Psi_H\Bigr) + \mathrm{c.c.} \, , \quad \dot{m}_a = \frac{1}{2}\Bigl(\ell_a\tho\edtp\bar{\Psi}_H-\bar{m}_a\tho^2\bar{\Psi}_H\Bigr)\, ,
\end{equation}
which follow by plugging \eqref{eq:recmet2} into eq.~\eqref{frame2}, and is automatically in the geodesic and parallel gauge.

\item \textit{Properly-weighted spin coefficients}:
\begin{equation}\label{eq:pertspincoefs}
    \dot{\kappa}' = \frac{1}{2} \eth'^{3}  \bar{\Psi}_H\, ,\quad \dot{\rho}' = \frac{1}{2} \eth'^{2} \tho \bar{\Psi}_H\,, \quad \dot{\sigma} = -\frac{1}{2} \tho^{3} \bar{\Psi}_H\,, \quad  \dot{\tau} = -\frac{1}{2} \eth' \tho^{2}  \bar{\Psi}_H\, , 
\end{equation}
which follow by simply perturbing their definitions (see Appendix \ref{app:GHPsum}) and plugging \eqref{eq:recmet2} and \eqref{eq:frame} in the resulting expressions. We also used the equation satisfied by the Hertz potential \eqref{eq:HertzEq} in order to simplify expressions and show that several perturbations of spin coefficients vanish. As expected, $\dot{\kappa}=\dot{\rho}=0$, signaling that $\ell$ remains a geodesic, twist-free congruence in the perturbed spacetime. 

\item \textit{GHP connection 1-form}:
\begin{equation}\label{eq:pertGHPConnection}
    \dot{\omega}_{a}=-\frac{1}{2}\left(\eth'^{2}\tho \bar{\Psi}_{H}\right)\ell_{a}+\frac{1}{2}\left(\eth'\tho^{2} \bar{\Psi}_{H}\right)\bar{m}_{a}\, ,
\end{equation}
which again follows by perturbing the definition of $\omega_{a}$ (see Appendix \ref{app:GHPsum}) and using \eqref{eq:recmet2} and \eqref{eq:frame}. 

\item \textit{Weyl Scalars}:
\begin{equation}\label{eq:Weylperts}
    \dot{\Psi}_{n}=-\frac{1}{2}\tho^{(4-n)}\eth'^{n}\bar{\Psi}_{H}\, ,
\end{equation}
obtained again by perturbing their definitions in Appendix \ref{app:GHPsum} and using \eqref{eq:recmet2} and \eqref{eq:frame}. In order to present the Weyl scalars in this simple form, we also used the Hertz potential equation \eqref{eq:HertzEq}. 
\end{itemize}

Before considering second order perturbations, here we highlight two aspects of our approach. First, from the compact expression \eqref{eq:Weylperts} one immediately obtains a version of the \textit{Teukolsky-Starobinsky identities} on our type $N$ spacetimes:
\begin{equation}\label{eq:TS}
    \tho^{m}\eth'^{4-m}\dot{\Psi}_{n}=\tho^{4-n}\eth'^{n}\dot{\Psi}_{4-m}\, , \quad \, \left(0\leq m,n\leq4\right)\, ,
\end{equation}
which relate the perturbations of the various Weyl scalars among themselves. Although these expressions have been obtained by working in the GPT gauge, we remark that the only ones that are not gauge-invariant in general are those involving $\{\dot{\Psi}_{3},\dot{\Psi}_{4}\}$ (see eqs.~\eqref{eq:psi3gauge}-\eqref{eq:psi4gauge}). 

The second aspect to highlight is that we work in terms of perturbations of the GHP connection $\omega_{a}$ instead of using spin coefficients without proper GHP weight. By doing so one can introduce the following GHP derivative operators acting on scalar quantities $\eta\GHPweight(p,q)$ as 
\begin{equation}
    \begin{aligned}\label{perturbation_ghp}
        \dot{\tho}\eta&\coloneqq\dot{\ell}^{a}\theta_{a}\eta-p\ell^{a}\dot{\omega}_{a}\eta-q\ell^{a}\dot{\bar{\omega}}_{a}\eta\,, & \quad \quad   \dot{\tho}'\eta&\coloneqq\dot{n}^{a}\theta_{a}\eta-p n^{a}\dot{\omega}_{a}\eta-q n^{a}\dot{\bar{\omega}}_{a}\eta\,, \\
        \dot{\eth}\eta&\coloneqq\dot{m}^{a}\theta_{a}\eta-p m^{a}\dot{\omega}_{a}\eta-q m^{a}\dot{\bar{\omega}}_{a}\eta\,, & \quad \quad    \dot{\eth}'\eta&\coloneqq\dot{\bar{m}}^{a}\theta_{a}\eta-p\bar{m}^{a}\dot{\omega}_{a}\eta-q\bar{m}^{a}\dot{\bar{\omega}}_{a}\eta \,,
\end{aligned}
\end{equation}
which satisfy Leibniz's rule relative to the variation operator,
\begin{equation}\label{eq:varLeibniz}
   \left(\mathcal{D}\eta\right)\dot{}=\mathcal{D}\dot{\eta}+\dot{\mathcal{D}}\eta\, , \quad \text{where} \quad   \mathcal{D}=\left\{\tho,\tho',\eth,\eth'\right\} \, .
\end{equation}
These operators have definite weights, although only relative to $\epsilon$-independent (but otherwise generic) GHP transformations, where $\epsilon$ is the expansion parameter. Indeed, while the connection form $\omega_a$ does not have a well defined GHP weight, the perturbation to the connection has $\dot{\omega}_{a}\GHPweight(0,0)$. Thus, 
\begin{equation}
    \dot{\tho}\eta\GHPweight(p+1,q+1)\, ,\quad \dot{\tho}'\eta\GHPweight(p-1,q-1)\, ,\quad \dot{\eth}\eta\GHPweight(p+1,q-1)\, ,\quad  \dot{\eth}'\eta\GHPweight(p-1,q+1)\, .
\end{equation}
This allows us to preserve GHP-covariance at each step when perturbing our equations, what is particularly useful in obtaining the first and second variations of equation \eqref{Nonlinear_Teukolsky_Equation}. 

%%%%%%%%%%%%%%%%%%%%%%%%%%%%%%%%%%%%%%%%%%%%%%%%%%%%%%%%%
\subsection{Second-order perturbations}\label{sec:second}
%%%%%%%%%%%%%%%%%%%%%%%%%%%%%%%%%%%%%%%%%%%%%%%%%%%%%%%%%

With the solution and operators introduced in Section \ref{sec:first}, it is straightforward to obtain the second variation of eq.~\eqref{Nonlinear_Teukolsky_Equation}. In particular, the second-order source term (the right-hand-side in eq.~\eqref{eq:formalsecond}) can be written immediately in a GHP-covariant guise, by using eq.~\eqref{perturbation_ghp} and Leibniz's rule. If, in addition, this is evaluated on the first-order solution of section~\ref{sec:first}, it only depends on the Hertz potential and its derivatives. We find
\begin{equation}\label{Master2nd}
\left(\tho'\tho-\eth'\eth\right)\ddot{\Psi}_{0}=-2\left(\dot{\mathcal{O}}_{0}\dot{\Psi}_{0}+\dot{\mathcal{O}}_{1}\dot{\Psi}_{1}+\dot{\mathcal{O}}_{2}\dot{\Psi}_{2}\right)\equiv\mathcal{S}_{+}+\mathcal{S}_{-} \, ,
\end{equation}
where 
\begin{align}
    2\mathcal{S}_{+} =& \Bigl(\tho^6 \bar{\Psi}_H\Bigr) \Bigl(\edtp^2\bar{\Psi}_H\Bigr) + 4 \Bigl(\tho^5\bar{\Psi}_H\Bigr) \Bigl(\edtp^2\tho \bar{\Psi}_H\Bigr) + 6\Bigl(\tho^2\edtp^2\bar{\Psi}_H\Bigr)\Bigl(\tho^4\bar{\Psi}_H\Bigr) \notag\\
    &+ 4\Bigl(\tho^3\bar{\Psi}_H\Bigr)\Bigl(\edtp^2\tho^3\bar{\Psi}_H\Bigr) +\Bigl(\tho^2\bar{\Psi}_H\Bigr)\Bigl(\edtp^2\tho^4\bar{\Psi}_H\Bigr)-2\Bigl(\edtp\tho\bar{\Psi}_H\Bigr)\Bigl(\edtp\tho^5\bar{\Psi}_H\Bigr)\notag\\
    &-6\Bigl(\edtp\tho^3\bar{\Psi}_H\Bigr)^2-8\Bigl(\edtp\tho^2\bar{\Psi}_H\Bigr)\Bigl(\edtp\tho^4\bar{\Psi}_H\Bigr)\, , \label{sourcePlus}\\ \notag \\ 
    2\mathcal{S}_{-} =& \Bigl(\tho^2\Psi_H\Bigr)\Bigl(\edt^2\tho^4\bar{\Psi}_H\Bigr) + \Bigl(\edt^2\Psi_H\Bigr)\Bigl(\tho^6\bar{\Psi}_H\Bigr) -2\Bigl(\edt\tho\Psi_H\Bigr)\Bigl(\edt\tho^5\bar{\Psi}_H\Bigr)\, . \label{sourceMin}
\end{align}
It is useful to distinguish these two terms depending on whether they involves products of $\bar{\Psi}_H$ with itself (in $\mathcal{S}_+$), or combinations of $\Psi_H$ with $\bar{\Psi}_H$ (in $\mathcal{S}_-$). These will lead to source terms which oscillate with different frequencies, so we refer to these as the $++$ and the $+-$ channels. The master equation~\eqref{Master2nd} describes generic second order curvature fluctuations on a non-homogeneous plane wave background, in terms of the Hertz potential which generates the first order metric fluctuations. We note that, when written in coordinates, our source term reduces to previous results in the literature~\cite{Kehagias:2025ntm}, although this agreement was not entirely expected given that frame gauges are different. 

%%%%%%%%%%%%%%%%%%%%%%%%%%%%%%%%%%%%%%%%%%%%%%%%
\section{Homogeneous plane waves}\label{sec:homogeneous}
%%%%%%%%%%%%%%%%%%%%%%%%%%%%%%%%%%%%%%%%%%%%%%%%

The previous discussion and results are valid for the general vacuum plane waves~\eqref{ppWave}, which have arbitrary symmetric $A_{ij}(u)$ of vanishing trace. From now on we will restrict to homogeneous plane waves of the form
\begin{equation}\label{symmetricwave}
ds^{2}= -\left(\Lambda^{2} y^{2} - \Omega ^{2} x^{2}\right) d u^{2}+2 du dv-dx^{2}-dy^{2}\, ,
\end{equation}
for which the analysis simplifies. However, as noted in Section \ref{sec:PenroseLimits}, such spacetimes still capture physical regimes of great interest -- they emerge as the Penrose limit along the equatorial LRs of a Kerr black hole. 

In more detail, the LR Penrose limit of a Kerr black hole with mass $M$ and dimensionless angular momentum $a/M$ yields the metric~\eqref{symmetricwave} with $\Lambda=\Omega$, and\footnote{Remark that, from the perspective of the plane wave \eqref{symmetricwave}, one can always scale $u \to c u$ and $v \to v/c$. From the perspective of the Penrose limit, this would amount to choosing a different affine time parameter. The choice \eqref{Omega_Lambda_Kerr} chooses this affine time to relate directly to the asymptotic Killing time with respect to which the quasinormal modes in the black hole spacetime are defined, to straightforwardly compare.}
\begin{equation}\label{Omega_Lambda_Kerr}
    \Omega_{\rm p,r} = \Lambda_{\rm p,r} = \frac{12M\Bigl(r_{\rm p,r}^2-2M r_{\rm p,r} + a^2\Bigr)}{(r_{\rm p,r}-M)^2 r_{\rm p,r}^3} \, , \qquad r_{\rm p,r} = 2M\Bigl[1+\cos\Bigl(\frac{2}{3}\arccos(\mp a/M)\Bigr)\Bigr] \, ,
\end{equation}
where $r_{\rm p,r}$ are the prograde and retrograde radii of the equatorial LRs. We notice that the vacuum condition $\Lambda=\Omega$ holds in \eqref{Omega_Lambda_Kerr}, since the Penrose limit of a Ricci-flat solution like Kerr is Ricci-flat, too. However, below we will retain some more generality by allowing $\Lambda\ne\Omega$ and assume $\Lambda,\Omega>0$, unless stated otherwise. 

%%%%%%%%%%%%%%%%%%%%%%%%%%%%%%%%%%%%%%%%%%%
\subsection{Scalar wave equation}\label{sec:ScalarWave}
%%%%%%%%%%%%%%%%%%%%%%%%%%%%%%%%%%%%%%%%%%%

The wave equation in a general plane wave $\Box\Phi = 0$ is simply given by 
\begin{equation}
    2\Phi_{,uv} -\Phi_{,xx}-\Phi_{,yy}+A_{ij}x^i x^j\Phi_{,vv} = 0 \, .
\end{equation}
Thus, if $A_{ij}x^i x^j = \Lambda^2 y^2-\Omega^2x^2$, we can find solutions of the form 
\begin{equation}\label{eqn:sep}
    \Phi = e^{ip_u u+ip_v v}X(x)Y(y) \, , 
\end{equation}
for (at this point) arbitrary values of $p_u,p_v$. Solutions of the form \eqref{eqn:sep} are just one of several ways to separate the wave equation on plane wave spacetimes. Notably, a separation in Rosen coordinates \eqref{eqn:Rosen} is also possible, which has a slightly different flavor. However, it will be solutions of the form \eqref{eqn:sep} that most explicitly have a close connection to quasinormal modes. Plugging in the ansatz \eqref{eqn:sep} leads to two separated equations
\begin{equation}
    \begin{aligned}
        X'' - p_v^2\Omega^2 x^2X =& (C-p_up_v)X \, , \\
        Y'' + p_v^2\Lambda^2 y^2Y =& -(C+p_up_v)Y \, ,
    \end{aligned} 
\end{equation}
where $C$ is a separation constant. We immediately recognize the equations for the simple and inverted harmonic oscillators (SHO and IHO, respectively), with an extra term. We comment on the well-known algebraic relation between modes on symmetric plane waves and the SHO and IHO below. For now, we simply remark that the general solution can be written in terms of parabolic cylindric functions.

If we see these plane waves as corresponding to the Penrose limit of a LR, then the $x$ direction corresponds to the periodic $\theta$ direction, whereas the $y$ direction is the radial direction with respect to the black hole. In~\cite{Fransen:2023eqj} it was argued that mode solutions that are regular (bounded) along the $x$ direction, and purely outgoing (and in fact asymptotically divergent) along the $y$ direction oscillate at the same quasinormal frequencies as eikonal QNMs in Kerr. This is the class of solutions we will be focusing on. To that aim, let us first introduce the following family of off-shell modes,
\begin{equation}\label{modes}
     \Phi^{(p_u, p_v,n_x,n_y)} \equiv e^{i u p_u + i v p_v - |p_v| \Omega \frac{x^2}{2}-i p_v \Lambda \frac{y^2}{2}}\mathcal{N}_{(p_v,n_x,n_y)} H_{n_x}\left(\sqrt{|p_v| \Omega} x\right) H_{n_y}\left(\sqrt{i p_v \Lambda} y\right)  \, ,
\end{equation} 
with $n_x,n_y$ non-negative integers, $H_{n_{x,y}}\left(\xi\right)$ denote Hermite polynomials, and we choose normalization factors $\mathcal{N}_{(p_v,n_x,n_y)}$ given by
\begin{equation}\label{normalization}
    \mathcal{N}_{(p_v,n_x,n_y)}=
    \begin{cases}
        \frac{1}{2^{n_x+n_y}\pi} \Gamma\Bigl(\frac{1-n_x}{2}\Bigr)\Gamma\Bigl(\frac{1-n_y}{2}\Bigr) \, , \qquad& n_x,n_y  \text{ even} \, , \\
        \frac{1}{2^{n_x+n_y}\pi n_y} \Gamma\Bigl(\frac{1-n_x}{2}\Bigr)\Gamma\Bigl(1-\frac{n_y}{2}\Bigr) \, , \qquad& n_x\text{ even, } n_y  \text{ odd} \, , \\
        \frac{1}{2^{n_x+n_y}\pi n_x} \Gamma\Bigl(1-\frac{n_x}{2}\Bigr)\Gamma\Bigl(\frac{1-n_y}{2}\Bigr) \, , \qquad& n_x\text{ odd, } n_y  \text{ even} \, , \\
        \frac{1}{2^{n_x+n_y}\pi n_x n_y }\Gamma\Bigl(1-\frac{n_x}{2}\Bigr)\Gamma\Bigl(1-\frac{n_y}{2}\Bigr) \, , \qquad& n_x,n_y  \text{ odd}\, . 
    \end{cases}
\end{equation}
This normalization guarantees that when evaluated at the location of the central geodesic $x=y=0$, corresponding to the original geodesic whenever the plane wave spacetime is a Penrose limit, the mode is normalized to $1$ when it is non-vanishing. If instead the mode vanishes there, the constant guarantees that its $x$- and $y$-derivatives at $x=y=0$ are given by a natural dimensionful quantity, independent of the mode number ($\sqrt{|p_v| \Omega}$ for the $x$-direction and $\sqrt{i p_v \Lambda}$ for the $y$-direction).
The modes~\eqref{modes} are eigenfunctions of the wave operator, 
\begin{equation}\label{eqn:scalarwave}
    \square \Phi^{(p_u,p_v,n_x,n_y)} = 2p_v \Biggl[\mathrm{sgn}(p_v)\Omega\Bigl(n_x+\frac{1}{2}\Bigr)+i\Lambda\Bigl(n_y+\frac{1}{2}\Bigr)-p_u\Biggr]\Phi^{(p_u,p_v,n_x,n_y)} \, ,
\end{equation}
so, solutions to the free wave equation (or similarly the equation for the Hertz potential \eqref{eq:HertzEq}) are simply eigenfunctions corresponding to the zero eigenvalue. This imposes a quantization condition for $p_u$:
\begin{equation}\label{onshell}
p_u = \mathrm{sgn}(p_v)  \Omega\left(n_x + \frac{1}{2}\right) + i \Lambda \left(n_y + \frac{1}{2}\right) \, .
\end{equation}
Since we are assuming $\Omega, \Lambda > 0$, these modes are purely outgoing in the $y$-direction and decay exponentially along $u$, while their real frequencies $p_v$ can have arbitrary sign. 

The quantization condition allows us to establish a dictionary between the modes of a plane wave which emerges as the Penrose limit of the equatorial LR of a Kerr black hole, with the high frequency or eikonal limit of the black hole QNMs. Indeed, let $p_v=m\Omega_{\rm p,r}$ correspond to the azimuthal harmonic number, and let $n_x=\ell-|m|$, $n_y=n$. Here, $\ell$ is the polar spherical harmonic index, and $n$ is used to denote the overtone index. Using this dictionary, and plugging in the value of $\Omega$ and $\Lambda$ given in Eq.~\eqref{Omega_Lambda_Kerr}, we find that $p_u=\omega_{\ell m n}$ becomes the subleading correction to the frequency of QNMs in the eikonal or high frequency regime~\cite{Fransen:2023eqj}.

A more algebraic approach to the modes \eqref{modes} starts from the observation that, in homogeneous plane waves, the isometry algebra generators are
\begin{align}
a_{\pm}=\frac{e^{\pm i \Omega u}}{\sqrt{2 \Omega}}\left(\pm i \Omega x \partial_{v}+\partial_{x}\right)\,,\quad b_{\pm}=\frac{e^{\mp \Lambda u}}{\sqrt{2 \Lambda}}\left(\mp \Lambda y \partial_{v}+\partial_{y}\right)\, , \quad \bold{1}=i\partial_{v}\, , \quad H_{\Omega,\Lambda}=\mp i \partial_{u}\, .
\end{align}
These span the harmonic oscillator algebra
\begin{equation}\label{eqn:HOalgebra}
[a_{-},a_{+}]=\bold{1}\, , \quad \left[H_{\Omega},a_{\pm}\right]=\pm \Omega a_{\pm}\, , \quad [b_{-},b_{+}]=i \bold{1}\, , \quad \left[H_{\Lambda},b_{\pm}\right]=\mp i \Lambda b_{\pm} \, .
\end{equation}
This implies in particular that $a_{\pm}$ and $b_{\pm}$ act, through the Lie derivative, as ``raising'' and ``lowering'' operators for the $n_x$ and $n_y$ modes \eqref{modes}. Specifically, defining the number operators $N_x =  -\pounds_{a_{+}} \pounds_{a_{-}}$ and $N_y =  i\pounds_{b_{+}} \pounds_{b_{-}}$ these modes are such that
\begin{equation}\label{eqn:Smodes}
\begin{aligned}
N_x  \Phi^{(p_u,p_v,n_x,n_y)} &= -\pounds_{a_{+}} \pounds_{a_{-}}\Phi^{(p_u,p_v,n_x,n_y)} = n_x p_v \Phi^{(p_u,p_v,n_x,n_y)} \, , \\ N_y \Phi^{(p_u,p_v,n_x,n_y)} &= i\pounds_{b_{+}} \pounds_{b_{-}}\Phi^{(p_u,p_v,n_x,n_y)} = n_y p_v \Phi^{(p_u,p_v,n_x,n_y)} \, , \\ \pounds_{\partial_u}\Phi^{(p_u,p_v,n_x,n_y)}  &= i p_u \Phi^{(p_u,p_v,n_x,n_y)} \, , \\ \quad \pounds_{\partial_v}\Phi^{(p_u,p_v,n_x,n_y)}  &= i p_v \Phi^{(p_u,p_v,n_x,n_y)} \, . 
\end{aligned}
\end{equation}
Up to the normalization, for $p_v > 0$, they could have thus been generated by acting with $\pounds_{a_{+}}$ and $\pounds_{b_{+}}$ on the fundamental (Gaussian) mode $\Phi^{(p_u,p_v,0,0)}$. However, for $p_v < 0$, it is in fact $a_{+}$ that acts as a lowering operator while $a_{-}$ is the raising operator. 

This algebraic approach to the mode functions \eqref{modes} is of course well-known, so we will not go into further details here. However, before using these modes, together with the simple form of the wave operator acting on them \eqref{eqn:scalarwave}, to go to higher order in perturbation theory by solving the master equation \eqref{Master2nd}, we will discuss how the same algebraic approach can just as well be used to approach the metric perturbations head-on. We should point out here that in a different (``coherent state'' as opposed to ``number'') basis, not particularly well-suited to the description of quasinormal-modes, tensor harmonics on plane wave spacetimes have also been described for instance in~\cite{mason1989ward,Adamo:2017nia}.

The above discussion is strictly only valid for the symmetric plane waves \eqref{symmetricwave}, which are a special case of homogeneous plane waves of special interest in relation to the Penrose limit of equatorial lightrings. However, any homogeneous plane wave has an additional Killing vector involving (though not necessarily identically to) $\partial_u$ to extend the Heisenberg algebra \eqref{eqn:HOalgebra} common to all plane waves. Thus, while not as simple as \eqref{eqn:Smodes}, we expect that the above discussion as well as the remainder of this section can be generalized beyond the symmetric plane waves to homogeneous plane waves. See also~\cite{Blau:2002js,Blau:2003rt,Figueroa-OFarrill:2005lhx}.

%%%%%%%%%%%%%%%%%%%%%%%%%%%%%%%%%%%%%%%%%%%%%%%%
\subsection{Metric perturbations}\label{sec:metric}
%%%%%%%%%%%%%%%%%%%%%%%%%%%%%%%%%%%%%%%%%%%%%%%%

Just as the scalar wave equation reduced to an algebraic expression when acting on the modes satisfying \eqref{eqn:Smodes}, the linearized Einstein equations around a homogeneous plane wave will turn out to reduce to algebraic equations in terms of (symmetric) tensor modes $h^{(p_u,p_v,n_x,n_y)}_{\mu \nu}$ which satisfy\footnote{Although here we can proceed formally with $\Lambda\ne\Omega$, we remark that gravitational fluctuations are only sensible if the background is on-shell, $\Lambda=\Omega$.}
\begin{equation}\label{eqn:hmodes}
\begin{aligned}
N_x  h^{(p_u,p_v,n_x,n_y)}_{\mu \nu} &= -\pounds_{a_{+}} \pounds_{a_{-}}h^{(p_u,p_v,n_x,n_y)}_{\mu \nu} = n_x p_v h^{(p_u,p_v,n_x,n_y)}_{\mu \nu} \, , \\ N_y  h^{(p_u,p_v,n_x,n_y)}_{\mu \nu} &= i\pounds_{b_{+}} \pounds_{b_{-}}h^{(p_u,p_v,n_x,n_y)}_{\mu \nu} = n_y p_v h^{(p_u,p_v,n_x,n_y)}_{\mu \nu} \, , \\ \pounds_{\partial_u}h^{(p_u,p_v,n_x,n_y)}_{\mu \nu}  &= i p_u h^{(p_u,p_v,n_x,n_y)}_{\mu \nu} \, , \\ \quad \pounds_{\partial_v}h^{(p_u,p_v,n_x,n_y)}_{\mu \nu}  &= i p_v h^{(p_u,p_v,n_x,n_y)}_{\mu \nu} \, . 
\end{aligned}
\end{equation}
For each set of mode numbers $(p_u,p_v,n_x,n_y)$, there will be ten independent such modes. Four of those, generated by similarly constructed vector modes $V^{(p_u,p_v,n_x,n_y)}_{\mu}$
\begin{equation}\label{eqn:Vmodes}
\begin{aligned}
N_x V^{(p_u,p_v,n_x,n_y)}_{\mu } &= -\pounds_{a_{+}} \pounds_{a_{-}}V^{(p_u,p_v,n_x,n_y)}_{\mu} = n_x p_v V^{(p_u,p_v,n_x,n_y)}_{\mu} \, , \\ N_y  V^{(p_u,p_v,n_x,n_y)}_{\mu} &= i\pounds_{b_{+}} \pounds_{b_{-}}V^{(p_u,p_v,n_x,n_y)}_{\mu} = n_y p_v V^{(p_u,p_v,n_x,n_y)}_{\mu} \, , \\ \pounds_{\partial_u}V^{(p_u,p_v,n_x,n_y)}_{\mu}  &= i p_u V^{(p_u,p_v,n_x,n_y)}_{\mu} \, , \\ \quad \pounds_{\partial_v}V^{(p_u,p_v,n_x,n_y)}_{\mu}  &= i p_v V^{(p_u,p_v,n_x,n_y)}_{\mu} \, , 
\end{aligned}
\end{equation}
are pure gauge. Four more will be eliminated by the vacuum Einstein equations, and the final two dynamical, gravitational degrees of freedom turn out to be generated by the Hertz potential method~\eqref{eq:recmet2}, when taking the Hertz potentials to be proportional to the scalar modes $\Phi^{(p_u,p_v,n_x,n_y)}$.

%%%%%%%%%%%%%%%%%%%%%%%%%%%%%%%%%%%%%%%%%
\subsubsection{First-order perturbations}
%%%%%%%%%%%%%%%%%%%%%%%%%%%%%%%%%%%%%%%%%

The diagonalization of $\pounds_{\partial_u}$ and $\pounds_{\partial_v}$ is readily achieved. Therefore, motivated additionally by our expectations from the scalar modes, define
\begin{equation}\label{eqn:Vansatz1}
\begin{aligned}
V^{(p_u,p_v,n_x,n_y,I)}_{\mu}  &= e^{i u p_u + i v p_v} e^{-\frac{|p_v| \Omega}{2} x^2-i \frac{p_v \Lambda}{2} y^2}\tilde{V}^{(p_u,p_v,n_x,n_y,I}_{\mu}(x,y) \, , \\
h^{(p_u,p_v,n_x,n_y,IJ)}_{\mu \nu}  &= e^{i u p_u + i v p_v} e^{-\frac{|p_v| \Omega}{2} x^2-i \frac{p_v \Lambda}{2} y^2}\tilde{h}^{(p_u,p_v,n_x,n_y,IJ)}_{\mu \nu}(x,y) \, .
\end{aligned}
\end{equation}
We expect to find four independent vector modes, which we therefore label by an additional index $I \in \left\lbrace n, \ell, x, y\right\rbrace$ as well as the labels of the representation $(p_u,p_v,n_x,n_y)$ defined by the (vector version of the) relations \eqref{eqn:Vmodes}. Similarly, we expect ten independent tensor modes for each $(p_u,p_v,n_x,n_y)$, such that in eq.~\eqref{eqn:Vansatz1} we have added a symmetric $IJ$ label to the tensor modes.

There are two complementary approaches that are useful for constructing modes satisfying eqs.~\eqref{eqn:Vmodes} and~\eqref{eqn:hmodes}. The first is to start by finding the ``ground states'' or fundamental modes and subsequently raising the mode number using the properties of the algebra~\eqref{eqn:HOalgebra}. The second approach simply relies on solving the differential equations implied by \eqref{eqn:Vmodes} and \eqref{eqn:hmodes}. In order to do this explicitly for the different vector and tensor modes, it is useful to decompose in components with respect the frame $(n, \ell, E^{(x)}, E^{(y)})$, where in terms of the principal null frame \eqref{NPframe}
\begin{equation}
E^{(x)} = \frac{1}{\sqrt{2}}\left(m + \bar{m}\right) = \partial_x \, , \quad 	E^{(y)} = \frac{1}{i \sqrt{2}}\left(m - \bar{m}\right) = \partial_y \, .
\end{equation}
Explicitly, we decompose
\begin{equation}
V_{\mu} = V_{\ell} n_{\mu} + V_{n} \ell_{\mu}- V_{x}E^{(x)}_{\mu}- V_{y}E^{(y)}_{\mu} \, ,
\end{equation}
and so on. Note in particular the sign choices.

To cross-check, we have applied both the algebraic and differential equations methods. However, as both methods are straightforward we do not enter into further details about these derivations. Instead, we simply quote the result
\begin{equation}%\label{eqn:Vmodes}
\begin{aligned}
\tilde{V}^{(p_u,p_v,n_x,n_y, n)}_{\mu} &= \mathcal{N}_{(p_v,n_x,n_y,n)} H_{n_x}(\sqrt{|p_v| \Omega} x) H_{n_y}(\sqrt{i p_v \Lambda} y) \ell_{\mu} \, ,  \\
\tilde{V}^{(p_u,p_v,n_x,n_y, x)}_{\mu} &=  \mathcal{N}_{(p_v,n_x,n_y,x)} H_{n_x}(\sqrt{|p_v| \Omega} x) H_{n_y}(\sqrt{i p_v \Lambda} y) E^{(x)}_{\mu}   \\ & - i \sqrt{\Omega/4|p_v|} \left( -2n_x  H_{n_x-1}(\sqrt{|p_v| \Omega} x) + H_{n_x+1}(\sqrt{|p_v| \Omega} x)   \right) H_{n_y}(\sqrt{i p_v \Lambda} y) \ell_{\mu}  \, ,  \\
\tilde{V}^{(p_u,p_v,n_x,n_y, y)}_{\mu} &= \mathcal{N}_{(p_v,n_x,n_y,y)} H_{n_x}(\sqrt{|p_v| \Omega} x) H_{n_y}(\sqrt{i p_v \Lambda} y) E^{(y)}_{\mu} \\ &  -i \sqrt{i\Lambda/4p_v}    H_{n_x}(\sqrt{|p_v| \Omega} x) \left( -2n_y  H_{n_y-1}(\sqrt{i p_v \Lambda} y)+H_{n_y+1}(\sqrt{i p_v \Lambda} y)  \right)  \ell_{\mu} \, , \\
\tilde{V}^{(p_u,p_v,n_x,n_y, \ell)}_{\mu} &=  \mathcal{N}_{(p_v,n_x,n_y,\ell)}  H_{n_x}(\sqrt{|p_v| \Omega} x) H_{n_y}(\sqrt{i p_v \Lambda} y) n_{\mu}  \\  & -i \sqrt{i\Lambda/4p_v}  H_{n_x}(\sqrt{|p_v| \Omega} x) \left( -2n_y  H_{n_y-1}(\sqrt{i p_v \Lambda} y)+H_{n_y+1}(\sqrt{i p_v \Lambda} y)  \right)  E^{(y)}_{\mu} \\  & -i \sqrt{\Omega/4|p_v|}    H_{n_y}(\sqrt{i p_v \Lambda} y) \left(- 2n_x  H_{n_x-1}(\sqrt{ |p_v| \Omega} x)+  H_{n_x+1}(\sqrt{|p_v| \Omega} x)  \right)  E^{(x)}_{\mu} \\  &   +\frac{1}{2 p^2_v}\left(2 p_u p_v + p^2_v (\Lambda^2 y^2 - \Omega^2 x^2)\right)H_{n_x}(\sqrt{|p_v| \Omega} x) H_{n_y}(\sqrt{i p_v \Lambda} y)\ell_{\mu}   \, . 
\end{aligned}
\end{equation}
Here we have kept the normalizations $\mathcal{N}_{(p_v,n_x,n_y,I)}$ arbitrary for now.

The mode $\tilde{V}^{(p_u,p_v,n_x,n_y, \ell)}$ is clearly the most complicated, it is also the only one which can contribute to $\ell^{\mu} V_{\mu} \neq 0$. Conveniently, it turns out that this mode is pure gauge (for $p_v \neq 0$)
\begin{equation}
	V^{(p_u,p_v,n_x,n_y, \ell)}_{\mu} =\frac{\mathcal{N}_{(p_v,n_x,n_y,\ell)}}{i p_v \mathcal{N}_{(p_v,n_x,n_y)}} \partial_{\mu}\left(  \Phi^{(p_u, p_v,n_x,n_y)}  \right) \, .
\end{equation}
As a result, in the above formulation, it is particularly convenient to impose the gauge $\ell^{\mu}V_{\mu} = 0$. For a explicit expression of the tensor modes, see \eqref{eqn:hmodesexpl} in Appendix \ref{app:tensorharmonics}.

The reason for introducing modes satisfying \eqref{eqn:hmodes} is that they do not mix under the linearized wave equation and, moreover, render them algebraic. Consider for the vector example the Maxwell equations\footnote{We use $A$ instead of $V$ to emphasize the difference between electromagnetic gauge potentials, for which we impose equations of motion, and general vector fields $V$ for which we do not.}
\begin{equation}\label{eqn:maxwell}
F^{(p_u,p_v,n_x,n_y)}_{\mu \nu} = \nabla_{\mu}A^{(p_u,p_v,n_x,n_y)}_{\nu}-\nabla_{\nu}A^{(p_u,p_v,n_x,n_y)}_{\mu} \, , \quad \nabla^{\mu}F^{(p_u,p_v,n_x,n_y)}_{\mu \nu} = 0 \, ,
\end{equation}
where we take $A^{(p_u,p_v,n_x,n_y)}$ to be a sum of the modes \eqref{eqn:Vmodes} with constant coefficients $A_I$ for the relevant mode labeled by $I \in \left\lbrace n, \ell, x, y\right\rbrace$. Just like the scalar wave equation, the Maxwell equations \eqref{eqn:maxwell} impose the mass-shell or dispersion relation \eqref{onshell} ($p_v \neq 0$)
\begin{equation}\label{Vonshell}
p_u = \mathrm{sgn}(p_v)  \Omega\left(n_x + \frac{1}{2}\right) + i \Lambda \left(n_y + \frac{1}{2}\right) \, .
\end{equation}
In addition, they eliminate one further (``longitudinal'') combination of the vector modes. We are left with the expected two physical ``transverse'' electromagnetic polarizations for each set of spacetime mode numbers:
\begin{equation}\label{eqn:Aphys}
A^{(p_u,p_v,n_x,n_y)}_{\mu} = A_x V^{(p_u,p_v,n_x,n_y, x)}_{\mu} + A_y V^{(p_u,p_v,n_x,n_y, y)}_{\mu} \, .
\end{equation}
These two physical degrees of freedom can be generated by a Hertz potential $\Phi_{H}$ satisfying the massless scalar wave equation as in the gravitational case.

The tensor version goes through identically. For brevity, we refer to Appendix \ref{app:tensorharmonics} for the explicit expressions. In particular, the full ten-dimensional set of symmetric tensor modes is given in \eqref{eqn:hmodesexpl}, up to the pure gauge modes which we define implicitly in terms of the vector modes in \eqref{eqn:hgauge}. We label them $h^{(p_u,p_v,n_x,n_y, IJ)}_{\mu \nu}$, with $I, J \in \left\lbrace n, \ell, x, y\right\rbrace$. 

Analogously to the Maxwell equations for the vector modes, the Einstein equations impose eq.~\eqref{Vonshell} and eliminate all but two gravitational polarizations. These polarizations are captured exactly by one scalar Hertz potential and its conjugate mode. It is of course precisely this description in terms of scalar Hertz potentials, and closely related Weyl scalars that were used in the GHP formulation in  Section \ref{sec:first}. We find that the GHP formulation is ultimately more efficient to discuss the non-linearities. Nevertheless, let us conclude our discussion of a more directly metric-based approach by discussing non-linearities in this formulation.

%%%%%%%%%%%%%%%%%%%%%%%%%%%%%%%%%%%%%%%%%
\subsubsection{Second-order perturbations}
%%%%%%%%%%%%%%%%%%%%%%%%%%%%%%%%%%%%%%%%%

The crux of perturbation theory on homogeneous plane waves at non-linear orders is that it trivializes as soon as the source terms can be decomposed into modes such as \eqref{modes}, \eqref{eqn:Vmodes}, or \eqref{eqn:hmodes}. While this is true regardless of whether we are dealing with scalars, vectors, or tensors, the actual decomposition naturally becomes more difficult at higher spin. However, \emph{if} the source is again in the form of a sum of eigenfunctions of the wave operator, solutions to the inhomogeneous equation become readily available. For the scalar wave equation, the above claim is immediate. Consider
\begin{equation}\label{eqn:Sscalarwave}
    \square \Phi = S \Phi^{(p_u,p_v,n_x,n_y)}\, ,
\end{equation}
for some constant $S$. Then using \eqref{eqn:scalarwave}, a particular solution is given by
 \begin{equation}\label{eqn:Sscalarwavesol}
   \Phi = \frac{S \Phi^{(p_u,p_v,n_x,n_y)}}{2p_v \Biggl[\mathrm{sgn}(p_v)\Omega\Bigl(n_x+\frac{1}{2}\Bigr)+i\Lambda\Bigl(n_y+\frac{1}{2}\Bigr)-p_u\Biggr]} \, .
\end{equation}
In the next section, we will discuss how a source of the form \eqref{eqn:Sscalarwave}, or in general a linear combination of such sources, can be found for a toy example and the master equation \eqref{Master2nd}.
 
Consider instead the sourced Maxwell equations
\begin{equation}\label{eqn:maxwells}
 \nabla^{\mu}\left(\nabla_{\mu}A_{\nu}-\nabla_{\nu}A_{\mu} \right) = \cA^{(\hat{I})} V^{(p_u,p_v,n_x,n_y,\hat{I})}_{\mu}\, ,
\end{equation}
where $\cA^{(\hat{I})}$ are constants and $\hat{I} \in \left\lbrace x, y, \parallel \right\rbrace$ runs over the three divergence free combinations~\footnote{We need the right-hand side to be a conserved current, or more directly, because of the divergence of the left-hand side $[\nabla_{\nu},\nabla_{\mu}]F^{\nu \mu} = 2 R_{\alpha \mu}F^{\alpha \mu} = 0$.} of the modes, which in particular for a non-resonant source (meaning $p_u$ does not satisfy eq.~\eqref{Vonshell})
\begin{equation}\label{eqn:Vparallel}
V^{(p_u,p_v,n_x,n_y,\parallel)}_{\mu} =  V^{(p_u,p_v,n_x,n_y,n)}_{\mu} - \frac{p_v}{2 p_u - (2 n_y +1)i \Lambda - (2 n_x +1)\Omega} V^{(p_u,p_v,n_x,n_y,\ell)}_{\mu} \, ,
\end{equation}
where we have assumed that $\mathcal{N}_{(p_v,n_x,n_y,n)} = \mathcal{N}_{(p_v,n_x,n_y,\ell)}$ in eq.~\eqref{eqn:Vmodes}. Up to the decomposition of the right-hand side into the mode sum, eq.~\eqref{eqn:maxwells} is the form that second-and higher order perturbation equations will take. The solution to \eqref{eqn:maxwells} is given by
\begin{equation}\label{eqn:maxwellssols}
A_{\mu} = \frac{\cA^{(\hat{I})}\left(V^{(p_u,p_v,n_x,n_y,\hat{I})}_{\mu} - c_{(p_u,p_v,n_x,n_y,\hat{I})}V^{(p_u,p_v,n_x,n_y,\ell)}_{\mu}\right)}{2p_v \Biggl[\mathrm{sgn}(p_v)\Omega\Bigl(n_x+\frac{1}{2}\Bigr)+i\Lambda\Bigl(n_y+\frac{1}{2}\Bigr)-p_u\Biggr]}  \, , 
\end{equation}
which is a simple generalization of the scalar case~\eqref{eqn:Sscalarwavesol}. The second term in \eqref{eqn:maxwellssols} comes from the gauge ambiguity in inverting the wave equation. To stay in our preferred gauge despite the appearance of the $V^{(p_u,p_v,n_x,n_y,\ell)}_{\mu}$ modes in divergence free combination $V^{(p_u,p_v,n_x,n_y,\parallel)}_{\mu}$ for the source, we would simply choose $c_{(p_u,p_v,n_x,n_y,\ell)}$ to cancel that component while  $c_{(p_u,p_v,n_x,n_y,y)} = c_{(p_u,p_v,n_x,n_y,x)} = 0$. For the case of electromagnetic perturbations, one may imagine non-linearities arising from an effective Heisenberg-Euler Lagrangian but our ultimate interest is of course in tensor perturbations and General Relavity, which we discuss next.

For the linearized Einstein equation\footnote{While not necessary in general, when discussing the Einstein equations, we will always set $\Lambda = \Omega$ in order for the vacuum Einstein equation of the background to be satisfied.} with a single mode source no new complications arise with respect to the Maxwell example. That is, the solution to
\begin{equation}\label{eqn:lineinstein}
	\dot{G}[h_{\mu \nu}] = \cA^{(\hat{K})} h^{(p_u,p_v,n_x,n_y,\hat{K})}_{\mu \nu} \, ,
\end{equation}
where $\dot{G}$ is the linearized Einstein tensor, is given by
\begin{equation}\label{eqn:linesinteinmodesol}
h_{\mu \nu} = -\frac{2 \cA^{(\hat{K})} h^{(p_u,p_v,n_x,n_y,\hat{K})}_{\mu \nu}}{2p_v \Biggl[\mathrm{sgn}(p_v)\Omega\Bigl(n_x+\frac{1}{2}\Bigr)+i\Lambda\Bigl(n_y+\frac{1}{2}\Bigr)-p_u\Biggr]} \, .
\end{equation}
Here we have left implicit a further projection associated to the gauge choice, resulting from the fact that divergence free modes labeled by $\hat{K}$, which are the allowed sources to \eqref{eqn:lineinstein} as in the Maxwell example, are not in our chosen gauge. In conclusion, as advertised, the core problem is to find the mode decomposition of any given source. Subsequently, the solution of the wave equation is trivial. 

As a simple illustration consider the example of two linear $n_x = n_y = 0$, $p_v >0$, on-shell gravitational modes sourcing a second order mode 
\begin{equation}\label{eqn:hEE00}
h_{\mu \nu} = \sum_{i\in \left\lbrace a,b \right\rbrace}\Biggl(h^{(i)}_{+}  h^{(p^{(i)}_u,p^{(i)}_v,0,0,+)}_{\mu \nu}+ h^{(i)}_{\times } h^{(p^{(i)}_u,p^{(i)}_v,0,0,\times)}_{\mu \nu}\Biggr)\, ,
\end{equation}
where, for the sake of this example, $p^{(a)}_v, p^{(b)}_v > 0$ and eq.~\eqref{Vonshell} is satisfied. In addition, for clarity, we introduce the following (divergence free) combinations of the tensor harmonics starting from the definitions in \eqref{eqn:hmodes}
\begin{equation}
\begin{aligned}
 h^{(p^{(i)}_u,p^{(i)}_v,n_x,n_y,+)}_{\mu \nu} &= h^{(p^{(i)}_u,p^{(i)}_v,n_x,n_y,xx)}_{\mu \nu}-h^{(p_u,p_v,n_x,n_y,yy)}_{\mu \nu}\\
 &+\frac{(2n_x +1)\Omega-i(2n_y+1)\Lambda}{p^{(i)}_v}h^{(p^{(i)}_u,p^{(i)}_v,n_x,n_y,nn)}_{\mu \nu} \, , \\
  h^{(p^{(i)}_u,p^{(i)}_v,n_x,n_y,\times)}_{\mu \nu} &= 2 h^{(p^{(i)}_u,p^{(i)}_v,n_x,n_y,xy)}_{\mu \nu} \, ,
 \end{aligned}
\end{equation}
which, combined with \eqref{Vonshell}, solve the homogeneous, linearized Einstein equations.

Now, by a direct computation one can find the second order Einstein tensor $\ddot{G}$ to be as given in \eqref{eqn:G2mode00} in Appendix \ref{app:tensorharmonics}. However, more important than the explicit expression is that it is decomposed solely in terms of modes $h^{(p'_u,p'_v,0,0)}_{\mu \nu}$ with $n_x = n_y = 0$. As a result $\pounds_{a_-} \ddot{G} = \pounds_{b_-} \ddot{G} = 0$. To see that this must have been the case, note that $\ddot{G}$ is built as a linear combination of terms including products of $g_{\mu \nu}$, $h_{\mu \nu}$, and its covariant derivatives. Now for isometries $\xi$, $[\pounds_{\xi} , \nabla_{\mu}] = 0$. Therefore, together with $\pounds_{a_-} h_{\mu \nu} = \pounds_{b_-} h_{\mu \nu} = 0$ for the parent mode, by our choice of the linear perturbations \eqref{eqn:hEE00}, it indeed follows that $\pounds_{a_-} \ddot{G} = \pounds_{b_-} \ddot{G} = 0$. In turn, this implies that $\ddot{G}$ satisfies the conditions of~\eqref{eqn:hmodes} with $n_x=n_y=0$ . Thus, it can be decomposed in such modes as shown explicitly in \eqref{eqn:G2mode00}. Finally, after performing this decomposition into the $n_x = n_y = 0$ modes \eqref{eqn:hmodesexpl}, the second order metric perturbations are immediately found using \eqref{eqn:linesinteinmodesol}.

While a similar direct computation is still possible for more complicated modes, it naturally becomes more tedious. Therefore, we avoid working with the linearized Einstein equations and tensor harmonics \ref{app:tensorharmonics} in favor of the master equation \eqref{Master2nd} of the GHP formulation and scalar harmonics \eqref{modes}. 

%%%%%%%%%%%%%%%%%%%%%%%%%%%%%%%%%%%%%%%%%%%%%%%%%%%%%%
\section{Quadratic quasinormal modes (QQNMs)}\label{sec:QQNMs}
%%%%%%%%%%%%%%%%%%%%%%%%%%%%%%%%%%%%%%%%%%%%%%%%%%%%%%

Linear perturbations on certain homogeneous plane waves can be neatly related to the high-frequency regime of the QNMs of a Kerr BH, through the Penrose limit identification. This motivates exploring a similar connection between second order perturbations on the plane wave and black hole QQNMs. The particular solution to the second order Teukolsky equation on the plane wave will contain source-driven modes, which oscillate with frequencies that correspond to the sum or the difference of the frequencies of linear QNMs. Recent works~\cite{Kehagias:2025ntm, Perrone:2025zhy} have taken some first steps towards analyzing these QQNMs and relate them to the QQNMs of Kerr BHs, albeit restricting to modes with $n_x=n_y=0$, and only those generated by the $\mathcal{S}_{+}$ term. In this section, we will discuss the QQNMs on the plane wave arising from all possible mode combinations, and identifying general selection rules. In order to do so, we first introduce the solutions of a scalar toy model that showcases most of the technical difficulties in a simpler setting, and then discuss our results in the gravitational case. 

%%%%%%%%%%%%%%%%%%%%%%%%%%%%%%%%%%%%%%%%%%%%%%%%%%%
\subsection{Scalar QQNMs}\label{sec:scalarmodel}
%%%%%%%%%%%%%%%%%%%%%%%%%%%%%%%%%%%%%%%%%%%%%%%%%%%

Before addressing the full gravitational case, it is illustrative to consider as a toy model a nonlinear wave equation for a scalar field. We consider the cubic self-interaction leading to the equation of motion
\begin{equation}\label{eqn:eomcubicscalar}
\square \Phi + \Phi^2 = 0 \, ,
\end{equation}
where we look for solutions of the form $\Phi =\epsilon \dot{\Phi} + \frac{1}{2}\epsilon^2 \ddot{\Phi}$. The leading order solution corresponds to modes of the homogeneous plane wave, as discussed before. Generically, we can assume the leading order solution to be a superposition of two modes (as defined in \eqref{modes})
\begin{equation}
    \dot{\Phi} = \mathcal{A}_a \Phi^{(p_v^{(a)},n_x^{(a)},n_y^{(a)})} + \mathcal{A}_b \Phi^{(p_v^{(b)},n_x^{(b)},n_y^{(b)})} \, ,
\end{equation}
where, when not listed, it is implied that $p_u$ is fixed by \eqref{Vonshell}, as required to be a leading order solutions to the wave equation (we also note that $a,b$ are merely labels, and not indices running in any set).

The next-to-leading order, which is the first order sensitive to the nonlinearity, satisfies the equation
\begin{equation}
    \begin{aligned}
        \square\ddot{\Phi} =&-2\mathcal{A}_a^2\Bigl(\Phi^{(p_v^{(a)},n_x^{(a)},n_y^{(a)})}\Bigr)^2 - 2\mathcal{A}_b^2\Bigl(\Phi^{(p_v^{(b)},n_x^{(b)},n_y^{(b)})}\Bigr)^2 -4\mathcal{A}_a\mathcal{A}_b\Phi^{(p_v^{(a)},n_x^{(a)},n_y^{(a)})}\Phi^{(p_v^{(b)},n_x^{(b)},n_y^{(b)})} \\
        =& \mathcal{S}_{aa} + \mathcal{S}_{bb}+2\mathcal{S}_{ab}\, .
    \end{aligned}
\end{equation}
The particular solution can be decomposed in three pieces $\ddot{\Phi} = \ddot{\Phi}_{aa} + \ddot{\Phi}_{bb}+2\ddot{\Phi}_{ab}$ where $\square\ddot{\Phi}_{IJ}=\mathcal{S}_{IJ}$ for $I,J=a,b$. It turns out that the source term can be generally written in terms of a linear combination of infinitely many eigenfunctions of the wave operator $\Phi^{(p_u,p_v,n_x,n_y)}$, as
\begin{equation}\label{toy_source_decomposition}
    \mathcal{S}_{IJ} = \sum_{n'_y=0}^{n_y^{(I)}+n_y^{(J)}}\sum_{n'_x=0}^\infty s_{IJ}^{(n'_x,n'_y)}\mathcal{A}_I\mathcal{A}_J \Phi^{(p'_u,p'_v,n'_x,n'_y)} \, , \qquad I,J=a,b \, ,
\end{equation}
where $p'_u=p_u^{(I)}+p_u^{(J)}$ and $p'_v=p_v^{(I)}+p_v^{(J)}$, and 
\begin{equation}\label{eqn:toyintegrals}
    \begin{aligned}
       & s_{IJ}^{(n'_x,n'_y)} = -2\mathcal{I}_x\mathcal{I}_y \, , \\
       & \mathcal{I}_x = \int^{\infty}_{-\infty} \frac{\sqrt{|p'_v|\Omega} dx }{2^{n'_x}n'_x!\sqrt{\pi}} H_{n_x^{(I)}}\left(\sqrt{|p_v^{(I)}|\Omega}x\right)H_{n_x^{(J)}}\left(\sqrt{|p_v^{(J)}|\Omega}x\right)H_{n'_x}\left(\sqrt{|p'_v|\Omega}x\right) e^{-(|p'_v|+\frac{\Delta}{2})\Omega x^2}  \, , \\
        & \mathcal{I}_y = \int_C  \frac{\sqrt{i p'_v \Omega}dy}{2^{n'_y}n'_y!\sqrt{\pi}}H_{n_y^{(I)}}\left(\sqrt{ip_v^{(I)}\Omega}y\right)H_{n_y^{(J)}}\left(\sqrt{ip_v^{(J)}\Omega}y\right)H_{n'_y}\left(\sqrt{ip'_v\Omega}y\right) e^{-ip'_v\Omega y^2} \, ,
    \end{aligned}
\end{equation}
where $\Delta = |p_v^{(I)}|+|p_v^{(J)}|-|p'_v|$ is only non-zero whenever $\mathrm{sign}(p_v^{(I)})\neq \mathrm{sign}(p_v^{(J)})$ and we will comment on the $\mathcal{I}_y$ integration contour $C$ later. As discussed in the previous section, by performing this decomposition in eigenfunctions of $\square$, the solution to the second order equation becomes trivial, since $\ddot{\Phi}$ can then be written as a superposition of eigenfunctions. We translate the difficulty from solving an inhomogeneous differential equation, to expanding the source term in terms of the $\Phi^{(p'_u,p'_v,n'_x,n'_y)}$, i.e., computing the integrals $\mathcal{I}_{x,y}$. 

The structure of these integrals shows why the sum in Eq.~\eqref{toy_source_decomposition} only includes a finite amount of terms in the $y$ direction, whereas it includes (in general) an infinite amount of terms in the $x$-direction -- if $n'_y > n_y^{(I)}+n_y^{(J)}$ then $\mathcal{I}_y=0$, since it is the projection of a lower degree polynomial onto a higher order degree Hermite polynomial. Moreover, as it is exactly this polynomial decomposition we need, we can avoid the subtleties in choosing the appropriate integration contour in \eqref{eqn:toyintegrals}, by defining $\mathcal{I}_y$ in terms of this polynomial decomposition. We cannot use such a finite polynomial decomposition for the $x$-integral $\mathcal{I}_x$ in general, but here the integral over the real line is always absolutely convergent.

The application of various identities of Hermite polynomials would allow us to write the integrals in \eqref{eqn:toyintegrals} more explicitly. Alternatively, we can observe that
\begin{equation}\label{gamma_integral}
    \int dx x^{k} e^{-a x^2} = \frac{1+(-1)^k}{2}a^{-\frac{1+k}{2}}\Gamma\Bigl(\frac{1+k}{2}\Bigr) \, .
\end{equation}
Thus, we can expand the integrands in \eqref{eqn:toyintegrals} in terms of expressions of the form~\eqref{gamma_integral}, and solve it analytically, once we know the Taylor coefficients of the product of Hermite polynomials. Abstractly we can write 
\begin{equation}\label{eqn:toyintegralsols}
    \begin{aligned}
        \mathcal{I}_x =& \frac{1}{2^{n_x}n_x!\sqrt{\pi}} \sum_{\alpha=0}^{n_x^{(I)}}\sum_{\beta=0}^{n_x^{(J)}}\sum_{\gamma=0}^{n'_x} \chi_{n_x^{(I)},\alpha}\chi_{n_x^{(J)},\beta}\chi_{n'_x,\gamma} \Bigl(\frac{1+(-1)^{\alpha+\beta+\gamma}}{2}\Bigr)\Gamma\Bigl(\frac{1+\alpha+\beta+\gamma}{2}\Bigr)\\
        &\hspace{4cm}\times\sqrt{\frac{|p_v^{(I)}|^\alpha|p_v^{(J)}|^\beta|p'_v|^{\gamma+1}}{(|p'_v|+\Delta/2)^{\alpha+\beta+\gamma+1}}} \, , \\
        \mathcal{I}_y =& \frac{1}{2^{n'_y}n'_y!\sqrt{i\pi\mathrm{sgn}(p'_v)}} \sum_{\alpha=0}^{n_y^{(I)}}\sum_{\beta=0}^{n_y^{(J)}}\sum_{\gamma=0}^{n'_y} \chi_{n_y^{(I)},\alpha}\chi_{n_y^{(J)},\beta}\chi_{n'_y,\gamma} \Bigl(\frac{1+(-1)^{\alpha+\beta+\gamma}}{2}\Bigr)\Gamma\Bigl(\frac{1+\alpha+\beta+\gamma}{2}\Bigr)\\
        &\hspace{5cm}\times\sqrt{\frac{(p_v^{(I)})^\alpha(p_v^{(J)})^\beta}{(p'_v)^{\alpha+\beta}}}  \, ,
    \end{aligned}
\end{equation}
where $\chi_{n,k}$ are simply the coefficients of the Hermite polynomials, i.e., $H_n(x)=\sum \chi_{n,k}x^k$. We will not make much use of more explicit, analytical results for these integrals but, as an illustration, we present some of these in Appendix \ref{app:scalartoy}.
Whenever $\Delta=0$ (which is the case for the $\mathcal{S}_{aa},\mathcal{S}_{bb}$ terms, and for $\mathcal{S}_{ab}$ whenever $p_v^{(a)}$ and $p_v^{(b)}$ have the same sign), the sum only contains a finite amount of terms in the $x$ direction as well. We will come back to the physical interpretation of this issue in terms of overtone excitations when discussing the gravitational case. However, another point of view from the perspective of the Hermite polynomials is that one can formally absorb a change in sign in $p_v$ in the mode number condition \eqref{eqn:Smodes} and the quantization condition \eqref{Vonshell} by analytically continuing in $x$ and $n_x$. While generalization of Hermite polynomials with non-integer $n_x$ can be made sense of, they are no longer polynomials.
In any case, once the $s_{IJ}^{(n'_x,n'_y)}$ have been determined, for instance using \eqref{eqn:toyintegrals} and \eqref{eqn:toyintegralsols}, the second order perturbation is found to be
\begin{equation}\label{scalar_ratios}
    \begin{aligned}
        \ddot{\Phi}_{IJ} =& \sum_{n'_x,n'_y}\mathcal{R}_{I\times J}^{(n'_x,n'_y)}\mathcal{A}_I\mathcal{A}_J  \Phi^{(p'_u,p'_v,n'_x,n'_y)} \, , \\
        \mathcal{R}_{I\times J}^{(n'_x,n'_y)} =& \frac{s_{IJ}^{(n'_x,n'_y)}}{|p'_v|\Omega(1+2n'_x)+ip'_v\Omega\left(1+2n'_y\right) -2 p'_v p'_u} \, ,
    \end{aligned}
\end{equation}
where the denominator in $\mathcal{R}_{I\times J}^{(n'_x,n'_y)}$ is just the eigenvalue of the wave operator corresponding to $\Phi^{(p'_u,p'_v,n'_x,n'_y)}$. Notice that there are no resonances at second order. A resonance would occur if $p'_u$ satisfies the on-shell condition~\eqref{onshell}, which would make the denominator in~\eqref{scalar_ratios} vanish. However, since $p'_u=p_u^{(a)}+p_u^{(b)}$, the resonance condition would correspond to 
\begin{equation}\label{scalar_resonance_condition}
    \mathrm{sgn}(p'_v)\Bigl(n'_x+\frac{1}{2}\Bigr) + i\Bigl(n'_y+\frac{1}{2}\Bigr) = \mathrm{sgn}(p_v^{(a)})\Bigl(n_x^{(a)}+\frac{1}{2}\Bigr)+\mathrm{sgn}(p_v^{(b)})\Bigl(n_x^{(b)}+\frac{1}{2}\Bigr)+i\Bigl(n_y^{(a)}+n_y^{(b)}+1\Bigr) \, , 
\end{equation}
which has no solution for any combination of positive integers $n_{x,y}^{(a,b)}, n'_{x,y}$. 
Above we have introduced the nonlinear ratios $\mathcal{R}_{I\times J}^{(n'_x,n'_y)}$. It is important to emphasize that these ratios are dependent on the normalization of the mode functions $\Phi^{(p_u,p_v,n_x,n_y)}$. If one were to choose a different normalization for the modes $\Phi^\lambda$, the ratios are also transformed,  
\begin{equation}\label{normalization_transformation}
    \tilde{\Phi}^\lambda =  \alpha_\lambda \Phi^\lambda \, \implies \tilde{\mathcal{R}}^{(n'_x,n'_y)}_{I\times J} = \frac{\alpha_{\lambda_I}\alpha_{\lambda_J}}{\alpha_{\lambda'}}\mathcal{R}_{I\times J}^{(n'_x,n'_y)} \, .
\end{equation}
However, despite this arbitrariness they still encode physically meaningful information. For example, consider a single parent mode $\dot{\Phi}_{a}=\dot{\Phi}_{b}=\mathcal{A}\Phi^{(p_v,n_x,n_y)}$ that does not vanish at the central geodesic $\gamma$ (the LR from the Penrose limit perspective). Then, the quantity $\ddot{\Phi}_{aa}\vert_{\gamma}/\left(\dot{\Phi}_{a}\vert_{\gamma}\right)^{2}$ which is independent of any normalisation choice, is given by
\begin{equation}\label{eqn:invariantR}
   \frac{\ddot{\Phi}_{aa}\vert_{\gamma}}{\left(\dot{\Phi}_{a}\vert_{\gamma}\right)^{2}}= \sum_{n_{x}',n_{y}'}\mathcal{R}_{a\times a}^{(2n'_x,2n'_y)}\, ,
\end{equation}
as follows from \eqref{scalar_ratios} (notice self-couplings only connect to even modes). We note that this can be extended straightforwardly to more general parent mode configurations. 

Further care is needed whenever $p'_v=0$. This can occur whenever $p_v^{(a)}=0$ or $p_v^{(b)}=0$, or whenever $p_v^{(a)}=-p_v^{(b)}\equiv p_v$. The first case is trivial, and never appears by our choice of linear modes. The second case is potentially interesting, as it corresponds to two oscillatory modes, whose non-linear combination is a non-oscillatory mode. This is clear since the source term does not depend on $v$, and is a purely damped contribution in the $u$ direction
\begin{equation}
    \begin{aligned}
        \mathcal{S}_{ab} =& e^{-2\Omega(n_y^{(a)}+n_y^{(b)}+1)u} e^{-|p_v|\Omega x^2}H_{n_x^{(a)}}\left(\sqrt{|p_v|\Omega}x\right)H_{n_x^{(b)}}\left(\sqrt{|p_v|\Omega}x\right)\\
        &\hspace{3.5cm}\times H_{n_y^{(a)}}\left(\sqrt{ip_v\Omega}y\right)H_{n_y^{(b)}}\left(\sqrt{-ip_v\Omega}y\right)  \, .
    \end{aligned}
\end{equation}
In this sense, these modes could be related to a nonlinear memory effect~\cite{Mitman:2024uss, Mitman:2025hgy}. If the source term does not depend on $v$, then the particular solution to the sourced wave equation does not depend on $v$ either. While the wave equation is readily solved, the solutions take on a very different character than the quadratic quasinormal modes, which are our main interest in the remainder. Moreover, they can take us out of the Penrose limit approximation (contribute at a higher eikonal order~\cite{Isaacson:1968hbi,Isaacson:1968zza}) and are best understood as renormalizing the background spacetime within a class of spacetimes which retain $\partial_{v}$ as an isometry. We leave a more complete discussion of these contributions to future work. In addition, the existence of such perturbations indicates that there are likely driven quasi-resonances at third-order in perturbation theory, which would also be interesting to investigate further. 

%%%%%%%%%%%%%%%%%%%%%%%%%%%%%%%%%%%%%%%%%%%%%%%%%%%%%%%%%%%%%%%%%%%%%%%
\subsection{Gravitational QQNMs}
%%%%%%%%%%%%%%%%%%%%%%%%%%%%%%%%%%%%%%%%%%%%%%%%%%%%%%%%%%%%%%%%%%%%%%%

The gravitational case shares many similarities with the scalar toy model discussed above. In precise terms we consider a linear fluctuation of $\dot{\Psi}_0$, which is the superposition of two distinct modes
\begin{equation}\label{eq:linear_psi_0}
    \dot{\Psi}_0 = \mathcal{A}_{a} \Phi^{(p^{(a)}_v,n^{(a)}_x,n^{(a)}_y)}+\mathcal{A}_{b} \Phi^{(p^{(b)}_v,n^{(b)}_x,n^{(b)}_y)} \, .
\end{equation}
Whenever $p_v^{(a/b)}\neq 0$, such perturbation is generated by the Hertz potential
\begin{equation}
    \Psi_H \equiv \Psi_H^{(a)} + \Psi_H^{(b)} =  -\frac{2}{\left(p_v^{(a)}\right)^4}\bar{\mathcal{A}}_{a} \bar{\Phi}^{(p^{(a)}_v,n^{(a)}_x,n^{(a)}_y)}-\frac{2}{\left(p_v^{(b)}\right)^4}\bar{\mathcal{A}}_{b} \bar{\Phi}^{(p^{(b)}_v,n^{(b)}_x,n^{(b)}_y)} \, , 
\end{equation}
by virtue of Eq.~\eqref{eq:Weylperts}.

The second order Teukolsky equation for $\ddot{\Psi}_0$ is, in terms of the Hertz potential
\begin{equation}
    \begin{aligned}
        \square \ddot{\Psi}_0 = 2(\mathcal{S}_+ + \mathcal{S}_-) \, , 
    \end{aligned}
\end{equation}
with $\mathcal{S}_\pm $ defined in Eqs.~\eqref{sourcePlus}-\eqref{sourceMin}, and the factor of $2$ appears since $\mathcal{O}_0 = \thop\tho-\edtp\edt=\frac{1}{2}\square$. Now, writing $\Psi_H^{(I)} = e^{-ip_v^{(I)}v}\psi^{(I)}_H$, and making use of the fact that $\tho\Psi_H^{(I)} = -ip_v^{(I)} \Psi_H^{(I)}$, with $I=a,b$, we can rewrite the equation as  
\begin{equation}\label{gravitational_source_expanded}
    \begin{aligned}
         \square &\ddot{\Psi}_0 \, = \sum_{I,J=a,b}\Bigl(\mathcal{S}_{IJ}^+ + \mathcal{S}_{IJ}^-\Bigr) \, , \\
        & \mathcal{S}_{IJ}^+ = -\frac{1}{2}e^{i(p_v^{(I)}+p_v^{(J)})v}\Biggl[-2\Bigl((p_v^{(I)})^5p_v^{(J)}+4(p_v^{(I)})^4(p_v^{(J)})^2 + 3(p_v^{(I)})^2(p_v^{(J)})^2\Bigr) \edt'\bar{\psi}_H^{(I)} \edt'\bar{\psi}_H^{(J)}\\
        &+\Bigl((p_v^{(I)})^6 + 4 (p_v^{(I)})^5p_v^{(J)}+6(p_v^{(I)})^4(p_v^{(J)})^2+4(p_v^{(I)})^3(p_v^{(J)})^3 + (p_v^{(I)})^2(p_v^{(J)})^4\Bigr)\bar{\psi}_H^{(I)} \edt'^2\bar{\psi}_H^{(J)}\Biggr] \\
        &+ \left(I \leftrightarrow J\right) \, ,  \\
        & \mathcal{S}_{IJ}^- = -(p_v^{(I)})^4e^{i(p_v^{(I)}-p_v^{(J)})v}\Bigl[(p_v^{(I)})^2\bar{\psi}_H^{(I)}\edt^2\psi_H^{(J)} + (p_v^{(J)})^2\psi_H^{(J)}\edt^2\bar{\psi}_H^{(I)} +2p_v^{(I)}p_v^{(J)} \edt\bar{\psi}_H^{(I)} \edt\psi_H^{(J)}\Bigr]\, .
    \end{aligned}
\end{equation}
The source, hence, decomposes into two possible channels: an additive channel ($+$) and a difference channel ($-$). Thus, if the linear fluctuations are excited with two different frequencies, these drive a total of $6$ different QQNM frequencies: $3$ in the additive channel ($aa$, $bb$, and $ab=ba$ channels), and $3$ in the difference channel ($ab$, $ba$, and the ``zero-frequency'' channel with contributions from both $aa$ and $bb$).

Notice that the source term hence involves quadratic combination of the linear solutions, with up to second order spatial derivatives, due to the action of the $\edt$ operator. Building upon the results of the scalar toy model, we postulate that each of the source terms can be written uniquely as
\begin{equation}
    \begin{aligned}
        \mathcal{S}_{IJ}^+ = \sum_{n'_y=0}^{n_y^{(I)}+n_y^{(J)}+2}
        \sum_{n'_x=0}^\infty s_{IJ}^{(n'_x,n'_y)} \mathcal{A}_I \mathcal{A}_J\Phi^{(p'_u,p'_v,n'_x,n'_y)} \, , \qquad p'_u = p_u^{(I)}+p_u^{(J)} \, ,\qquad p'_v=p_v^{(I)}+p_v^{(J)} \, , \\ \mathcal{S}_{IJ}^- = \sum_{n'_y=0}^{n_y^{(I)}+n_y^{(J)}+2}
        \sum_{n'_x=0}^\infty t_{IJ}^{(n'_x,n'_y)} \mathcal{A}_I \bar{\mathcal{A}}_J\Phi^{(p'_u,p'_v,n'_x,n'_y)} \, , \qquad p'_u = p_u^{(I)}-p_u^{(J)} \, ,\qquad p'_v=p_v^{(I)}-p_v^{(J)} \, .
    \end{aligned}
\end{equation}
Each of the source coefficients $s_{IJ}^{(n'_x,n'_y)}, t_{IJ}^{(n'_x,n'_y)}$ can be written as a product of two integrals in the $x$ and $y$ direction, respectively. However, the integrals will involve more general polynomials than the product of two Hermite polynomials -- indeed notice that the maximum degree in the $y$ direction is now $n_y^{(I)}+n_y^{(J)}+2$, since there are up to two derivatives. Additional care is needed to deal with the Gaussian exponential factor which is present whenever $p_v^{(a)}$ and $p_v^{(b)}$ have the same (different) sign in $t_{ab}^{(n'_x,n'_y)}$ (respectively $s_{ab}^{(n'_x,n'_y)}$). 

The solution is then readily written in terms of the source coefficients as
\begin{equation}\label{gravitational_ratios_decomposed}
    \begin{aligned}
        \ddot{\Psi}_0 =& \sum_{I,J=a,b}\sum_{n'_y=0}^{n_y^{(I)}+n_y^{(J)}+2}\sum_{n'_x=0}^{\infty}\Biggl(\Rpp_{I\times J}^{(n'_x,n'_y)}\mathcal{A}_I\mathcal{A}_J \Phi^{(p_u^{(I)}+p_u^{(J)}, p_v^{(I)}+p_v^{(J)},n'_x,n'_y)} \\
        &\hspace{3.5cm}+ \Rpm_{I\times J}^{(n'_x,n'_y)}\mathcal{A}_I \bar{\mathcal{A}}_J\Phi^{(p_u^{(I)}-p_u^{(J)}, p_v^{(I)}-p_v^{(J)},n'_x,n'_y)} \Biggr) \, ,  \\ 
        \Rpp_{I\times J}^{(n'_x,n'_y)} =&  \frac{s_{IJ}^{(n'_x,n'_y)}\mathcal{N}^{-1}_{(p_v^{(I)}+p_V^{(J)}, n'_x, n'_y)}}{|p^{(I)}_v+p_v^{(J)}|\Omega(1+2n'_x)+i(p_v^{(I)}+p_v^{(J)})\Bigl[2i(p_u^{(I)}+p_u^{(J)})+\Omega(1+2n'_y)\Bigr]} \, , \\
        \Rpm_{I\times J}^{(n'_x,n'_y)} =&  \frac{t_{IJ}^{(n'_x,n'_y)}\mathcal{N}^{-1}_{(p_v^{(I)}-p_V^{(J)}, n'_x, n'_y)}}{|p^{(I)}_v-p_v^{(J)}|\Omega(1+2n'_x)+i(p_v^{(I)}-p_v^{(J)})\Bigl[2i(p_u^{(I)}-p_u^{(J)})+\Omega(1+2n'_y)\Bigr]} \, .
    \end{aligned}
\end{equation}
As in the black hole case, there are no contributions which oscillate with a frequency $\propto e^{-i(p_u^{(a)}+p_u^{(b)})u}$, i.e., modes that could be associated to a $--$ channel. Notice that the fact that there is no $--$ channel follows from Eq.~\eqref{Nonlinear_Teukolsky_Equation} and $\dot{\Psi}_i \sim \bar{\Psi}_H$ (Eq.~\eqref{eq:Weylperts}). Additionally we highlight that the denominator never vanishes, i.e., the system is never resonant up to second order in perturbation theory away from $p_v' = 0$ (see discussion above for the scalar case). 

Here it is worth to emphasize two aspects about the physical significance of our quadratic ratios. First, even though they transform non-trivially under a different choice of mode normalisation (see eq.\eqref{normalization_transformation}), they still measure invariantly the quadratic excitations of the curvature scalar $\Psi_0$ at the LR, i.e., $x=y=0$, as compared to the first order fluctuations. Just as in the scalar case (see discussion below eq.\eqref{normalization_transformation}), one can define
\begin{equation}\label{eqn:invariantRgrav}
    \ddot{\Psi}_0\lvert_{x=y=0} = \sum_{I,J=a,b} \Bigl(\Rpp_{I\times J}\left(\dot{\Psi}_{0,I}\lvert_{x=y=0}\right)\left(\dot{\Psi}_{0,J}\lvert_{x=y=0}\right) + \Rpm_{I\times J}\left(\dot{\Psi}_{0,I}\lvert_{x=y=0}\right)\left(\bar{\dot{\Psi}}_{0,J}\lvert_{x=y=0}\right)\, ,
\end{equation}
where the LR ratios are 
\begin{equation}
    \Rpp_{I\times J} = \sum_{n'_x,n'_y} \Rpp_{I\times J}^{(2n'_x,2n'_y)} \, , \qquad \Rpm_{I\times J} = \sum_{n'_x,n'_y} \Rpm_{I\times J}^{(2n'_x,2n'_y)} \, . 
\end{equation}
These expressions are a priori particular to our choice of normalization \eqref{normalization}, which was specifically taken for the (even-even) modes to be unit at $x = y = 0$. However, taking as an example the oscillating second order component related to a single parent mode, one has\footnote{Note that only parent modes with even overtone indices contribute to \eqref{eqn:invariantRgrav}, as these are non-vanishing on the LR, and similarly \eqref{eqn:invariantRratio} is only defined for such even mode, but similar ratios could be defined for spatial derivatives at $x = y = 0$ that would be sensitive to odd overtones.}
\begin{equation}\label{eqn:invariantRratio}
  \frac{\ddot{\Psi}^{++}_{0, aa}\lvert_{x=y=0}}{\left( \dot{\Psi}_{0, a}\lvert_{x=y=0}\right)^2}  = \Rpp_{a \times a} = \sum_{n'_x,n'_y} \Rpp_{a\times a}^{(2n'_x,2n'_y)} \, .
\end{equation}
The left-hand-side is a physically meaningful quantity, independent of any mode-normalisation choice, and is given merely by the sum of our quadratic ratios. In fact, this is what motivates our choice of normalisation. 
Had we chosen a different one, then the right-hand-side of \eqref{eqn:invariantRratio} would be an artificially-weighted sum of ratios instead.

The second aspect to emphasize is the behaviour of these ratios under gauge transformations. Our GPT gauge, introduced in Section \ref{sec:gauge}, is adapted to a null frame that is parallelly-propagated along a twist-free, null geodesic congruence. This choice of ``inertial'' null frame ensures that the associated curvature scalars are physically meaningful. In addition, the quadratic curvature ratios are invariant under the residual gauge symmetry within the GPT gauge. To see this, we recall that at second order $\ddot{\Psi}_{0}$ transforms, in general, according to \eqref{eq:ddPsi0gauge}. However, in order to respect the GPT gauge, the gauge parameters must be fixed by the $v$-independent (but otherwise arbitrary) functions $\dot{\xi}_{v},\dot{C}_{I},\dot{a},\dot{\theta}$, via equations~\eqref{eq:parametersgeodesicconditions} and~\eqref{eq:parameterstransverscond}. If at linear level one has two excited QNMs, so that $\dot{\Psi}_{0}$ is of the form 
\begin{equation}
    \dot{\Psi}_{0}= \mathcal{A}_a \Phi^{(p_v^{(a)},n_x^{(a)},n_y^{(a)})} + \mathcal{A}_b \Phi^{(p_v^{(b)},n_x^{(b)},n_y^{(b)})}=\mathcal{A}_{a} e^{i p_{v}^{(a)}v} \psi_{a}(u,x,y) + \mathcal{A}_{b} e^{i p_{v}^{(b)}v}  \psi_{b}(u,x,y)\, ,
\end{equation}
then none of the residual gauge transformations acting on $\ddot{\Psi}_{0}$ via \eqref{eq:ddPsi0gauge} can have a $v$-dependence of the forms $e^{i(p_v^{(a)}+p_v^{(b)})v}$ or $e^{\pm i(p_v^{(a)}-p_v^{(b)})v}$, so they leave the $++$ and $+-$ channels invariant. The only exception is the fine-tuned configuration $2 p_v^{(I)}= p_v^{(J)}$, since in that case the QQNM frequency $p_v^{(J)}-p_v^{(I)}$ in the $+-$ channel is precisely $p_v^{(I)}$, one of the parent-mode frequencies, so that particular QQNM would be affected by the residual gauge transformations.  

The computation of the ratios proceeds in the same fashion as for the scalar toy model. Our code, written in \texttt{Mathematica}, is available in~\cite{web:CoG}.

%%%%%%%%%%%%%%%%%%%%%%%%%%%%%%%%%%%%%%%%%%%%%%%%%%%%
\subsection{Selection rules}
%%%%%%%%%%%%%%%%%%%%%%%%%%%%%%%%%%%%%%%%%%%%%%%%%%%%

Before directly calculating the excitation of QQNMs we comment on certain \emph{selection rules} that determine which modes are excited for any given mode combination. Let us consider the quadratic combination of two modes, $(p_v^{(a)},n_x^{(a)},n_y^{(a)})\times (p_v^{(b)},n_x^{(b)},n_y^{(b)})$. We must distinguish two cases, depending on the relative sign between $p_v^{(a)}$ and $p_v^{(b)}$. As discussed in the previous section, the QQNMs (the source-driven solution to the second order Teukolsky equation) can be decomposed in eigenvalues of the wave operator~\eqref{eqn:Sscalarwavesol}, labelled by $(p'_u,p'_v,n'_x,n'_y)$. The allowed values for these labels are:
\begin{itemize}
    \item $\mathrm{sgn}(p_v^{(a)}) = \mathrm{sgn}(p_v^{(b)})$. In this case we have 
    \begin{equation}\label{eqn:selection_rules_samesign}
        \begin{aligned}
            &++\text{ : }\quad p'_u = p_u^{(a)}+p_u^{(b)}, \quad p'_v =p_v^{(a)}+p_v^{(b)} \, , \quad 0\leq n'_\bullet \leq n_\bullet^{(a)}+n_\bullet^{(b)} \, ,\qquad  \bullet=\{x,y\} \, , \\
            &+-\text{ : }\quad p'_u = p_u^{(a)}-p_u^{(b)}, \quad p'_v =p_v^{(a)}-p_v^{(b)} \, , \quad 0\leq n'_x <\infty \, , \quad 0\leq n'_y \leq n_y^{(a)}+n_y^{(b)} \, , \\
            &-+\text{ : }\quad p'_u = p_u^{(b)}-p_u^{(a)}, \quad p'_v =p_v^{(b)}-p_v^{(a)} \, , \quad 0\leq n'_x <\infty \, , \quad 0\leq n'_y \leq n_y^{(a)}+n_y^{(b)} \, ,
        \end{aligned}
    \end{equation}
    where the $-+$ channel is obtained by swapping $a\leftrightarrow b$ in the $+-$ channel. There is an additional selection rule related to the parity of the overtones, which applies in all channels,
    \begin{equation}\label{eqn:parity_selection_rule}
        n'_\bullet \mod 2 = n_\bullet^{(a)}+n_\bullet^{(b)}\mod 2 \, ,\qquad  \bullet=\{x,y\} \, ,
    \end{equation}
    which simply follows from the well-defined parity of the linear modes. 

    \item $\mathrm{sgn}(p_v^{(a)}) \neq \mathrm{sgn}(p_v^{(b)})$. In this case, the overtone selection rules for the $++$ and $+-$ channels flip:
    \begin{equation}\label{eqn:selection_rules_oppsign}
        \begin{aligned}
            &++\text{ : }\quad p'_u = p_u^{(a)}+p_u^{(b)}, \quad p'_v =p_v^{(a)}+p_v^{(b)} \, , \quad 0\leq n'_x <\infty \, , \quad 0\leq n'_y \leq n_y^{(a)}+n_y^{(b)} \, , \\
            &+-\text{ : }\quad p'_u = p_u^{(a)}-p_u^{(b)}, \quad p'_v =p_v^{(a)}-p_v^{(b)} \, , \quad 0\leq n'_\bullet \leq n_\bullet^{(a)}+n_\bullet^{(b)}   \, ,
        \end{aligned}
    \end{equation}
    where we don't write the $-+$ channel for brevity, and the parity selection rule~\eqref{eqn:parity_selection_rule} still applies. 
\end{itemize}

The selection rules for $p'_u,p'_v$ trivially follow from the structure of linear modes~\eqref{modes}. The parity rule, once again, simply follows from the fact that these are quadratic combinations of functions with well-defined parities. The only selection rules which are not trivial are those affecting the overtone excitation. 

Already in the scalar toy model~\ref{sec:scalarmodel} we emphasized that there are two distinct cases which could lead to the excitation of either a finite or an infinite number of $n'_x$ overtones, which we refer to respectively as ``finite (excitation) channels'' or ``infinite (excitation) channels''. This is related to whether the factor $\Delta$ in eq.~\eqref{eqn:toyintegrals} vanishes, or not. Whenever $\Delta \neq 0$, which occurs whenever $\mathrm{sgn}(p_v^{(a)})=\mathrm{sgn}(p_v^{(b)})$ in the $+-$ channel, or whenever $\mathrm{sgn}(p_v^{(a)})\neq\mathrm{sgn}(p_v^{(b)})$ in the $++$ channel, the absolute value in the $x$-dependent Gaussian exponential acts as an obstruction, and the decomposition of the source term in Hermite polynomials now requires and infinite amount of terms. This explains why $n'_x$ can be arbitrarily large in the $+-$ and $-+$ cases in~\eqref{eqn:selection_rules_samesign}, and in the $++$ case in~\eqref{eqn:selection_rules_oppsign}.

Focusing now on the finite excitation channels,
a naive counting of the polynomial powers gives that an overtone of mode number up to $n' = n^{(a)} + n^{(b)} + 2$ could be excited, where the $+2$ is a consequence of the spatial derivatives in~\eqref{sourcePlus}--\eqref{sourceMin}. On the other hand, we find that the maximal overtone number is, instead, $n' \leq n^{(a)} + n^{(b)}$.  A special case of this selection rule is that the coupling between fundamental modes does not excite overtones. This was already observed and explained on symmetry grounds in metric perturbation approach of Section~\ref{sec:metric}. 
Next, we extend this argument, in terms of the Einstein tensor, to the Weyl scalar and subsequently generalize it.

The argument we used to show that the second order metric sourced by fundamental modes was itself composed only of fundamental tensor modes made use of the following statements,
\begin{equation}\label{eqn:arginput}
    \pounds_{a_-}\ddot{G} = \pounds_{b_-}\ddot{G} = 0 \, , \qquad [\pounds_\xi, \nabla_\mu] = 0 \, , \quad \xi \text{ an isometry.}
\end{equation}
Before continuing, we should emphasize that we considered the case where $p^{(a)}_v, p^{(b)}_v >0$ and complex metric perturbations or, equivalently, just the $++$ channel. If the signs of both $p_v$ are flipped, the role of $a_+$ and $a_-$ is reversed, but the argument still holds. On the other hand, if one of the contributing linear modes is lowered by $a_-$ while the other is lowered by $a_+$, such as when $\mathrm{sgn}(p_v^{(a)})\neq \mathrm{sgn}(p_v^{(b)})$ in the $++$ channel, the argument is bound to fail, as expected given that this would correspond to an infinite excitation channel. For brevity, we will not continuously emphasize these different mode combinations. Rather we again just present the argument for the $++$ channel with  $p^{(a)}_v, p^{(b)}_v >0$, leaving implicit the analysis with $a_+ \leftrightarrow a_-$ for the other finite excitation channels.

First, we want to show that the $a_{-},b_-$ operators anhilate the quadratic $\ddot{\Psi}_0$~\footnote{Here we assume that we can always isolate the $++$ contribution in second-order quantities with well-defined $p_u,p_v$, such as $\ddot{\Psi}_0$.}. The Weyl tensor $\ddot{C}_{abcd}$ is so anhilated by the same argument that we applied to the Einstein tensor. Therefore the only remaining question is to establish how do $\pounds_{a_-}$ and $\pounds_{b_-}$ act on the background and perturbed frames. We find, by a direct computation
\begin{align}\label{eqn:Lieframe}
    \pounds_{a_{\pm}} \ell =& 	\pounds_{b_{\pm}} \ell  = 0\, , \quad 	\pounds_{a_{\pm}} m = \pm \frac{-i}{2} e^{\pm i \Omega u} \ell \, ,   \quad 
     \pounds_{a_{\pm}} \bar{m} = \pm \frac{-i}{2} e^{\pm i \Omega u}\ell \, , \notag\\
     \pounds_{b_{\pm}} m =& \pm \frac{i}{2} e^{\pm \Lambda u} \ell \, , 
    \, ,   \quad 	\pounds_{b_{\pm}} \bar{m} = \pm \frac{-i}{2} e^{\pm \Lambda u} \ell \, .
\end{align}
Therefore $\pounds_{a_-}$ and $\pounds_{b_-}$ clearly annihilate $\ddot{C}_{abcd} \ell^{a}m^{b}\ell^{c}m^{d}$ by the symmetries of the Weyl tensor. On the other hand, noting $\dot{\ell} = 0$ in the GPT gauge we adopted and, using eq.~\eqref{eq:frame}
\begin{equation}
    \dot{C}_{\alpha \beta \gamma \delta} \ell^{\alpha}\dot{m}^{\beta}\ell^{\gamma}m^{\delta} =     -\frac{1}{2}\dot{C}_{\alpha \beta \gamma \delta} \ell^{\alpha}\bar{m}^{\beta}\ell^{\gamma}m^{\delta}\tho^2 \bar{\Psi}_H \, .
\end{equation}
This too is annihilated by $\pounds_{a_-}$ or $\pounds_{b_-}$ since both  $\tho^2 \bar{\Psi}_H$ and $\dot{C}_{\alpha \beta \gamma \delta} \ell^{\alpha}\bar{m}^{\beta}\ell^{\gamma}m^{\delta}$ are (again when restricted to the $p^{(a)}_v+p^{(b)}_v$ modes and using \eqref{eqn:Lieframe} as well as symmetries of the Weyl tensor). Finally, 
\begin{equation}
    C_{\alpha \beta \gamma \delta}\ddot{\ell}^{\alpha}m^{\beta}\ell^{\gamma}m^{\delta} =     C_{\alpha \beta \gamma \delta}\ell^{\alpha}\ddot{m}^{\beta}\ell^{\gamma}m^{\delta} =   C_{\alpha \beta \gamma \delta}\ell^{\alpha}\dot{m}^{\beta}\ell^{\gamma}\dot{m}^{\delta} = 0 \, ,
\end{equation}
because, for the background, only $\Psi_4 \neq 0$. We conclude that two parent modes with $n^{(a)}_y = n^{(b)}_y = 0$ can only couple to a mode of $n'_y = 0$ and, similarly, two parent modes with $n^{(a)}_x = n^{(b)}_x = 0$ in a finite excitation channel can only couple to a mode of $n'_x = 0$.

Running through the same logic for $n^{(a)}_y, n^{(b)}_y \neq 0$ and acting with more than $n^{(a)}_y + n^{(b)}_y$ derivatives, we similarly conclude that $n'_y \leq n^{(a)}_y + n^{(b)}_y$ and analogously for $n^{(a)}_x, n^{(b)}_x \neq 0$ in a finite excitation channel. This represents a rather remarkable cancellation. In fact, minor changes of the source terms~\eqref{sourcePlus}-\eqref{sourceMin} would break this selection rule. In fact, it is a prime example of a feature that is more readily understood from the metric perturbation approach of Section~\ref{sec:metric}, even though the master equation \eqref{Master2nd} is typically more convenient for calculations. Confirming that this selection rule is satisfied serves as a consistency check both in the derivation of the source term for the second order master equation, as well as for our calculation of the excitation of QQNMs. 

Let us now connect these selection rules with the selection rules for QQNMs for Kerr black holes. Recall that QNMs are labelled by three indices $(\ell, m, n)$, which are related to the plane wave representations via $p_v=m\Omega$, $n_x=\ell-|m|$, and $n_y=n$, where the QNM frequencies are $\omega_{\ell m n} = p_u$, with $p_u$ satisfying~\eqref{Vonshell}. Moreover the plane wave approximation requires $|\ell-|m||\ll \ell$. For QQNMs we have~\cite{Redondo-Yuste:2023seq, Khera:2024bjs} that $(\ell^{(a)},m^{(a)}, n^{(a)})\times (\ell^{(b)},m^{(b)},n^{(b)}) \to (\ell', m')$, with $m' = m^{(a)} \pm m^{(b)}$ (in the $++$ and $+-$ channels), and $\ell' \geq |m'|$ is arbitrary. The QQNM has frequency $\omega' = \omega^{(a)} \pm \omega^{(b)}$. Notice that QQNMs in Kerr cannot be easily decomposed in radial overtones, unlike the decomposition we achieve here. Our results hence suggest that such a decomposition may be possible, likely following the construction of a radial scalar product following~\cite{London:2023aeo, London:2023idh}, or the bilinear form of~\cite{Green:2022htq}. Leaving aside the issue of the overtone decomposition, our selection rules for $p'_u,p'_v$ are identical to the selection rules for $\omega', m'$, as expected. 

Interpreting the selection rules for the overtones is more subtle. First, as noted, no decomposition of the black hole QQNMs is usually made in the radial overtones, corresponding to $n_y$ in the Penrose limit. Therefore, these selection rules are likely an emergent phenomenon in the high-frequency limit, although it would be interesting to investigate this further. Similarly, at least for a Kerr black hole, with potentially unbounded $\ell'$, the selection rules for $n_x$ are likely emergent. 
On the other hand, for a Schwarzschild black hole, the situation of the $n_x$ modes is more transparent. In that case, $\ell'$ is additionally constrained to satisfy $|m'|, |l^{(a)} - l^{(b)}| \leq \ell' \leq l^{(a)} + l^{(b)}$. Therefore, it follows that at most $l^{(a)} + l^{(b)}-\max{\left(|m^{(a)} \pm m^{(b)}|, |l^{(a)} - l^{(b)}|\right)}+1$ modes are excited. In the ++ channel, by the relation $n_x=\ell-|m|$, this becomes exactly the $n_x^{(a)}+n_x^{(b)}+1$ modes that are excited in the finite excitation channels. On the other hand, for the +- channel, this becomes instead  $n_x^{(a)}+n_x^{(b)}+1+\left(|m^{(a)}|+|m^{(b)}|-\max{\left(|m^{(a)} - m^{(b)}|, |l^{(a)} - l^{(b)}|\right)}\right)$ number of modes that are excited with $|m^{(a)} - m^{(b)}| \leq \max{(m^{(a)},m^{(b)})}$ and $|l^{(a)} - l^{(b)}| \leq \max{(l^{(a)},l^{(b)})}$. Therefore, because by assumption $l^{(a)} \sim |m^{(a)}| \gg 1$ and $l^{(b)} \sim |m^{(b)}| \gg 1$, it follows that the number of modes excited is large, or in fact infinite in the geometrical optics approximation. This is indeed what is found for the corresponding infinite excitation channels.

%%%%%%%%%%%%%%%%%%%%%%%%%%%%
\subsection{QQNM Excitation}
%%%%%%%%%%%%%%%%%%%%%%%%%%%%

The simplicity of the homogeneous plane wave background allows us to provide analytical formulas for the ratios describing the QQNM excitation in quite general cases. We first examine the coupling of fundamental modes with different values of $p_v^{(a)}\neq p_v^{(b)}$, and then discuss the self-coupling of overtones. 

\subsubsection{Fundamental modes}

We begin by examining the coupling between fundamental modes, i.e., couplings of the form $(p_v^{(a)},0,0)\times (p_v^{(b)},0,0)$, where $p_v^{(a)} \neq p_v^{(b)}$, generically. We must distinguish two cases, depending on whether the signs of the eigenvalues in $v$ coincide or differ. Our results can be summarized in the following concise formulas. For the $++$ sector, we have 
\begin{equation}\label{ratio_fundamental_pp}
    \begin{aligned}
        \Rpp_{(p_v^{(a)},0,0)\times(p_v^{(b)},0,0)}^{(0,0)} =& \frac{i (p_v^{(a)}+p_v^{(b)})^4}{(p_v^{(a)})^3 (p_v^{(b)})^3} \, \qquad \left(\text{with}\quad p_v^{(a)}\geq p_v^{(b)} > 0 \right)\, , \\
        \Rpp_{(p_v^{(a)},0,0)\times(p_v^{(b)},0,0)}^{(2n,0)} =&  -\binom{-1/2}{n}\Biggl(\frac{i+2(1+i)n}{1+2(1+i)n}\Biggr)(p_v^{(a)})^{1-n}(p_v^{(b)})^{n-3}\Biggl(1+\frac{p_v^{(b)}}{p_v^{(a)}}\Biggr)^{9/2} \, ,\\
        &\left(\text{with}\quad p_v^{(a)}\geq |p_v^{(b)}| > 0 > p_v^{(b)}\right) \, , 
    \end{aligned}
\end{equation}
where $n$ is any non-negative integer. For the $+-$ sector instead we find
\begin{equation}\label{ratio_fundamental_pm}
    \begin{aligned}
        \Rpm_{(p_v^{(a)},0,0)\times(p_v^{(b)},0,0)}^{(2n,0)} =&  2^{1-2n}\binom{2n}{n}\Biggl(\frac{i+2(1+i)n}{1+2(1+i)n}\Biggr)(p_v^{(a)})^{1-n}(p_v^{(b)})^{n-3}\Biggl(1-\frac{p_v^{(b)}}{p_v^{(a)}}\Biggr)^{1/2} \, , \\
        &\left(\text{with}\quad p_v^{(a)}\geq p_v^{(b)} > 0 \right)\, , \\ 
        \Rpm_{(p_v^{(a)},0,0)\times(p_v^{(b)},0,0)}^{(0,0)} =&- \frac{2ip_v^{(a)}}{(p_v^{(b)})^3} \, , \qquad  \left(\text{with}\quad p_v^{(a)}>|p_v^{(b)}|>0>p_v^{(b)}\right) \, .
    \end{aligned}
\end{equation}
\begin{figure}
    \centering
    \includegraphics[width=0.95\linewidth]{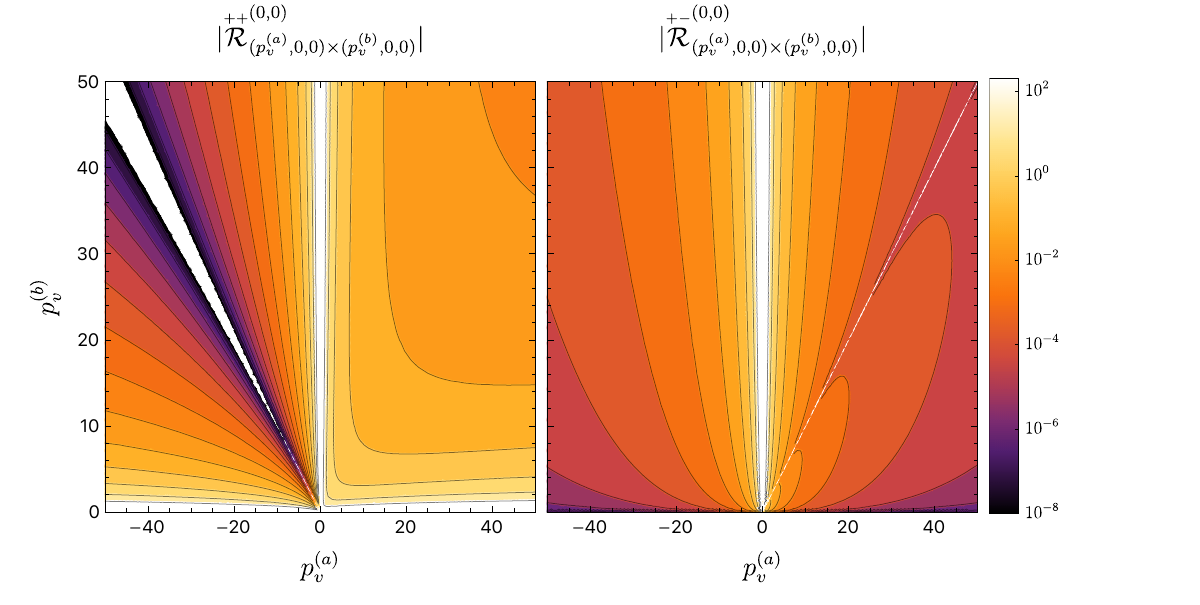}
    \caption{Absolute value of the ratios $\Rpp,\Rpm$ for the coupling between $(p_v^{(a)},0,0)\times(p_v^{(b)},0,0)\to (0,0)$, exciting the fundamental mode, in logarithmic scale. We exclude the cases where $p_v^{(a)}+p_v^{(b)}=0$ (for $++$) and $p_v^{(a)}-p_v^{(b)}=0$ (for $+-$), since those correspond to zero frequency modes, as well as the zero-frequency parent modes. }
    \label{fig:ratio_fundamental}
\end{figure}
The scaling for the ratios leading to the fundamental mode is shown in figure~\ref{fig:ratio_fundamental}. We find that generically these ratios decay as $p_v^{(a)}\sim |p_v^{(b)}| \to \infty$, but they can grow to be large whenever one of the two eigenvalues becomes large, while the other one remains finite. We examine the asymptotic behavior later.

Notice that as required by the selection rules, the overtone number is conserved in the $y$-direction. Therefore, the combination of fundamental modes only leads to the excitation of overtones in the $x$-direction within the infinite-excitation channels. As $n \gg 1$, the amplitude of the $n$--th overtone decays as 
\begin{equation}
    \left|\Rpp^{(2n,0)}_{(p_v^{(a)},0,0)\times(-|p_v^{(b)}|,0,0)}\right|\sim \left|\Rpm^{(2n,0)}_{(p_v^{(a)},0,0)\times(|p_v^{(b)}|,0,0)}\right| \sim n^{-1/2}\Biggl(\frac{\left|p_v^{(b)}\right|}{p_v^{(a)}}\Biggr)^n \, .
\end{equation}
In the high-frequency regime, i.e., taking $p_v^{(a)},|p_v^{(b)}|\sim p_v \to \infty$, but keeping $p_v^{(a)}-|p_v^{(b)}|=\delta$ fixed and finite, the ratios scale as
\begin{equation}\label{ratio_fundamental_asymptotics}
    \begin{aligned}
        \Rpp_{(p_v,0,0)\times(p_v-\delta,0,0)}^{(0,0)} =& \frac{16i}{p_v^2}\Bigl(1+\frac{\delta}{p_v} \Bigr) + \mathscr{O}(p_v^{-4}) \, , \\  
        \Rpm_{(p_v,0,0)\times(\delta-p_v,0,0)}^{(0,0)} =& -\frac{2i}{p_v^2}\Bigl(1+\frac{3\delta}{p_v}\Bigr)+\mathscr{O}(p_v^{-4}) \, ,  \\
        \Rpp_{(p_v,0,0)\times(\delta-p_v,0,0)}^{(2n,0)} =&  (-1)^{n}\binom{-1/2}{n}\Biggl(\frac{i+2(1+i)n}{1+2(1+i)n}\Biggr)\sqrt{\frac{\delta^{9}}{p_v^{13}}}+\mathscr{O}(p_v^{-15/2}) \, , \\
        \Rpm_{(p_v,0,0)\times(p_v-\delta,0,0)}^{(2n,0)} =& 2^{1-2n}\binom{2n}{n}\Biggl(\frac{i+2(1+i)n}{1+2(1+i)n}\Biggr)\sqrt{\frac{\delta}{p_v^5}} + \mathscr{O}(p_v^{-7/2}) \, ,
    \end{aligned}
\end{equation}
where $p_v\gg \delta > 0$. Thus, in the high-frequency limit all of these ratios vanish, albeit not at the same rate. Moreover, whenever we take $\delta \ll p_v$, the contributions to one of the channels -- the one responsible for the infinite excitation channels -- become highly suppressed. In the limit $\delta \to 0$, corresponding to $p_v^{(a)} \pm p_v^{(b)}=0$, the ratio in this channel vanishes identically. As discussed previously, additional care is needed to treat these modes, since they do not depend on $v$. In a way, they play a similar role as the metric-completion piece when reconstructing the second-order metric perturbation in Kerr and we leave their study to future work. 

Finally we note that if $p_v^{(a)} \to \infty$ keeping $p_v^{(b)}$ fixed, the ratios grow linearly with $p_v^{(a)}$. This is reminiscent of the result for a similar mode combination found for Schwarzschild black holes in the high-frequency regime in~\cite{Bucciotti:2025rxa}. This reinforces the expectation that the exploration of QQNMs in plane waves may shed light towards the structure and excitation of QQNMs in black holes, although further work is necessary to connect these two in a precise manner.

\subsubsection{Self-excitation}

\begin{figure}
    \centering
    \includegraphics[width=0.99\linewidth]{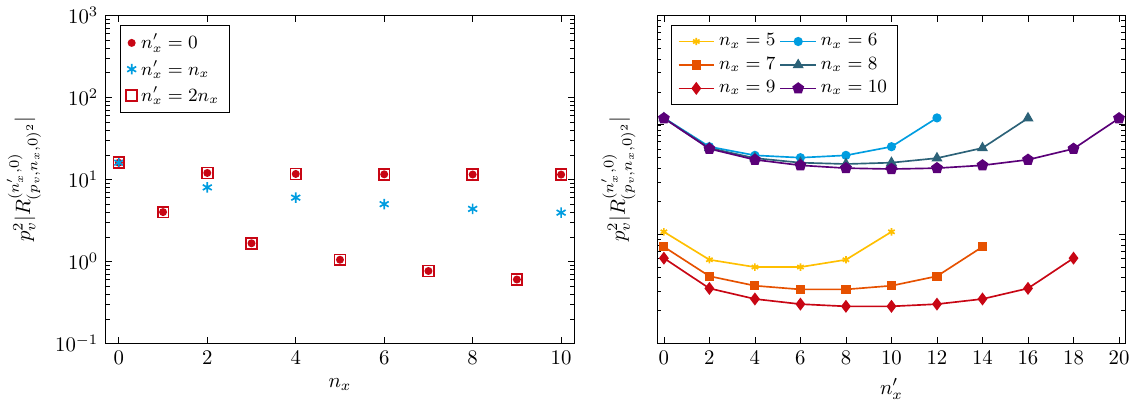}
    \caption{\textbf{Left.} Scaling of the self-coupling ratios $|\Rpp_{(p_v,n_x,0)^2}^{(n'_x,0)}|$ as a function of the \emph{parent} overtone $n_x$, for different values of $n'_x$. Red squares, blue stars, and red points correspond to $\alpha_{n_x},\beta_{n_x},\gamma_{n_x}$ in Eq.~\eqref{self_coupling_overtones_x} respectively. Remark $|\gamma_{n_x}| = \alpha_{n_x}$. \textbf{Right. } Scaling of the same self-coupling ratios as a function of the \emph{child} overtone index $n'_x$, for different values of the parent overtone number $n_x$. The same behavior is found when swapping the $x$ and $y$ directions, see also \eqref{eqn:crossing}.}
    \label{fig:spectra}
\end{figure}

Next we discuss the self-excitation of overtones, i.e., couplings with $p_v^{(a)}=p_v^{(b)}=p_v$,  $n_x^{(a)}=n_x^{(b)}=n_x$, and $n_y^{(a)}=n_y^{(b)}=n_y$. We focus on the $++$ sector and assume for simplicity that $p_v>0$. This means that we restrict to study the coupling of overtones within finite excitation channels. We emphasize that the code that we make available~\cite{web:CoG} can compute arbitrary couplings. We find 
\begin{equation}\label{self_coupling_overtones_x}
     \Rpp_{(p_v,n_x,0)^2}^{(2n_x,0)} = \frac{(-1)^{n_x}i\alpha_{n_x}}{p_v^2} \, , \qquad \Rpp_{(p_v,n_x,0)^2}^{(n_x,0)} = (1+(-1)^{n_x})\frac{\beta_{n_x}}{2p_v^2}\, , \qquad  \Rpp_{(p_v,n_x,0)^2}^{(0,0)} = \frac{\gamma_{n_x}}{p_v^2} \, , 
\end{equation}
where $\alpha_n,\beta_n,\gamma_n$ are coefficients whose behavior is illustrated  in figure~\ref{fig:spectra}. We notice that $|\gamma_{n_x}| = \alpha_{n_x}$, while $|\beta_{n_x}| < \alpha_{n_x}$ for all $n_x>0$. This hints that the excitation of different overtones occurs in a symmetric way -- the minimum and maximum overtone number possible are excited with the same amplitude up to a phase. Figure~\ref{fig:spectra} also shows that $\alpha_{n_x}, |\beta_{n_x}| \to \mathrm{const}.$ seemingly as $n_x\to\infty$ for odd and even $n_x$ separately. Notice also that these ratios scale as $p_v^{-2}$, which is the same scaling found for the fundamental modes. This scaling depends on the fact that we are normalizing the Weyl scalar $\dot{\Psi}_0$ of the modes at the LR, and different normalizations would lead to somewhat different scalings, as discussed in Eq.~\eqref{normalization_transformation}, although see again also the ratio \eqref{eqn:invariantRratio} for the invariant physical implication of this result.

For the excitation of overtones in the $y$ direction, we note the following crossing symmetry between $x \leftrightarrow y$:
\begin{equation}\label{eqn:crossing}
\Rpp_{(p_v,n_x,n_y)^2}^{(n'_x,n'_y)} = -\left(\Rpp_{(p_v,n_y,n_x)^2}^{(n'_y, n'_x)}\right)^*   \, .
\end{equation}
In fact, this holds even if the two parent modes have different $p_v$, as long as they are in the finite excitation channel. Clearly, the $x$ and $y$ directions behave differently in the infinite excitation channel. Regardless \eqref{eqn:crossing} implies that for the couplings in  \eqref{self_coupling_overtones_x}, we can restrict our discussion to self-excitation of overtones in the $x$-direction. These are analyzed in more detail in figure~\ref{fig:spectra}. The panels in this figure show the absolute values of the coefficients $\alpha_n,\beta_n,\gamma_n$ from \eqref{self_coupling_overtones_x} as a function of $n_x$ (left) and the excitation of different overtones, as given by their value of $n'_x$, due to the self-coupling $(p_v,n_x,0)\times (p_v,n_x,0) \to (n'_x,0)$ (right). We observe that the spectrum of overtone excitations is symmetric with respect to the parent overtone. It leads to a somewhat flat spectrum, slightly favoring the excitation of the most extreme overtone numbers $n'_x = 0$ and $n'_x = 2 n_x$. The odd and even parity overtones group themselves, yielding somewhat quantitatively different values. Again, this difference is sensitive to our choice of normalization of the modes. We have opted for a ``local'' choice of normalization which is physically based on the values the Weyl scalar (even) or its derivatives (odd) takes at the reference geodesic ($x=y=0$).

%%%%%%%%%%%%%%%%%%%%%%%%%%%%%%%%%%%%%%%%%%%
\section{Conclusions}\label{sec:conclusion}
%%%%%%%%%%%%%%%%%%%%%%%%%%%%%%%%%%%%%%%%%%%

Probing the non-linear regime of gravity is a challenging task. In this work we have taken one further step in this direction by studying the back-reaction of perturbations on plane wave spacetimes. Plane wave geometries are not only interesting in their own right, they also appear ubiquitously as approximate geometries near null geodesics. Specifically, near equatorial light-rings of Kerr black holes, a special class of Lorentzian symmetric (Cahen-Wallach) plane waves emerge, which capture an asymptotic branch of quasinormal modes. The non-linearities around such plane waves can thus be interpreted in terms of quadratic quasinormal modes.  

We have first presented a GHP-framework to study second-order perturbations suitable for general plane wave spacetimes. In doing so we introduce a novel set of gauge and frame-fixing conditions, built upon the geometry of plane waves, which we dub geodesic, parallel and transverse (GPT) gauge. We reduce the problem of studying second-order perturbations to solving a sourced master wave equation, where the source is written in terms of a Hertz potential generating the first-order metric perturbations. Along the way, we also derive Teukolsky-Starobinsky identities for generic plane-wave spacetimes. 

Additionally, we present a complementary framework built upon metric perturbations for the case of plane waves with additional Killing vectors. While we expect this approach to work for the entire class of  homogeneous plane waves, we work it out explicitly for the previously mentioned Lorentzian symmetric plane waves. Here, the algebraic structure of the plane waves reduces to that of simple and inverse harmonic oscillators, and the Killing vectors can be used to generate tensor harmonics, which in turn reduce Einstein's equations to mere algebraic equations. Moreover, such an approach based on metric perturbations, evades entirely the difficulty of reconstructing the metric from curvature perturbations.

Finally, we use our perturbation frameworks to define and compute the excitation of quadratic quasinormal modes, and measure their excitation in the second-order fluctuations of the Weyl scalar $\Psi_0$. Our work here extends previous work~\cite{Kehagias:2025ntm, Perrone:2025zhy}, which came out while our work was already in progress,  by computing the excitation of QQNMs for arbitrary combinations of linear modes. The second-order fluctuations in $\Psi_0$ can always be written as a (possibly infinite) sum of eigenfunctions of the wave operator -- this allows us to uncover emergent selection rules in the overtone numbers of the quadratic quasinormal modes. We provide analytical formulas for some relevant mode couplings as well as providing a code in~\cite{web:CoG}, which can compute them more generally. 

The excitation of QQNMs presented here is motivated by and related to the high-frequency limit of QQNMs of rotating BHs. The linear modes of symmetric plane waves correspond to the high frequency limit of Kerr QNMs through the identification, for $\ell \sim |m|$: $p_v \propto m$, $n_x=\ell-m$, and $n_y=n$, where $(\ell,m,n)$ are the polar, azimuthal, and overtone numbers. This identification makes $p_u$ correspond to the Kerr QNM frequencies, up to a mode independent proportionality constant depending on the affine time used as $u$. Therefore, we can put in (approximate) correspondence the QQNMs excited in the plane wave spacetime to certain QQNMs in a Kerr black hole -- indeed, their frequencies would match. 

Our results capture the second order fluctuations of $\Psi_0$ evaluated exactly at the equatorial LR of the Kerr black hole. Moreover, as discussed in previous sections, our results depend on a gauge (and frame) which builds upon the geometric properties of homogeneous plane waves. Therefore, the main steps that should be taken to perform a precise match with the black hole quadratic quasinormal modes are: (i) matching the gauge and frame used in describing perturbations of Kerr to the GPT gauge presented here, and (ii) translating the ratio extracted at the lightring to the ratio extracted at future null infinity. We leave this for future work but note that Refs.~\cite{Perrone:2025zhy, Kehagias:2025ntm} have put forward proposals to partially carry out this matching. 

Another aspect of the second order perturbations which we leave to future work are zero-frequency modes, which appear in certain quadratic combinations. These zero-frequency modes likely cannot be interpreted as QQNMs of a black hole. Nevertheless, it would be interesting to better understand these contributions simply from the point of view of plane wave perturbation theory in its own right as well as in relation to the effective gravitational stress-energy tensor for the background that should be solved self-consistently with geometrical optics perturbations, as discussed already in~\cite{Isaacson:1968hbi,Isaacson:1968zza}.

In addition to the immediate extensions of the present work suggested above, there is significant potential to bring more of the existing literature on plane wave spacetimes, sometimes already containing higher order perturbative results, to bear on or reinterpret them in light of the non-linear ringdown. To be sure, such extensions would only represent a highly particular limit of the ringdown problem and, in many instances, it will only be a toy model. Yet, it may not be entirely unfair to say that the Cahen-Wallach plane wave stands to the ringdown as the de Sitter space does to the cosmic microwave background. Thus, we believe that it would be a worthwhile pursuit.    

\acknowledgments
We acknowledge support by VILLUM Foundation (grant no. VIL37766) and the DNRF Chair program (grant no. DNRF162) by the Danish National Research Foundation. 
The Center of Gravity is a Center of Excellence funded by the Danish National Research Foundation under grant No. 184.
This project has received funding from the European Union's Horizon 2020 research and innovation programme under the Marie Sklodowska-Curie grant agreement No 101007855 and No 101131233.
K.~F. is supported by the Heising-Simons Foundation ``Observational Signatures of Quantum Gravity'' collaboration grant 2021-2817, the U.S. Department of Energy, Office of Science, Office of High Energy Physics, under Award No. DE-SC0011632, and the Walter Burke Institute for Theoretical Physics. D.~P. is supported by NSF Grants No. AST-2307146, PHY-2513337, PHY-090003, and PHY-20043, by NASA Grant No. 21-ATP21-0010, by John Templeton Foundation Grant No. 62840, by the Simons Foundation, and by Italian Ministry of Foreign Affairs and International Cooperation Grant No. PGR01167.

\bibliographystyle{JHEP}
\bibliography{ref}
\clearpage 

\appendix 

\section{(Inverted) Harmonic oscillators}\label{app:SHOIHO}

Here we briefly review the simple harmonic oscilator (SHO) and inverted harmonic oscillator (IHO), in part to fix our notation and also because it will be useful for constructing and interpreting solutions of gravitational fluctuations (for the IHO we will follow~\cite{Subramanyan:2020fmx}).

\paragraph{SHO.} The Hamiltonian of the SHO is ($\hbar=1$)
\begin{equation}
    H=\frac{1}{2m}\left(-\partial_{x}^{2}+ m^{2} \omega^{2} x^{2}\right)=\omega\left(a_+ a_-+\frac{1}{2}\right)
\end{equation}
with $m,\omega>0$, and we introduced
\begin{equation}
a_-=(1/\sqrt{2m\omega})\left(m \omega x + \partial_{x}\right) \, , \quad a_+=(1/\sqrt{2m\omega})\left(m \omega x - \partial_{x}\right) \, ,
\end{equation}
which form the Heisenberg algebra. 
\begin{equation}
    [a_{-},a_{+}]=1\, , \quad \left[H,a_{-}\right]=-\omega a_{-}\, , \quad \left[H,a_{+}\right]=\omega a_{+}\,.
\end{equation}
The ground-state $\psi_0$ and energy eigenstates $\psi_n$ are defined by 
\begin{equation}
    a_{-} \left( \psi_{0} \right) =0\rightarrow \left\{\psi_{0}= \mathcal{A}_{0} e^{- \frac{m \omega}{2}x^{2}}\, ,\quad \psi_{n}={a_{+}}^{n}\left(\psi_{0}\right)=\mathcal{A}_{n}  e^{- \frac{m \omega}{2}x^{2}}H_{n}\left(\sqrt{m \omega} x\right)\right\}
\end{equation}
where $\mathcal{A}_{n}$ are constants and $H_{n}(x)$ are Hermite polynomials. The energy levels of these states are
\begin{equation}
    H\psi_{n}=E_{n}\psi_{n}\, , \quad E_{n}=\omega\left(n+\frac{1}{2}\right)\, , \quad n=0,1,2, ...
\end{equation}
\paragraph{IHO.} In this case the Hamiltonian is
\begin{equation}
   H=\frac{1}{2m}\left(-\partial_{x}^{2}- m^{2} \lambda^{2} x^{2}\right)=-\lambda\left(b_+ b_- + \frac{i}{2}\right)\, , 
\end{equation}
with $m,\lambda>0$. Now it leads to scattered states, instead of bounded ones, and we introduced $b_{\pm}=(1/\sqrt{2m\lambda})\left( m\lambda x\mp i\partial_{x}\right)$, which satisfy the modified Heisenberg algebra
\begin{equation}
    [b_{-},b_{+}]=i\, , \quad \left[H,b_{\pm}\right]=\mp i \lambda b_{\pm} \, .
\end{equation}
We do not have ground states in this case, but we can define a ``purely ingoing/outgoing fundamental state'' and their excitations by 
\begin{equation}
     b_{\mp} \left( \phi^{\pm}_{0} \right) =0\rightarrow \left\{\phi^{\pm}_{0}= \mathcal{B}^{\pm}_{0} e^{\pm i \frac{m \lambda}{2}x^{2}}\, ,\quad \phi^{\pm}_{n}={b_{\pm}}^{n}\left(\phi^{\pm}_{0}\right)=\mathcal{B}^{\pm}_{n} e^{\pm i \frac{m \lambda}{2}x^{2}} H^{\pm}_{n}\left(\sqrt{m \lambda} x\right)\right\}
\end{equation}
where $\mathcal{B}_{n}$ are constants and $H^{\pm}_{n}(x)=e^{\mp i x^{2}}\frac{d^{n}}{dx^{n}}e^{\pm i x^{2}}$. With this, the energy eigenstates satisfy
\begin{equation}
    H \phi^{\pm}_{n}=\tilde{E}^{\pm}_{n}\phi^{\pm}_{n}\, , \quad \tilde{E}^{\pm}_{n}=\mp i\lambda\left(n+\frac{1}{2}\right)\, , \quad n=0,1,2, ...
\end{equation}

\section{GHP formalism}\label{app:GHPsum}

The Geroch--Held--Penrose (GHP) formalism~\cite{Geroch:1973am} (see also~\cite{Stewart:1974uz,Aksteiner:2010rh,Aksteiner:2016pjt,Green:2022htq}) allows a compact and covariant translation of spinor equations into scalar ones. While the formalism is most naturally introduced through its close relation to spinor dyads (see~\cite{Geroch:1973am,Penrose:1985bww}), here we will take an approach directly based on tensorial methods. The starting point is the choice of two globally-defined null directions, $\ell^{a}$ and $n^{a}$ with $\ell^{a}\ell_{a}=n^{a}n_{a}=0$ and $\ell^{a}n_{a}=1$. Then, one considers the bundle of null frames $(\ell_{a},n_{a},m_{a},\bar{m}_{a})$ compatible with that choice, so $m_{a}m^{a}=m_{a}\ell^{a}=m_{a}n^{a}=0$ and $m_{a}\bar{m}^{a}=-1$. This has the structure of a principal bundle with group $\mathbb{C}_{\times}$ (complex numbers without the origin), which acts on the null frames by
\begin{equation}\label{gaugeaction}
    (\ell_{a},n_{a},m_{a}; \lambda)\mapsto  (\lambda \bar{\lambda}\ell_{a},\lambda^{-1}\bar{\lambda}^{-1}n_{a},\lambda\bar{\lambda}^{-1}m_{a}) \quad \forall \lambda\in\mathbb{C}_{\times}\,.
\end{equation}
In other words, the transformation \eqref{gaugeaction} maps null frames into null frames while preserving our choice of null directions $\ell^{a},n^{a}$. 
Two important symmetries of the GHP equations, the so-called \textit{priming and starring operations}, consist of the formal replacements  
\begin{align}\label{eq:primingstarring}
\text{Priming}\ ('):&\quad \ell^{a}\rightarrow n^{a}\, , \quad n^{a}\rightarrow\ell^{a}\, ,\quad m^{a}\rightarrow \bar{m}^{a}\,, \quad  \bar{m}^{a}\rightarrow m^{a}\, ,\\
\text{Starring}\ (^*):&\quad  \ell^{a}\rightarrow m^{a}\, , \quad n^{a}\rightarrow -\bar{m}^{a}\, ,\quad m^{a}\rightarrow -\ell^{a}\,, \quad  \bar{m}^{a}\rightarrow n^{a}\, .
\end{align}
The independent components of the spacetime covariant derivative $\nabla_{a}$ are encoded in the \textit{spin coefficients}, defined as
 \begin{equation}\label{eq:NPFrameSpinCoefficients}
    \begin{aligned}
            \kappa&=m^{a}\ell^{b}\nabla_{b}\ell_{a}\, , \quad \rho=m^{a}\bar{m}^{b}\nabla_{b}\ell_{a}\, , \quad \sigma=m^{a}m^{b}\nabla_{b}\ell_{a}\, ,\quad \tau=m^{a}n^{b}\nabla_{b}\ell_{a}\, , \\
              \beta&=\frac{1}{2}\left(n^{a}m^{b}\nabla_{b}\ell_{a}+m^{a}m^{b}\nabla_{b}\bar{m}_{a}\right)\, , \quad \epsilon=\frac{1}{2}\left(n^{a}\ell^{b}\nabla_{b}\ell_{a}-\bar{m}^{a}\ell^{b}\nabla_{b}m_{a}\right)\, , 
    \end{aligned}
\end{equation}
together with their primed versions (e.g. $\kappa'=\bar{m}^{a}n^{b}\nabla_{b}n_{a}$). In the following we restrict to the case of vacuum GR, where the spacetime curvature is entirely given by Weyl's tensor $C_{abcd}$. Its independent components are encoded in the five complex Weyl scalars, defined as (taking into account our sign conventions for the curvature) 
\begin{equation}\label{eq:NPFramePsi}
    \begin{aligned}
        &\Psi_{0}=-C_{abcd}\ell^{a}m^{b}\ell^{c}m^{d}\, , \quad \Psi_{1}=-C_{abcd}\ell^{a}m^{b}\ell^{c}n^{d}\, ,\quad \Psi_{2}=-C_{abcd}\ell^{a}m^{b}\bar{m}^{c}n^{d} \, , \\
        &\Psi_{3}=-C_{abcd}\ell^{a}n^{b}\bar{m}^{c}n^{d}\, ,\quad \Psi_{4}= -C_{abcd}\bar{m}^{a}n^{b}\bar{m}^{c}n^{d}\, .
    \end{aligned}
\end{equation}
The scalars introduced above can be classified according to their transformation properties under \eqref{gaugeaction}. We say that a quantity $\eta$ has weights $(p,q)$ if under \eqref{gaugeaction} it transforms as
\begin{equation}\label{eq:etatrans}
    \eta\mapsto\lambda^{p}\bar{\lambda}^{q}\eta\, ,
\end{equation}
and we write $\eta\overset{\circ}{=}(p,q)$. For example,
\begin{equation}
    \ell^{a}\overset{\circ}{=}(1,1)\, , \quad n^{a}\overset{\circ}{=}(-1,-1)\, , \quad m^{a}\overset{\circ}{=}(1,-1)\, , \quad \bar{m}^{a}\overset{\circ}{=}(-1,1)\, ,
\end{equation}
and consequently
\begin{equation}
        \begin{aligned}
            &\kappa\overset{\circ}{=}(3,1)\, ,\quad \sigma\overset{\circ}{=}(3,-1)\, ,\quad \rho\overset{\circ}{=}(1,1)\, ,\quad  \tau\overset{\circ}{=}(1,-1)\, ,\\
            &\Psi_{i}\overset{\circ}{=}(4-2i,0)\quad (i=0,...,4)\, . \\
        \end{aligned}
\end{equation}
The GHP weights of the complex conjugate, primed and starred versions of a quantity follow from the fact that if $\eta\overset{\circ}{=}(p,q)$ then $\bar{\eta}\overset{\circ}{=}(q,p)$, $\eta'\overset{\circ}{=}(-p,-q)$ and $\eta^{*}\overset{\circ}{=}(p,-q)$ (these rules can be readily checked for quantities constructed out of contractions of the null frame with a tensor). The spin coefficients $\epsilon,\beta$ and their primes do not transform as \eqref{eq:etatrans}, and we say they are not properly weighted. Instead, they lead to the notion of GHP covariant derivative $\Theta_{a}$. This acts within properly weighted quantities $\eta\overset{\circ}{=}(p,q)$ as 
\begin{equation}\label{eq:GHPcovD}
    \Theta_{a}\eta=\nabla_{a}\eta-p\omega_{a}\eta-q\bar{\omega}_{a}\eta\, ,
\end{equation}
where the GHP connection 1-form $\omega_{a}$ is defined as
\begin{equation}\label{eq:GHPConnection1F}
 \omega_{a}=-\epsilon'\ell_{a}+\epsilon n_{a}+\beta'm_{a}-\beta \bar{m}_{a}=\frac{1}{2}\left(n^{b}\nabla_{a}\ell_{b}+m^{b}\nabla_{a}\bar{m}_{b}\right)\, .    
\end{equation}
The fact that it transforms correctly as a connection under $\mathbb{C}_{\times}$, that is, 
\begin{equation}\label{eq:omegarule}
    \omega_{a}\mapsto \omega_{a}+\lambda^{-1}\nabla_{a}\lambda\, ,
\end{equation}
guarantees that if $\eta\overset{\circ}{=}(p,q)$ then $\Theta_{a}\eta\overset{\circ}{=}(p,q)$, so in particular we have $\Theta_{a}\overset{\circ}{=}(0,0)$. Its action is also defined on tensor-valued quantities that are properly weighted (e.g. $\Theta_{a}\left(C_{bcde}n^{e}\right)$), and it satisfies the property $(\Theta_{a}\eta)'=\Theta_{a}\eta'$. More formally, we say that properly weighted GHP quantities belong to the associated vector bundles $E_{p,q}$ of representations with weights $(p,q)$, and $\Theta_{a}$ is the covariant derivative on such vector bundles. Finally, the directional GHP covariant derivatives are defined as
\begin{equation}
    \tho=\ell^{a}\Theta_{a}\,, \ \ \thop=n^{a}\Theta_{a}\,, \ \ \eth=m^{a}\Theta_{a}\,, \ \ \eth'=\bar{m}^{a}\Theta_{a}\, .
\end{equation}
In particular, their action on GHP scalars $\eta\overset{\circ}{=}(p,q)$ reads
\begin{equation}
    \begin{aligned}
        \tho \eta=& \left(\ell^{a}\nabla_{a}-p\epsilon-q\bar{\epsilon}\right)\eta\, , \ \  \thop \eta= \left(n^{a}\nabla_{a}+p\epsilon'+q\bar{\epsilon}'\right)\eta\, ,\\ \eth \eta =&\left(m^{a}\nabla_{a}-p\beta+q\bar{\beta}'\right)\eta \, , \ \  \eth' \eta  =\left(\bar{m}^{a}\nabla_{a}+p\beta'-q\bar{\beta}\right)\eta \, .
    \end{aligned}
\end{equation}
With this, one can obtain the GHP projections of several geometric relations. The GHP directional derivatives of the null frame are 
\begin{equation}\label{eq:GHPDyadDerivs}
    \begin{aligned}
\tho \ell_a &= -\bar{\kappa} m_a - \kappa \bar{m}_a, &
\tho m_a &= -\bar{\tau}' \ell_a - \kappa n_a, \\
\thop \ell_a &= -\bar{\tau} m_a - \tau \bar{m}_a, &
\thop m_a &= -\bar{\kappa}' \ell_a - \tau n_a, \\
\eth \ell_a &= -\bar{\rho} m_a - \sigma \bar{m}_a, &
\eth m_a &= -\bar{\sigma}' \ell_a - \sigma n_a, \\
\eth' \ell_a &= -\bar{\sigma} m_a - \rho \bar{m}_a, &
\eth' m_a &= -\bar{\rho}' \ell_a - \rho n_a\, ,
\end{aligned}
\end{equation}
together with their primes and complex conjugates. The Ricci identities yield
\begin{equation}\label{eq:GHPEinstein}
    \begin{aligned}
        \eth \rho - \eth' \sigma &= (\rho - \bar{\rho}) \tau + (\bar{\rho}' - \rho') \kappa - \Psi_1, \\
\tho\rho - \eth' \kappa &= \rho^2 + \sigma \bar{\sigma} - \bar{\kappa} \tau - \kappa \tau', \\
\tho\sigma - \eth \kappa &= (\rho + \bar{\rho}) \sigma - (\tau + \bar{\tau}') \kappa + \Psi_0, \\
\tho\rho' - \eth \tau' &= \rho' \bar{\rho} + \sigma \sigma' - \tau' \bar{\tau}' - \kappa \kappa' - \Psi_2,
    \end{aligned}
\end{equation}
which, together with their primed and starred versions, are equivalent to the vacuum Einstein equations. The Bianchi identities yield 
\begin{equation}\label{eq:GHPBianchi}
\begin{aligned}
    (\tho- 4\rho)\Psi_1 - (\eth' - \tau')\Psi_0 &= -3\kappa \Psi_2, \\
(\tho- 3\rho)\Psi_2 - (\eth' - 2\tau')\Psi_1 &= \sigma' \Psi_0 - 2\kappa \Psi_3,
\end{aligned}
\end{equation}
together with their primed and starred versions, and finally one can also verify that the commutator between GHP derivatives acting on a scalar with weights $\eta\overset{\circ}{=}(p,q)$ is given by
\begin{equation}\label{eq:GHPCommutators}
    \begin{aligned}  
[\tho, \thop] \eta &= \left[ (\bar{\tau} - \tau')\eth + (\tau - \bar{\tau}')\eth' 
- p(\kappa \kappa' - \tau \tau' + \Psi_2) 
- q(\bar{\kappa} \bar{\kappa}' - \bar{\tau} \bar{\tau}' + \bar{\Psi}_2) \right] \eta,  \\
[\tho, \eth] \eta &= \left[ -\bar{\tau}' \tho- \kappa \thop + \bar{\rho} \eth + \sigma \eth' 
- p(\rho' \kappa - \tau' \sigma + \Psi_1) 
- q(\bar{\sigma}' \bar{\kappa} - \bar{\rho} \bar{\tau}') \right] \eta.
    \end{aligned}
\end{equation}
These are the basic GHP equations, which can be found in the original work~\cite{Geroch:1973am} and are implemented in the \texttt{SpinFrames} package of the \texttt{xAct} bundle of \texttt{Mathematica}, used in this work. 

%%%%%%%%%%%%%%%%%%%%%%%%%%%%
\section{Type $N$ metric reconstruction}\label{app:recmetN}

The starting point in order to recover a metric perturbation from a perturbation of one of the Weyl scalars satisfying a Teukolsky-like equation is the following identity amongst linear operators on the background \eqref{ppWave},
\begin{equation}\label{id}
    \mathcal{O}_{0}\left[\mathcal{T}_{0}(h_{ab})\right]=\mathcal{S}_{0}\left[G^{(1)} (h_{ab})\right]\, ,
\end{equation}
where $h_{ab}$ is any symmetric tensor field, $G^{(1)}[h_{ab}]$ is the linearized Einstein tensor, and the remaining operators and their adjoints (in the sense of~\cite{Wald:1978vm}) are given by
\begin{equation}\label{eqn:Waldoperators}
    \begin{aligned}
    \mathcal{T}_{0}(h_{ab})&=4\ell^{[b}m^{c]}\ell^{[d}m^{a]}\nabla_{a}\nabla_{b}h_{cd} & \quad \quad  \mathcal{T}^{\dagger a b}_{0}(\psi)&=-4\nabla_{d}\nabla_{c}\left(\ell^{[d}m^{a]}\ell^{[b}m^{c]}\psi\right) \\
    \mathcal{O}_{0}(\varphi)&=\frac{1}{2}\nabla^{a}\nabla_{a}\varphi & \quad \quad \mathcal{O}^{\dagger}_{0}(\psi)&=\frac{1}{2}\nabla^{a}\nabla_{a}\psi \\
    \mathcal{S}_{0}(h_{ab})&=4 \ell^{[b}m^{c]}\ell^{[d}m^{a]}\nabla_{a}\nabla_{b}h_{cd} & \quad \quad  \mathcal{S}_{0 \, ab}^{\dagger}(\psi)&=4\nabla^{d}\nabla^{c}\left(\ell_{[d}m_{a]}\ell_{[b}m_{c]}\psi\right) 
    \end{aligned}
\end{equation}
where $\varphi\GHPweight(4,0)$ and $\psi\GHPweight(-4,0)$ are GHP scalars. This identity, which is associated to the repeated PND $\ell$, has an analogue in vacuum type $D$ spaces~\cite{Wald:1978vm, Green:2019nam, Hollands:2024iqp}, where an additional identity exists associated to the second PND. Given that $G^{(1)}$ is self-adjoint, $G^{(1)\dagger}=G^{(1)}$, and using the property $\left(\mathcal{A}\mathcal{B}\right)^{\dagger}=\mathcal{B}^{\dagger}\mathcal{A}^{\dagger}$ of any two differential operators $\mathcal{A}$ and $\mathcal{B}$, one can write the identity
\begin{equation}
\mathcal{T}_{0}^{\dagger}\left[\mathcal{O}^{\dagger}_{0}\left(\psi\right)\right]=G^{(1)}\left[\mathcal{S}_{0}^{\dagger}\left(\psi\right)\right]\, .
\end{equation}
Then, it follows immediately that GHP scalars $\Psi_{H}\GHPweight(-4,0)$ with $\Psi_{H}\in \text{Ker}\mathcal{O}^{\dagger}_{0}$ (that is, $\mathcal{O}^{\dagger}_{0}\Psi_{H}=0$) generate solutions of the vacuum, linearised Einstein equations upon the action of $\mathcal{S}_{0}^{\dagger}$, that is,
\begin{equation}
    \mathcal{O}^{\dagger}_{0}\Psi_{H}=0\quad \Longrightarrow \quad G^{(1)}\left[h_{ab}\right]=0\, ,\ \  \text{with} \ \ h_{ab}=\mathcal{S}_{0 \, ab}^{\dagger}(\Psi_{H})+\mathrm{c.c.}\, .
\end{equation}

%%%%%%%%%%%%%%%%%%%%%%%%%%%%
\section{Scalar QQNMs}\label{app:scalartoy}

In Section~\ref{sec:scalarmodel}, we discussed a scalar toy model of the problem under consideration, governed by the non-linear equation~\eqref{eqn:eomcubicscalar}
\begin{equation}\label{eqn:app:eomcubicscalar}
\square \Phi + \Phi^2 = 0 \, .
\end{equation}
The purpose of the discussion in the main text was to introduce the quadratic quasinormal mode ratios and further illustrate that the only non-trivial step in their determination is the decomposition of the source term in terms of the modes $\Phi^{(p_v^{(a)},n_x^{(a)},n_y^{(a)})}$. However, as can be expected and as is an important motivation for the limit we are considering, this decomposition problem can nevertheless be solved analytically as we shall show in more detail here. 

To show that the integrals appearing in the scalar toy model~\eqref{eqn:toyintegrals} can be done analytically, all we need are the identities~\cite{Gradshteyn:1943cpj}
\begin{equation}\label{eqn:Hidentity1}
H_n(\gamma x) = \sum^{\left \lfloor{n/2}\right \rfloor }_{k = 0} \gamma^{n-2k}(\gamma^2-1)^k \binom{n}{2k} \frac{(2k)!}{k!}H_{n-2k}(x) \, ,
\end{equation}
and
\begin{equation}\label{eqn:Hidentity2}
H_{n}(z)H_{m}(z) = \sum^{\min(m,n)}_{r=0} 2^r r! \binom{m}{r}\binom{n}{r} H_{m+n-2r}(z) \, .
\end{equation}
Let us first consider the $\mathcal{I}_x$ integral
\begin{equation}
      \mathcal{I}_x = \int^{\infty}_{-\infty} \frac{\sqrt{|p'_v|\Omega} dx}{2^{n'_x}n'_x!\sqrt{\pi}}  H_{n_x^{(I)}}\left(\sqrt{|p_v^{(I)}|\Omega}x\right)H_{n_x^{(J)}}\left(\sqrt{|p_v^{(J)}|\Omega}x\right)H_{n'_x}\left(\sqrt{|p'_v|\Omega}x\right) e^{-(|p'_v|+\frac{\Delta}{2})\Omega x^2} \, , 
      \end{equation}
 with $p_v^{(a)},p_v^{(b)}>0$ or $p_v^{(a)},p_v^{(b)}<0$. This implies that $\Delta = |p_v^{(a)}| +  |p_v^{(b)}| - |p_v'| =0$ and $|p'_v|^{k+l}\left(\frac{|p^{(a)}_v|}{|p'_v|}-1\right)^k\left(\frac{|p^{(b)}_v|}{|p'_v|}-1\right)^l=\left(-|p^{(b)}_v|\right)^k\left(-|p^{(a)}_v|\right)^l$. We can use \eqref{eqn:Hidentity1} to obtain 
\begin{equation}\label{eqn:Hprodstep1}
\begin{aligned}
&H_{n^{(a)}_x}\left(\sqrt{|p^{(a)}_v| \Omega} x\right) H_{n^{(b)}_x}\left(\sqrt{|p^{(b)}_v| \Omega} x\right)  = \frac{\left|p^{(b)}_v\right|^{n^{(b)}_x/2}\left|p^{(a)}_v\right|^{n^{(a)}_x/2}}{|p'_v|^{(n^{(a)}_x+n^{(b)}_x)/2}}\sum^{\left \lfloor{n^{(a)}_x/2}\right \rfloor }_{k = 0}\sum^{\left \lfloor{n^{(b)}_x/2}\right \rfloor }_{l = 0}  \\
& \binom{n^{(a)}_x}{2k} \binom{n^{(b)}_x}{2l} \frac{(2k)!}{k!} \frac{(2l)!}{l!} \left(-\frac{|p^{(a)}_v|}{|p^{(b)}_v|}\right)^{l-k}    H_{n^{(a)}_x-2k}(\sqrt{|p'_v| \Omega} x) H_{n^{(b)}_x-2l}(\sqrt{|p'_v| \Omega} x) \, .
\end{aligned}
\end{equation}
Then, using the addition theorem~\eqref{eqn:Hidentity2}, and the orthogonality of Hermite polynomials, we find
\begin{equation}\label{eqn:Ix}
    \begin{aligned}
        \mathcal{I}_x =&  \frac{n^{(a)}_x! n^{(b)}_x!}{n'_x!}\frac{r^{\frac{n^{(b)}_x-n'_x}{2}}}{(1+r)^{\frac{n^{(b)}_x+n^{(a)}_x}{2}}}\frac{(-1)^{\frac{n^{(a)}_x + n^{(b)}_x-n_x'}{2}}}{\Gamma\left(1+\frac{n^{(a)}_x + n^{(b)}_x-n_x'}{2}\right)} \cK_{n_x^{(a)}}\left(n_x^{(a)}+n_x^{(b)}-n'_x;n_x^{(a)}+n_x^{(b)},1+r\right) \, , \\   r =& \frac{|p^{(a)}_v|}{|p^{(b)}_v|} \, ,  \quad  n_x^{(a)}+n_x^{(b)}-n'_x = 2 N'_x \,  , \quad 0 \leq N'_x \leq  n_x^{(a)}+n_x^{(b)} \, ,
   \end{aligned}
\end{equation}
where $\cK_{n_x^{a}}$ is a Krawtchouk polynomial
\begin{equation}\label{eqn:Krawtchouk}
\cK_{k}\left(x;n,q\right)  = \sum^k_{j=0} (-1)^j (q-1)^{k-j} \binom{x}{j} \binom{n-x}{k-j} \, . 
\end{equation}
Based on a parity selection rule, it is clear that the daughter modes will all have either only even mode numbers if $n^{(a)}_x = n^{(b)}_x \mod 2$ or odd quantum numbers otherwise; the cases not covered by \eqref{eqn:Ix} are identically zero. For the modes allowed by this parity selection, $n_x^{a}+n_x^{b}-n'_x$ will in particular always be even. 
Let us again emphasize that we have simply used certain ways to rewrite polynomials, and the integrals \eqref{eqn:toyintegrals} represent a way to pick up coefficients of such rewritings. Therefore, especially for the unstable $y$-direction, they are not fundamental and we do not need to make more precise the contour $C$ or statements about completeness of the mode functions. As a result, the integral $\mathcal{I}_y$ can be derived in an identically to \eqref{eqn:Ix}.

Consider now instead $\Delta\neq 0$. We expect to still be able to give a fairly explicit expression, on account of the integral\footnote{Taking into account already the parity selection rule and using $(1+\frac{\Delta}{2 |p'_v|})>0$.}~\cite{Gradshteyn:1943cpj}\footnote{The relevant integral from~\cite{Gradshteyn:1943cpj} is 7.374 (5.) but remark there is a sign missing in that formula, in the second argument of the hypergeometric function ${}_2F_1$.}
\begin{equation}\label{eqn:expHintegral}
\begin{aligned}
 \frac{1}{2^{n'_x}n'_x!\sqrt{\pi}}&\int^{\infty}_{-\infty} dz \, H_{N}(z) H_{n_x'}(z)  e^{-\left(1+\frac{\Delta}{2 |p'_v|}\right) z^2} =  \left(1+\frac{\Delta}{2 |p'_v|} \right)^{\frac{-N-n_x'-1}{2}}\left(-\frac{\Delta}{2 |p'_v|} \right)^{\frac{N+n_x'}{2}}\\ &\times \frac{2^{N}\Gamma(\frac{N+n_x'+1}{2})}{n'_x!\sqrt{\pi}} {}_2F_1\left(-N,-n_x';\frac{1-N-n_x'}{2};\frac{|p'_v|+\frac{\Delta}{2}}{\Delta}\right)\, , \quad  N \equiv n_x' \mod 2\, .
 \end{aligned}
\end{equation}
The first step is still to use \eqref{eqn:Hidentity1} although now
\begin{equation}
    |p'_v|^{k+l}\left(\frac{|p^{(a)}_v|}{|p'_v|}-1\right)^k\left(\frac{|p^{(b)}_v|}{|p'_v|}-1\right)^l=\left(\Delta-|p^{(b)}_v|\right)^k\left(\Delta-|p^{(a)}_v|\right)^l
\end{equation}
such that
\begin{equation}\label{eqn:Hprodstep1}
\begin{aligned}
&H_{n^{(a)}_x}\left(\sqrt{|p^{(a)}_v| \Omega} x\right) H_{n^{(b)}_x}\left(\sqrt{|p^{(b)}_v| \Omega} x\right)  = \frac{\left|p^{(b)}_v\right|^{n^{(b)}_x/2}\left|p^{(a)}_v\right|^{n^{(a)}_x/2}}{|p'_v|^{(n^{(a)}_x+n^{(b)}_x)/2}}\\
&\times\sum^{\left \lfloor{n^{(a)}_x/2}\right \rfloor }_{k = 0}\sum^{\left \lfloor{n^{(b)}_x/2}\right \rfloor }_{l = 0}   \binom{n^{(a)}_x}{2k} \binom{n^{(b)}_x}{2l} \frac{(2k)!}{k!} \frac{(2l)!}{l!} \left(\frac{\Delta-|p^{(a)}_v|}{|p^{(b)}_v|}\right)^{l}\left(\frac{\Delta-|p^{(b)}_v|}{|p^{(a)}_v|}\right)^{k}   \\
&\hspace{3cm}\times H_{n^{(a)}_x-2k}(\sqrt{|p'_v| \Omega} x) H_{n^{(b)}_x-2l}(\sqrt{|p'_v| \Omega} x) \, .
\end{aligned}
\end{equation}
Using the addition theorem~\eqref{eqn:Hidentity2} and the integral \eqref{eqn:expHintegral}, we find an explicit albeit fairly uninsightful expression, as could have already been expected from the simpler case \eqref{eqn:Ix}. Therefore, instead of presenting this expression, let us simplify to $n_x^{(a)} = n_x^{(b)} = 0$ and $p_v^{(b)} p_v^{(a)}<0$ (such that $\Delta \neq 0$), for which in the main text we also find an analytical expression in the gravitational case:
\begin{equation}
\mathcal{I}_x(n_x^{(a)} = n_x^{(b)} = 0)  =  
    \frac{2^{-n_x'} }{\left(\frac{n_x'}{2}\right)!} |p'_v|^{1/2} |p_v^{(>)}|^{-\frac{1+n_x'}{2}} |p_v^{(<)}|^{\frac{n_x'}{2}} \, \quad \quad p_v^{(b)} p_v^{(a)}<0\, , 
\end{equation}
where $p_v^{(>)}$ and $p_v^{(<)}$ denote whichever of $p_v^{(a)}$ or $p_v^{(b)}$ have respectively the largest and smallest absolute value.

\newpage

\section{Tensor harmonics}\label{app:tensorharmonics}

Entirely analogous to the vector modes in \eqref{eqn:Vmodes}, the tensor modes satisfying \eqref{eqn:hmodes} are given by
\begin{equation}\label{eqn:hmodesexpl}
\begin{aligned}
\tilde{h}^{(p_u,p_v,n_x,n_y, n n)}_{\mu \nu} &=  \mathcal{N}_{(p_v,n_x,n_y,n n)} H_{n_x}(\sqrt{|p_v| \Omega} x) H_{n_y}(\sqrt{i p_v \Lambda} y) \ell_{\mu} \ell_{\nu}\, ,  \\
%%%%%%%%%%%%%%%%%%%%%%%%%%%%%
\tilde{h}^{(p_u,p_v,n_x,n_y, x n)}_{\mu \nu} &=  \mathcal{N}_{(p_v,n_x,n_y, x n)} H_{n_x}(\sqrt{|p_v| \Omega} x) H_{n_y}(\sqrt{i p_v \Lambda} y) E^{(x)}_{(\mu}\ell_{\nu)}   \\ & - i \sqrt{\frac{\Omega}{4|p_v|}} \left(H_{n_x+1}(\sqrt{|p_v| \Omega} x) -2n_x  H_{n_x-1}(\sqrt{|p_v| \Omega} x)    \right) H_{n_y}(\sqrt{i p_v \Lambda} y) \ell_{\mu}\ell_{\nu}   \, ,  \\
%%%%%%%%%%%%%%%%%%%%%%%%%%%%%
\tilde{h}^{(p_u,p_v,n_x,n_y, y n)}_{\mu \nu} &= \mathcal{N}_{(p_v,n_x,n_y,y n)} H_{n_x}(\sqrt{|p_v| \Omega} x) H_{n_y}(\sqrt{i p_v \Lambda} y) E^{(y)}_{(\mu}\ell_{\nu)} \\ &  -i \sqrt{\frac{i\Lambda}{4p_v}}    H_{n_x}(\sqrt{|p_v| \Omega} x) \left(H_{n_y+1}(\sqrt{i p_v \Lambda} y) -2n_y  H_{n_y-1}(\sqrt{i p_v \Lambda} y)  \right)  \ell_{\mu}\ell_{\nu}  \, , \\
%%%%%%%%%%%%%%%%%%%%%%%%%%%%%%%%%%%
\tilde{h}^{(p_u,p_v,n_x,n_y, xx)}_{\mu \nu} &=  \mathcal{N}_{(p_v,n_x,n_y,xx)} H_{n_x}(\sqrt{|p_v| \Omega} x) H_{n_y}(\sqrt{i p_v \Lambda} y) E^{(x)}_{\mu}  E^{(x)}_{\nu}   \\  & -2i \sqrt{\frac{\Omega}{4|p_v|}}  \left(H_{n_x+1}(\sqrt{|p_v| \Omega} x) -2n_x  H_{n_x-1}(\sqrt{|p_v| \Omega} x)    \right) H_{n_y}(\sqrt{i p_v \Lambda} y)  E^{(x)}_{(\mu}\ell_{\nu)} \\  &- \Omega^2 x^2 H_{n_x}(\sqrt{|p_v| \Omega} x) H_{n_y}(\sqrt{i p_v \Lambda} y) \ell_{\mu}\ell_{\nu}      \, , \\
%%%%%%%%%%%%%%%%%%%%%%%%%%%%%%%%%%%%%%
\tilde{h}^{(p_u,p_v,n_x,n_y, xy)}_{\mu \nu} &=    \mathcal{N}_{(p_v,n_x,n_y,xy)} H_{n_x}(\sqrt{|p_v| \Omega} x) H_{n_y}(\sqrt{i p_v \Lambda} y) E^{(x)}_{(\mu}  E^{(y)}_{\nu)}   \\  & -i \sqrt{\frac{i\Lambda}{4p_v}}   H_{n_x}(\sqrt{|p_v| \Omega} x) \left(H_{n_y+1}(\sqrt{i p_v \Lambda} y)  -2n_y  H_{n_y-1}(\sqrt{i p_v \Lambda} y) \right) E^{(x)}_{(\mu}\ell_{\nu)}  \\  & -i \sqrt{\frac{\Omega}{4|p_v|}}  \left(  H_{n_x+1}(\sqrt{|p_v| \Omega} x) -2n_x  H_{n_x-1}(\sqrt{|p_v| \Omega} x)  \right) H_{n_y}(\sqrt{i p_v \Lambda} y)  E^{(y)}_{(\mu}\ell_{\nu)}  \\  &- \sqrt{\frac{i\Lambda}{4p_v}}    H_{n_x}(\sqrt{|p_v| \Omega} x)\left( H_{n_y+1}(\sqrt{i p_v \Lambda} y) -2n_y  H_{n_y-1}(\sqrt{i p_v \Lambda} y) \right) \\ &\times \sqrt{\frac{\Omega}{4|p_v|}}\left( H_{n_x+1}(\sqrt{|p_v| \Omega} x) -2n_x  H_{n_x-1}(\sqrt{|p_v| \Omega} x)  \right)\ell_{\mu}\ell_{\nu}  \, ,  \\
%%%%%%%%%%%%%%%%%%%%%%%%%%%%%%%%%
\tilde{h}^{(p_u,p_v,n_x,n_y, yy)}_{\mu \nu} &=   \mathcal{N}_{(p_v,n_x,n_y,yy)} H_{n_x}(\sqrt{|p_v| \Omega} x) H_{n_y}(\sqrt{i p_v \Lambda} y) E^{(y)}_{\mu}  E^{(y)}_{\nu}   \\  & -2i \sqrt{\frac{i\Lambda}{4p_v}}   H_{n_x}(\sqrt{|p_v| \Omega} x) \left( H_{n_y+1}(\sqrt{i p_v \Lambda} y) -2n_y  H_{n_y-1}(\sqrt{i p_v \Lambda} y) \right) E^{(y)}_{(\mu}\ell_{\nu)} \\  &+ \Lambda^2 y^2 H_{n_x}(\sqrt{|p_v| \Omega} x) H_{n_y}(\sqrt{i p_v \Lambda} y) \ell_{\mu}\ell_{\nu}        \, . 
\end{aligned}
\end{equation}
We do not display explicitly the pure gauge modes, which we define instead in terms of vector modes as 
\begin{equation}\label{eqn:hgauge}
	h^{(p_u,p_v,n_x,n_y, I\ell)}_{\mu \nu} = 2\nabla_{(\mu}V^{(p_u,p_v,n_x,n_y, I)}_{\nu)} \, .
\end{equation} 
Just as for the vector modes, these pure gauge modes are the only ones with non-vanishing components $\ell^{\mu}h_{\mu \nu}$ and it is thus convenient to impose a gauge where these vanish. We also leave open the choice of normalization $\mathcal{N}_{(p_v,n_x,n_y,IJ)}$.

When imposing the linearized Einstein equation, in addition to a relation between the mode numbers identical to those for the massless scalar and Maxwell field \eqref{Vonshell}, we find that for each set of spacetime mode numbers we are left with the following degrees of freedom
\begin{equation}\label{eqn:hEE}
    \begin{aligned}
        h^{(p_u,p_v,n_x,n_y)}_{\mu \nu} =& h_{+}\left( h^{(p_u,p_v,n_x,n_y,xx)}_{\mu \nu} -h^{(p_u,p_v,n_x,n_y,yy)}_{\mu \nu} \right. \\
        & \qquad \left. \, + \, \frac{(2 n_x +1 )\Omega-i(2 n_y +1 )\Lambda}{p_v}h^{(p_u,p_v,n_x,n_y,nn)}_{\mu \nu}\right) + 2 h_{\times } h^{(p_u,p_v,n_x,n_y,xy)}_{\mu \nu}\, ,
    \end{aligned}
\end{equation}
assuming $\mathcal{N}_{(p_v,n_x,n_y,nn)} = \mathcal{N}_{(p_v,n_x,n_y,xx)} = \mathcal{N}_{(p_v,n_x,n_y,yy)}$. We abbreviate the modes \eqref{eqn:hEE} as
\begin{equation}
\begin{aligned}
 h^{(+)}_{\mu \nu} &= h^{(xx)}_{\mu \nu}-h^{(yy)}_{\mu \nu}+\frac{(2n_x +1)\Omega-i(2n_y+1)\Lambda}{p_v}h^{(nn)}_{\mu \nu} \, , \\
  h^{(\times)}_{\mu \nu} &= 2 h^{(xy)}_{\mu \nu} \, ,
 \end{aligned}
\end{equation}
where for brevity we have kept implicit the labels $p_u, p_v, n_x$, and $n_y$. 

The modes \eqref{eqn:hEE} are generated by a scalar Hertz potential as in \eqref{eq:recmet2}
\begin{equation}\label{eqn:app:hfromHertz}
\begin{aligned}
h^{(p_u,p_v,n_x,n_y)}_{\mu \nu} &= \bar{h}\left( -\ell_{\mu}\ell_{\nu}\eth^{2}+ 2  \ell_{(\mu}m_{\nu)} \th\eth  - m_{\mu} m_{\nu} \th^2 \right)\Phi^{(p_u,p_v,n_x,n_y)}_{H}  \\
&+ h \left(-\ell_{\mu}\ell_{\nu}\eth'^{2} + 2 \ell_{(\mu}\bar{m}_{\nu)} \th\eth'  - \bar{m}_{\mu} \bar{m}_{\nu} \th^{2}\right)\Phi^{(p_u,p_v,n_x,n_y)}_{H}  \, .
\end{aligned}
\end{equation}
In order to recover \eqref{eqn:Aphys} identically with
\begin{equation}
\Phi^{(p_u,p_v,n_x,n_y)}_H  =  H_{n_x}\left(\sqrt{|p_v| \Omega} x\right) H_{n_y}\left(\sqrt{i p_v \Lambda} y\right) e^{i u p_u + i v p_v - |p_v| \Omega \frac{x^2}{2}-i p_v \Lambda \frac{y^2}{2}} \ ,
\end{equation}
and $\mathcal{N}_{(p_v,n_x,n_y,IJ)} = 1$, we need
\begin{equation} 
h_+ = h_{xx} = -h_{yy} = \frac{p_v h_{nn}}{(2 n_x +1 )\Omega-i(2 n_y +1 )\Lambda} = \frac{p_v^2}{2} (h + \bar{h})  \, , \, \, h_{\times} = \frac{h_{xy}}{2}=-\frac{ ip_v^2 }{2}(h-\bar{h}) \, ,
\end{equation}
where $h_{IJ}$ is again the coefficient of the mode $h^{(p_u,p_v,n_x,n_y,IJ)}_{\mu \nu}$ as defined in \eqref{eqn:hmodesexpl}.

The Weyl scalars $\Psi_0$ are particularly simple for the linearized metric perturbations on a plane wave background and closely related to the Hertz potentials. Specifically, the only non-zero Weyl scalars $\Psi_0$ associated to the tensor modes  \eqref{eqn:hmodesexpl} in the frame \eqref{NPframe} are associated to the on-shell degrees of freedom and are given by
\begin{equation}\label{eqn:Weylscalars}
\begin{aligned}
\Psi^{(xx)}_0 &= -\Psi^{(yy)}_0 = -i\Psi^{(xy)}_0 =-\frac{p_v^2}{4} H_{n_x}\left(\sqrt{|p_v| \Omega} x\right) H_{n_y}\left(\sqrt{i p_v \Lambda} y\right) e^{i u p_u +  i p_v( v- \Lambda \frac{y^2}{2}) - |p_v| \Omega \frac{x^2}{2}} \, .
\end{aligned}
\end{equation}
In particular, for a mode generated by a Hertz potential as from \eqref{eqn:app:hfromHertz} above, we have
\begin{equation}
\begin{aligned}
	\Psi_0[\Phi^{(p_u,p_v,n_x,n_y)}_H] &= -\frac{p_v^4}{2} h H_{n_x}\left(\sqrt{|p_v| \Omega} x\right) H_{n_y}\left(\sqrt{i p_v \Lambda} y\right) e^{i u p_u + i v p_v - |p_v| \Omega \frac{x^2}{2}-i p_v \Lambda \frac{y^2}{2}}\, , \\
	\bar{\Psi}_0[\Phi^{(p_u,p_v,n_x,n_y)}_H] &= -\frac{p_v^4}{2}\bar{h} H_{n_x}\left(\sqrt{|p_v| \Omega} x\right) H_{n_y}\left(\sqrt{i p_v \Lambda} y\right) e^{i u p_u + i v p_v - |p_v| \Omega \frac{x^2}{2}-i p_v \Lambda \frac{y^2}{2}}	\, .
	\end{aligned}
\end{equation}
Note that, because we did not consider real metric perturbations here, $\Psi_0$ and $\bar{\Psi}_0$ are not complex conjugates.

In the main text, we considered as an example calculation in the direct metric perturbation approach, a linear solution of the form
\begin{equation}\label{eqn:hEE00}
h_{\mu \nu} = \sum_{i\in \left\lbrace a,b \right\rbrace}h^{(i)}_{+}  h^{(p^{(i)}_u,p^{(i)}_v,0,0,+)}_{\mu \nu}+ h^{(i)}_{\times } h^{(p^{(i)}_u,p^{(i)}_v,0,0,\times)}_{\mu \nu}\, ,
\end{equation}
where, for the sake of the example, $p^{(a)}_v, p^{(b)}_v > 0$ and \eqref{Vonshell} is satisfied. A direct calculation led to the following second order Einstein tensor 
\begin{equation}\label{eqn:G2mode00}
\begin{aligned}
	& \ddot{G}_{\mu \nu}[h^{(a)}_{+},h^{(b)}_{+}] = \frac{i h^{(a)}_{+} h^{(b)}_{+}}{2(p_v^{(a)} + p_v^{(b)})} \left\lbrace \frac{\Omega \left(1-i\right)}{2 p_v^{(a)} p_v^{(b)}} \left\lbrack \phantom{\frac{\Omega}{\Omega}} \right. \right. \\ & \left(2 i (p_v^{(a)})^4  + (1+4i)(p_v^{(a)})^3p_v^{(b)}+6i (p_v^{(a)})^2(p_v^{(b)})^2 + (1+4i)(p_v^{(b)})^3p_v^{(a)}+2i (p_v^{(b)})^4  \right)h^{(xx)}_{\mu \nu}  \\ & \left. -\phantom{\frac{\Omega}{\Omega}}\left(2 i (p_v^{(a)})^4  - (1-4i)(p_v^{(a)})^3p_v^{(b)}+6i (p_v^{(a)})^2(p_v^{(b)})^2 - (1-4i)(p_v^{(b)})^3p_v^{(a)}+2i (p_v^{(b)})^4  \right)h^{(yy)}_{\mu \nu} \right\rbrack \\
	&+ \frac{\Omega^2}{p_v^{(a)} p_v^{(b)}(p_v^{(a)} + p_v^{(b)})}\left( 2(p_v^{(a)})^4 + (p_v^{(a)})^3p_v^{(b)} + 2 (p_v^{(a)})^2 (p_v^{(b)})^2 + p_v^{(a)} (p_v^{(b)})^3 +2 (p_v^{(b)})^4\right) h^{(nn)}_{\mu \nu}  \\ &+ \left. 
	\left( (p_v^{(a)})^2 +  p_v^{(a)} p_v^{(b)} +  (p_v^{(b)})^2\right) \left(h^{(\ell \ell)}_{\mu \nu} - \frac{2\Omega (1+i)}{p_v^{(a)} + p_v^{(b)}}h^{(n \ell)}_{\mu \nu} \right)\right\rbrace \, ,\\
	%%%%%%%%%%%%%%%%%%%%%%%%%%%%%%%%%%%%%%%%%%%%%%%%%%%%%%%%%%%%%%%%%%%%%%%%%%%%%%%%%%%%%
	& \ddot{G}_{\mu \nu}[h^{(a)}_{+},h^{(b)}_{\times}] = \frac{h^{(a)}_{+} h^{(b)}_{\times} (1+i)\Omega}{p^{(a)}_v} \left(2(p_v^{(a)})^2 +  p_v^{(a)} p_v^{(b)} +  (p_v^{(b)})^2\right)h^{(xy)}_{\mu \nu} \, , \\
	%%%%%%%%%%%%%%%%%%%%%%%%%%%%%%%%%%%%%%%%%%%%%%%%%%%%%%%%%%%%%%%%%%%%%%%%%%%%%%%%%%%%
	& \ddot{G}_{\mu \nu}[h^{(a)}_{\times},h^{(b)}_{\times}]  = \frac{i h^{(a)}_{\times} h^{(b)}_{\times}}{2(p_v^{(a)} + p_v^{(b)})} \\ &\times \left\lbrace \frac{\Omega \left(1-i\right)}{2}\left( (1-4i)(p_v^{(a)})^2-6i p_v^{(a)}p_v^{(b)} + (1-4i)(p_v^{(b)})^2\right)h^{(xx)}_{\mu \nu} \right.   \\ &+ \frac{\Omega \left(1-i\right)}{2}\left( (1+4i)(p_v^{(a)})^2+6i p_v^{(a)}p^{(b)}_v + (1+4i)(p_v^{(b)})^2  \right)h^{(yy)}_{\mu \nu} \\
	&- \frac{\Omega^2}{p_v^{(a)} + p_v^{(b)}}\left(7 (p_v^{(a)})^2 + 10 p_v^{(a)}p_v^{(b)} + 7 (p_v^{(b)})^2\right) h^{(nn)}_{\mu \nu} \\ &+ \left. 
	\left( (p_v^{(a)})^2 +  p_v^{(a)} p_v^{(b)} +  (p_v^{(b)})^2\right) \left(h^{(\ell \ell)}_{\mu \nu} - \frac{2\Omega (1+i)}{p_v^{(a)} + p_v^{(b)}}h^{(n \ell)}_{\mu \nu} \right)\right\rbrace \, ,\\
\end{aligned}
\end{equation}
where for definiteness we have fixed the normalizations $\mathcal{N}_{(p_v,n_x,n_y,IJ)}=1$ and for simplicity we distinguish the following three contributions from different polarizations: $h^{(a)}_{+} h^{(b)}_{+}$, $h^{(a)}_{+} h^{(b)}_{\times}$ (yielding also $(a) \leftrightarrow (b)$), and $h^{(a)}_{\times} h^{(b)}_{\times}$ as well as labeling the modes only with their tensor index, not including the common mode label which are implicitly $(p_u^{(a)}+p_u^{(b)},p_v^{(a)}+p_v^{(b)},0,0)$. In addition, we have used $\Omega = \Lambda$ and \eqref{Vonshell} for both $p^{(a)}_u$ and $p^{(b)}_u$ to ensure the zeroth and first order Einstein equations are satisfied. Finally, as discussed in the main text, \eqref{eqn:G2mode00} itself consists only of $n_x = n_y = 0$ modes. It is therefore readily decomposed and solved by \eqref{eqn:linesinteinmodesol}. 

Remark that, contrary to the component sourced by $\ddot{G}_{\mu \nu}[h^{(a)}_{+},h^{(b)}_{\times}]$, the quadratic modes sourced by $\ddot{G}_{\mu \nu}[h^{(a)}_{+},h^{(b)}_{+}]$ and $\ddot{G}_{\mu \nu}[h^{(a)}_{\times},h^{(b)}_{\times}]$ do \emph{not} simply take the naive form of a mode sourced by a linear Hertz potential. Therefore, quantifying these modes as simple scalar ratios of amplitudes between the linear and quadratic modes is more delicate and is most physically done on the level of the Weyl scalars. However, the proper definition of the Weyl scalars for the second order modes require a frame for the linearly perturbed metric, which is also naturally discussed in a higher-order description of the GHP-approach to the perturbations.

\end{document}